\documentclass[twocolumn]{article}
\usepackage{graphicx}
\usepackage{subfig}
\usepackage{vmargin}
\usepackage{enumitem}
\usepackage[font={small,it}]{caption}
\usepackage[export]{adjustbox}
\usepackage[utf8]{inputenc}
\usepackage[english]{babel}
\usepackage[
backend=bibtex,
style=nature,
citestyle=numeric-comp,
sorting=none
]{biblatex}
\usepackage{amsmath}
\usepackage[ruled,vlined]{algorithm2e}
\usepackage{xcolor}
\usepackage{csquotes}

\usepackage{xr-hyper}
\usepackage{hyperref}

\makeatletter
\newcommand*{\addFileDependency}[1]{
  \typeout{(#1)}
  \@addtofilelist{#1}
  \IfFileExists{#1}{}{\typeout{No file #1.}}
}
\makeatother

\newcommand*{\myexternaldocument}[1]{
    \externaldocument{#1}
    \addFileDependency{#1.tex}
    \addFileDependency{#1.aux}
}

\myexternaldocument{SupCovidMobility}

\setmarginsrb{1.5cm}{1.5cm}{1.5cm}{1.5cm}{0cm}{0cm}{1.5cm}{1.5cm}
 
\begin{document}
\twocolumn[
\begin{@twocolumnfalse}
\begin{center}
\Large{\sc{Effects of mobility restrictions during COVID19 in Italy}}
\vspace{0.3cm}

\small{Alex Smolyak$^{1,*}$, Giovanni Bonaccorsi$^{2}$, Andrea Flori$^{2}$, Fabio Pammolli$^{2,3}$  Shlomo Havlin$^1$}

\vspace{0.3cm}

\footnotesize{$^1$ Department of Physics, Bar-Ilan University, Ramat-Gan 52900, Israel; \\
\footnotesize{$^2$ Impact, Department of Management, Economics and Industrial Engineering, Politecnico di Milano}\\
\footnotesize{$^3$ SIT, Schaffhausen Institute of Technology, Schaffhausen}\\
                    $^*$ Correspondence to alex.smolyak@gmail.com\\}
\begin{abstract}
To reduce the spread and the effect of the COVID-19 global pandemic, non-pharmaceutical interventions (NPIs) have been adopted on multiple occasions by governments. In particular lockdown policies, i.e., generalized mobility restrictions, have been employed to fight the first wave of the pandemic. We analyze data reflecting mobility levels over time in Italy before, during and after the national lockdown, in order to assess some direct and indirect effects. By applying methodologies based on percolation and network science approaches, we find that the typical network characteristics, while very revealing, do not tell the whole story. In particular, the Italian mobility network during lockdown has been damaged much more than node- and edge-level metrics indicate. Additionally, many of the main Provinces of Italy are affected by the lockdown in a surprisingly similar fashion, despite their geographical and economic dissimilarity. Based on our findings we offer an approach to estimate unavailable high-resolution economic dimensions, such as real time Province-level GDP, based on easily measurable mobility information.\\
\end{abstract}

\end{center}
\end{@twocolumnfalse}
]
 \noindent
 
Among non-pharmaceutical interventions (NPIs), limitations to mobility of various degree have proven to be a successful strategy towards mitigating the spread of COVID-19 in populations across the world~\cite{schlosser2020covid,haug2020ranking,gross2020spatio}. In particular, starting with the first epidemic cluster in Wuhan, China, lockdown restrictions, i.e. full limitations of mobility on an entire territory, have been widely and effectively adopted~\cite{kraemer_2020_effect,chinazzi_2020_science}. 
 
Lockdowns are particularly effective since they incorporate a wide range of more targeted interventions, among which two are worth mentioning here - the limitation on large gatherings, and limitation on long range travel~\cite{gross2020spatio}. After the initial widespread adoption of lockdown restrictions, several studies~\cite{haug2020ranking,li2021temporal,dehning_inferring_2020,brauner_inferring_2021} have focused on disentangling the specific effect of each NPI, fostering the debate on the appropriateness and effectiveness of lockdowns as policy interventions.

In fact, in addition to the direct effect of aiding in the containment of COVID-19, one major side effect of the limitations was a severe economic impact induced by the lockdown that has been felt globally. While online commerce bloomed~\cite{oecd}, traditional brick-and-mortar businesses took a severe hit~\cite{Bartik17656}. As a consequence, governments have been criticized both for implementing lockdown too fast~\cite{scrollin,toi} and too slow~\cite{guardian}, with restrictions too harsh and too lax. 
 
Early studies on the economic effect of lockdown restrictions have highlighted their impact on consumption \cite{carvalho_tracking_nodate, cox_initial_nodate, martin2020socio}, on the supply chain \cite{guan_global_2020}, on inequality \cite{chang_mobility_2020, bonaccorsi2020economic} and on the overall economy \cite{chetty_economic_nodate}. The totality of the effect of lockdown restrictions on the economy, however, is still difficult to ascertain for several reasons. For one, the extent of the economic impact on various layers of the population is unknown at this point as the crisis is still ongoing. For another, different countries implement various economic stimuli to offset and ease the economic downturn, both on the personal and business levels. Lastly, with systems being as interconnected as they are, efficient trade-off evaluation of all possible effects (businesses closing, impact on education of all levels from primary school through higher education and research, strain on healthcare systems and the higher-order impact due to lack of doctors or equipment) is almost impossible in the short term.\\
Network Science~\cite{newman2018networks,barabasi2016network,cohen2010complex} had frequently been employed over the past two decades to investigate the behaviour of complex systems in multiple domains, ranging from ecosystems~\cite{bascompte2003nested,williams2000simple,gao2016universal} and biology~\cite{jeong2000large,barabasi2004network}, through engineering~\cite{buldyrev2010catastrophic}, urban traffic~\cite{li2015percolation} and computers~\cite{shao2015percolation} to economics and finance~\cite{gai2010contagion,battiston2012debtrank,buldyrev2020rise, smolyak2020mitigation}. Network tools have also proven useful in analyzing the spreading of the current pandemic in different contexts: from the direct estimation of the epidemiological evolution of the pandemic ~\cite{vespignani2020modelling,liu2020new}, 
to the development of new modeling approaches ~\cite{aleta_modelling_2020, block2020social}, 
to the analysis of mobility evolution~\cite{schlosser2020covid, pullano_evaluating_2020} and finally in relation to economic conditions and backlashes~\cite{bonaccorsi2020economic, chang_mobility_2020,spelta2020after}. 

Following this approach, in this work we develop a framework based on methodologies from network science and percolation theory \cite{barabasi2016network,Zhang8673} to explore the effects of the COVID-19 pandemic on the network structure of human mobility in Italy before and during the epidemic of 2020, from its very beginning to the second wave. We employ rich mobility data of Italy provided through Facebook's Data for Good program (available at \url{https://dataforgood.fb.com/tools/disease-prevention-maps/}) capturing movements of more than 3 millions of Italian individuals at a sub-municipal level and on a daily basis.

We use mobility data to construct the network of movements of individuals across territories, measuring how lockdown has affected the number and the intensity of connections among different areas of the country. In so doing, we aim to investigate the evolution of the Italian mobility network, measure the extent, beyond the immediately apparent one, of the mobility reduction after lockdown and draw conclusions regarding the existence and significance of an impact on the economy of lockdown policies.

Our hypothesis is that lockdown restrictions have increased the fragility of the Italian mobility network. What we want to further investigate is if this process has affected not only the network in general, which is plausible, but, more importantly, also the core of the network (i.e. the Giant  Connected  Component (GCC)). Furthermore we want to assess how long the disruptive effects of lockdown on the Italian mobility network have lasted even after restrictions were lifted. With respect to these objectives, the percolation methodology represents the ideal tool because it allows us to focus on the resilience profile of the network distinguishing between low impact and high impact effects. 

Finally, we investigate the economic consequences of mobility restrictions. Using official statistics about the geographical distribution of income per capita, we are able to observe a significant correlation between mobility and levels of income, suggesting that one can be used as a proxy of the other. Moreover, the percolation analysis allows us to investigate if the most resilient part of the network, after lockdown restrictions, is also the richest one. By observing that the average income per capita in the GCC has increased after lockdown, we are able to uncover that restrictions have heightened the economic segregation of the country, leaving poorer territories disconnected from the mobility network. 

While this work is not designed to make causal claims regarding the actual economic effect of lockdown restrictions, which, as we said, need to be disentangled from several concurrent phenomena, we believe that our analysis represents a useful tool to assess the potential economic impact of disruptive events such as the recent COVID-19 pandemic.

\section{Network structure before and under lockdown}\label{struct}

\begin{figure*}[!ht]
   \centering
   \hspace{0.00mm}
   \subfloat{\includegraphics[width=.65\textwidth,valign=t,left]{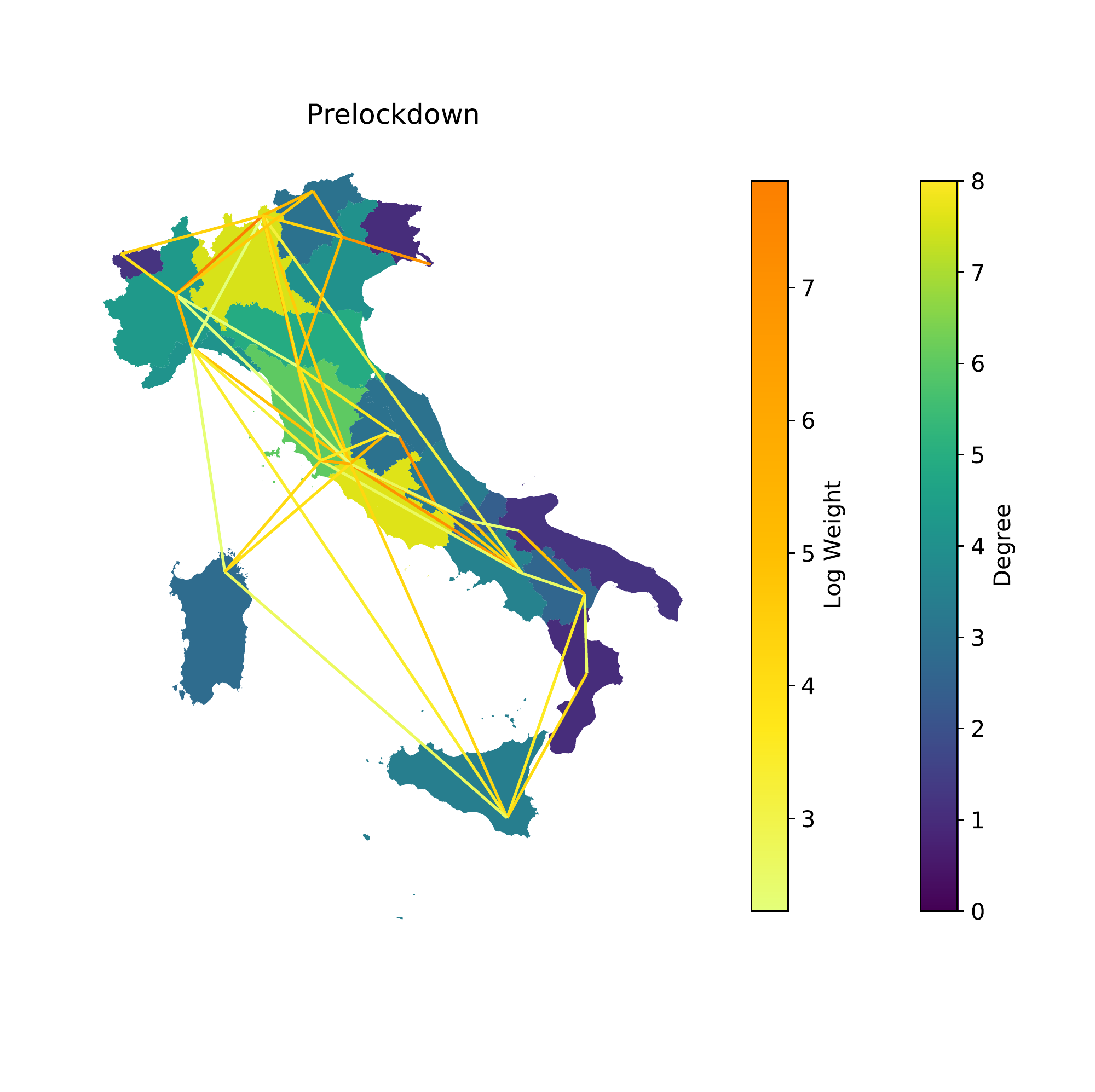}\label{1a}
   \begin{picture}(0,0)
\put(-500,-240){\subref{1a}}\quad
\end{picture}}
   \vspace{-10.00mm}
   \subfloat{\includegraphics[width=.65\textwidth,valign=t,left]{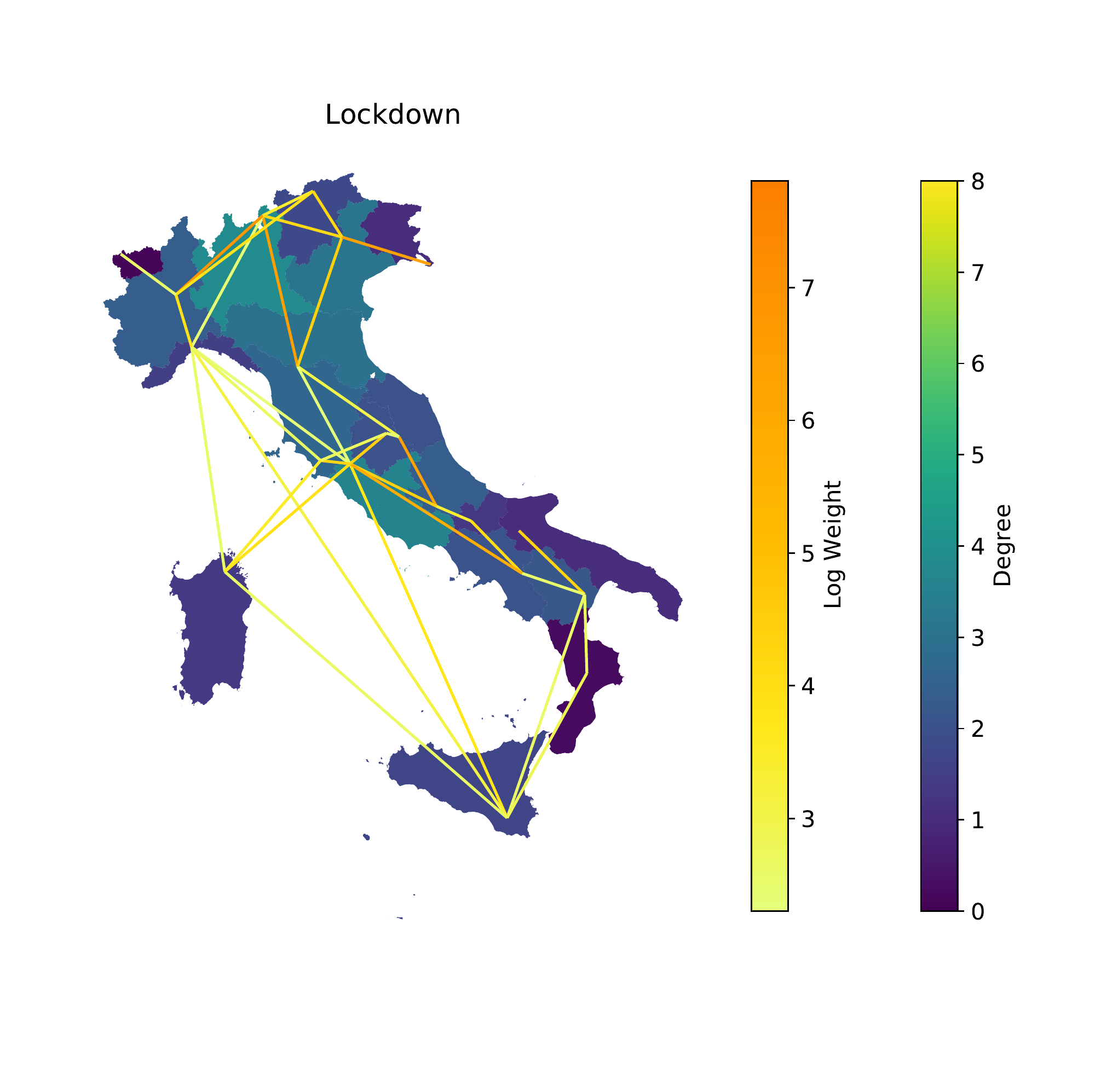}\label{1b}\begin{picture}(0,0)
\put(-500,-240){\subref{1b}}\quad
\end{picture}}
\vspace{-210mm}
   \subfloat{\includegraphics[width=.35\textwidth,valign=t,right]{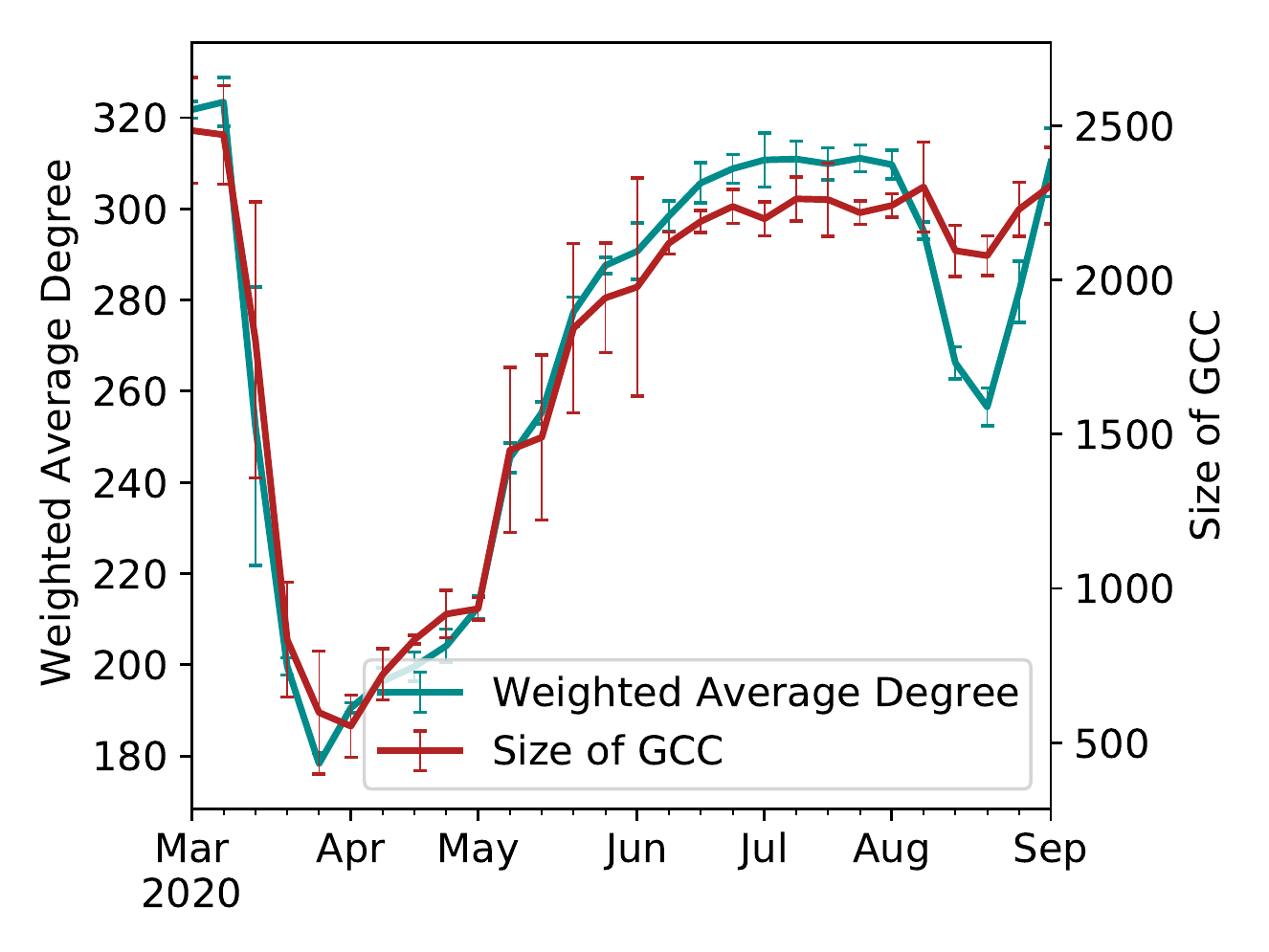}\label{1c}\begin{picture}(0,0)
\put(-180,-110){\subref{1c}}\quad
\end{picture}}
   \hspace{-25.00mm}
   \subfloat{\includegraphics[width=.35\textwidth,valign=t,right]{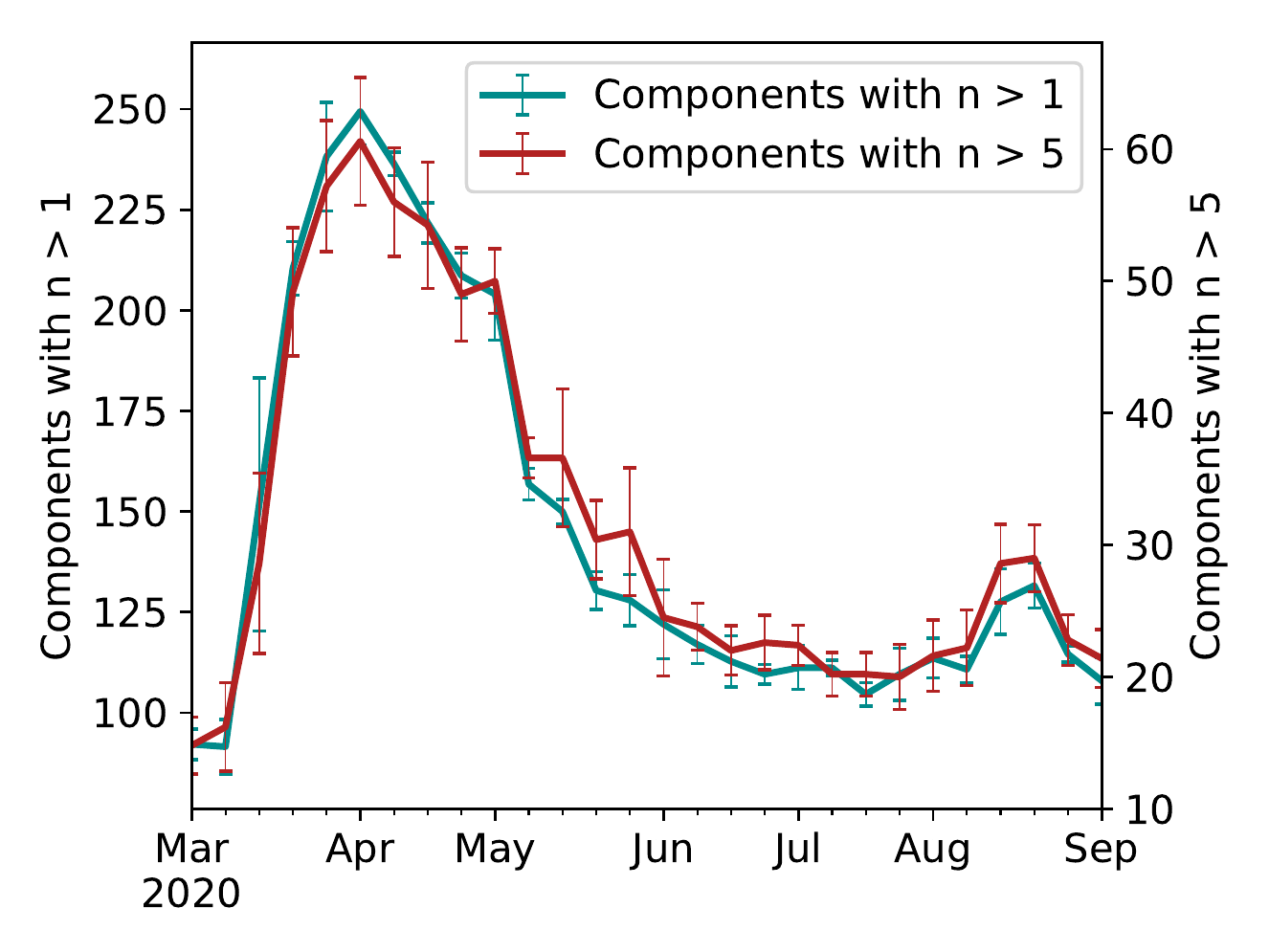}\label{1d}\begin{picture}(0,0)
\put(-180,-110){\subref{1d}}\quad
\end{picture}}\quad
   \hspace{-25.00mm}
   \subfloat{\includegraphics[width=.35\textwidth,valign=t,right]{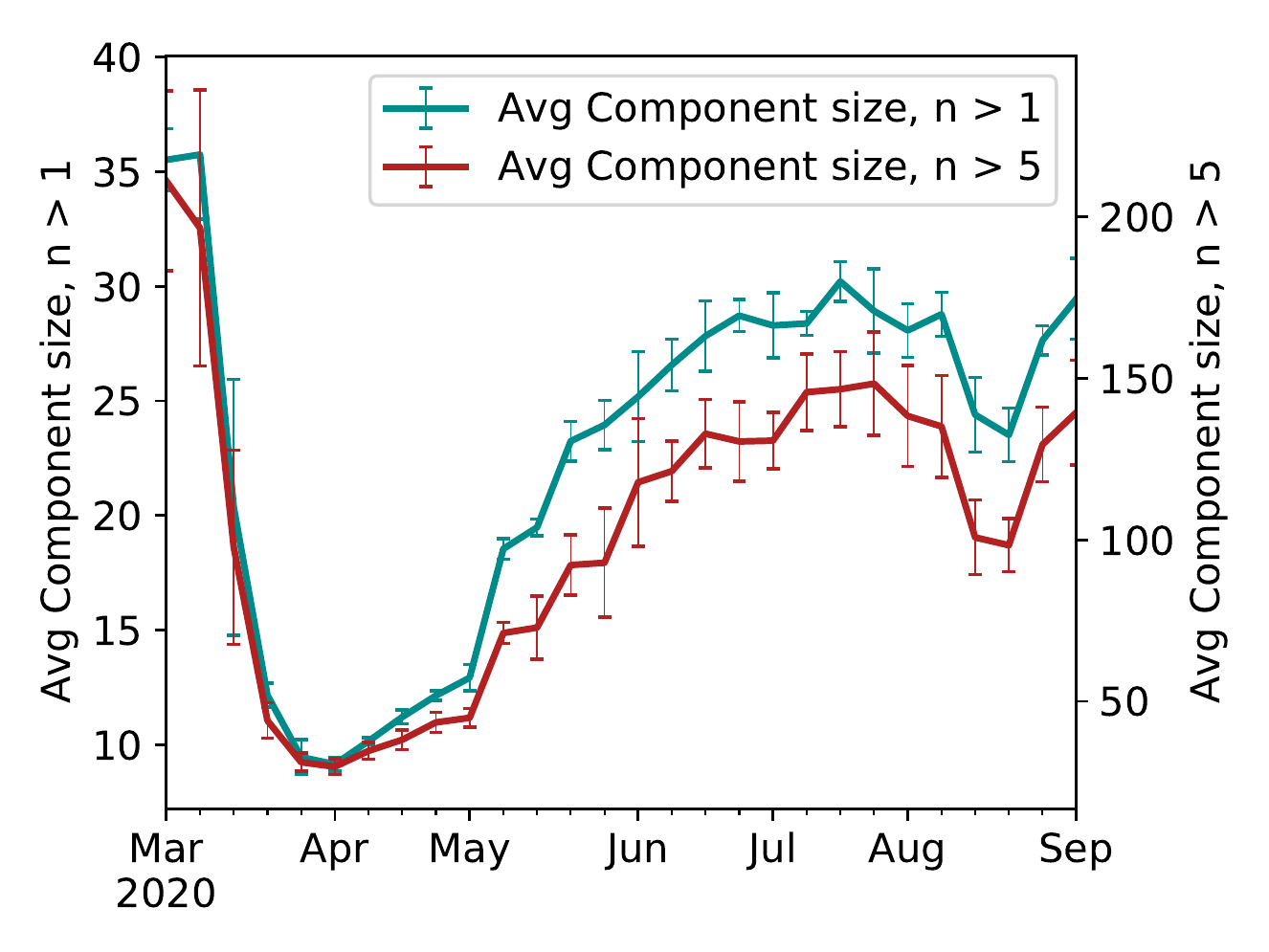}\label{1e}\begin{picture}(0,0)
\put(-180,-110){\subref{1e}}\quad
\end{picture}}\quad
\hspace{-25.00mm}
   \subfloat{\includegraphics[width=.35\textwidth,valign=t,right]{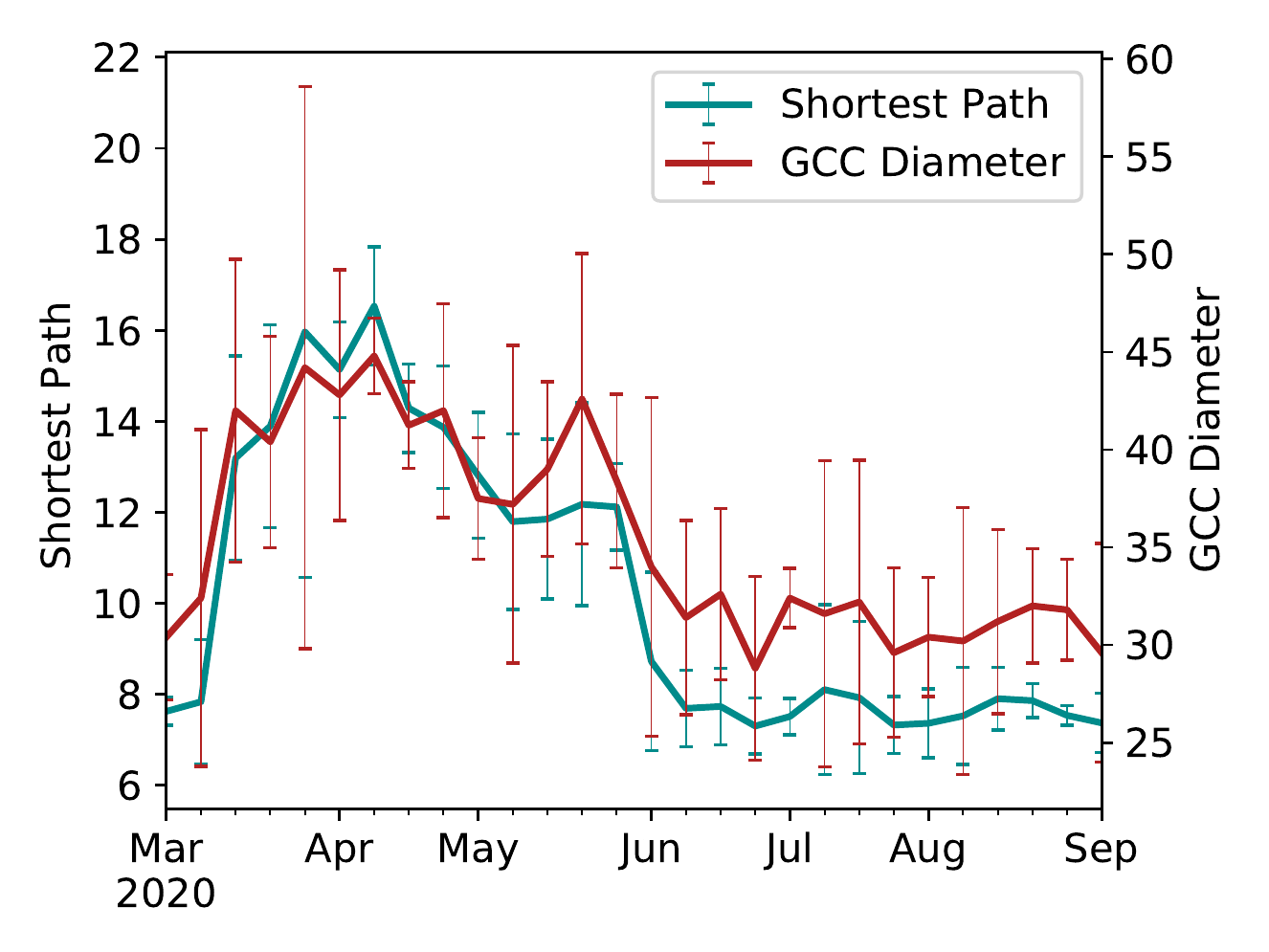}\label{1f}\begin{picture}(0,0)
\put(-180,-110){\subref{1f}}\quad
\end{picture}}\quad

   \caption{\footnotesize Network characteristics of traffic before and during lockdown for the coarsest (left) and finest (right) resolutions. (a) Region-level mobility before lockdown. Region color shows the daily average simple degree for business days before lockdown measure were implemented. The edge colors show the log of the weighted link, i.e. the daily number of people traversing each edge. (b) Mobility during lockdown, color scales matching (a). Many of the existing edges have failed, and those that remain are much weaker. (c) Average weighted degree (teal) and size of the Giant Connected Component (GCC, red) averaged over a week, error bars showing the standard deviation for the period. (d) Number of non-isolated components (teal) and number of components with more than 5 nodes in them (red). (e) Average size of: non-isolated components (teal), and components larger than 5 nodes (red). (f) Average shortest path length (teal) and the diameter of the GCC (red). }
   \label{fig:NetworkBehavior}
\end{figure*}

A network consists of elements, or nodes, connected to each other via links. In our case, the fundamental units (nodes) are rectangular tiles of areas in Italy (see~\nameref{sec:MatMet}). We set an edge between two tiles if a sufficient number of people travel from one tile to another in a given period of time. The smallest window of time in our data is 8 hours, however we aggregate them at the daily, weekly and monthly levels when necessary. Because tiles typically do not represent any particular entities in the real world, we apply a grouping procedure to them, aggregating multiple tiles to match them with their corresponding administrative units. In the case of Italy three such groupings exist: Municipality-, Province- and Region-scale, where the former is a small geographic area (of the order of a township), the second is a larger area aggregating several Municipalities and the latter is a relatively large one. Thus, Italy consists of about 8000 Municipalities, over 100 Provinces and 20 Regions. Since the number of tiles in our data set is of the order of the number of Municipalities, we do not analyze that administrative division and the most granular analysis is conducted directly on the tiles. The left panels of Fig.~\ref{fig:NetworkBehavior} show the connectivity of Italy's Regions (the coarsest level of aggregation) before (Fig.~\ref{1a}) and during (Fig.~\ref{1b}) the lockdown, while the right panel shows tile-level network characteristics. Throughout the analysis we identify four distinct periods, corresponding to different behavior of the metrics under analysis, as will be detailed below. Those are:
\begin{enumerate}
\item Pre-lockdown, baseline period, starting end of February, 2020 until March 8th.
\item Full lockdown, starting March 9th and lasting through May 18th
\item Partial restriction, whereby some limitations remained in place but others were lifted, May 19th till June 3rd
\item Removal of all limitations, June 4th until Mid July.
\end{enumerate}
One fundamental characterization of a network is its degree distribution, \textit{P(k)}, where \textit{k}, the degree of a node, is the number of other nodes it is connected with. The degree distribution shows the probability \textit{P} to have a node of degree \textit{k}. For random graphs, known also as Erdos-Renyi (ER) graphs, in which edges between nodes exist with some fixed probability \textit{p}, this distribution follows the Poisson one, $P(K)=\frac{(np)^ke^{-np}}{k!}$ where \textit{n} is the number of nodes in the network which is assumed large~\cite{bollobas2013modern}. In that case, the product \textit{np} is the average degree of the nodes in the network. Importantly, the distribution of degree in ER graphs is relatively narrow around the mean and most nodes have a degree close to the mean, i.e., large deviations are exceedingly rare. Many real-world networks exhibit instead much broader degree distributions, often characterized by a power law, $P(k)\propto k^{-\beta}$, with or without exponential truncation to accommodate cutoffs, finite size effects and other features which separate them from random graphs~\cite{albert2002statistical}. In such a distribution, the average degree $\langle k \rangle$ is not representative of the entire network degree distribution, due to the presence of relatively few hubs that are connected to many more nodes than the typical node. 

Edges connecting pairs of nodes need not be identical and may carry information with them. This information is typically regarded as an edge ``weight'' and may be useful in determining shortest paths, estimating travel time and assess the fragility of the network. In our case, edges carry two types of information. One is geographical, namely the geometric distance between two connected nodes, and the other is demographic, specifically, the number of people traveling along the edge. When edges weights are available, the simple degree may be generalized and augmented by a weighted degree, where the sum of edges weights incident on a node is taken, rather than the number of edges. Below, when discussing weighted degrees we refer to aggregating all people traveling to and from a certain node. As discussed in \nameref{sec:MatMet}, our edges are undirected.

We explore the global connectivity (i.e., the possibility to get from one node to another via existing edges) through the functional and the pure network perspectives (Fig.~\ref{fig:NetworkBehavior}). From the functional perspective, we note the connectivity of the largest-level aggregation of our data, namely, Italian Regions. This view allows us to appreciate the significant drop in the connectivity between large parts of the country. The simple unweighted degree of the Regions (as seen in the colored maps in panels~\subref{1a} and~\subref{1b} of Fig.~\ref{fig:NetworkBehavior}) can be viewed as a qualitative effect of the mobility restrictions. During lockdown connectivity dropped across the country such that interactions between different Regions (the coarsest aggregation level) almost disappeared. Advancing to more quantitative characteristics of the mobility network, we note the drop in the average daily weighted degree, shown by the color of the links, which measures the daily number of people moving between locations. 

Another important feature of a network is the extent to which nodes are connected globally to each other, which can be investigated by measuring the size of the network components. A component of a network is a set of nodes reachable from one another. Among components it is particularly informative to analyze the largest one, typically called the Giant Connected Component (GCC)~\cite{cohen2010complex}. This is because in complex networks above a certain degree of connectivity (an average degree of 1 in the case of ER networks, known as the percolation threshold) most nodes of the network are reachable from any other node, that is, most of the network belongs to the GCC. When the network becomes fragmented (in our case via the removal of edges due to mobility restrictions), less nodes tend to belong to the GCC and the number of smaller components increases, as may be observed in Fig.~\ref{1c} and~\ref{1d}. Interestingly, comparing the behavior of the mean weighted degree with that of the GCC of the network, we find a striking similarity, which can also be seen in the component sizes (SI) and will surface in other aspects of our analysis (resilience). There is no particular mechanism driving that - and yet the drop in average weighted degree traces very closely that of the component size (Fig.~\ref{1c}). As the GCC decreases, the network becomes more fragmented, leading to an increase in number of components. We report values for the numbers of non-isolated components and components larger than five nodes (Fig.~\ref{1d}), as well as their sizes (Fig.~\ref{1e}). Finally, we measure the reachability within the changing GCC. Interestingly, as the GCC shrinks during lockdown, the different nodes effectively become more distant as the average path length and diameter of the component increase (Fig.~\ref{1f}).\\

\section{Scaling}\label{sec:Scaling}
\begin{figure*}[h!]
   \centering
   \hspace{0.00mm}
   \subfloat[][]{\includegraphics[width=.48\textwidth,valign=t]{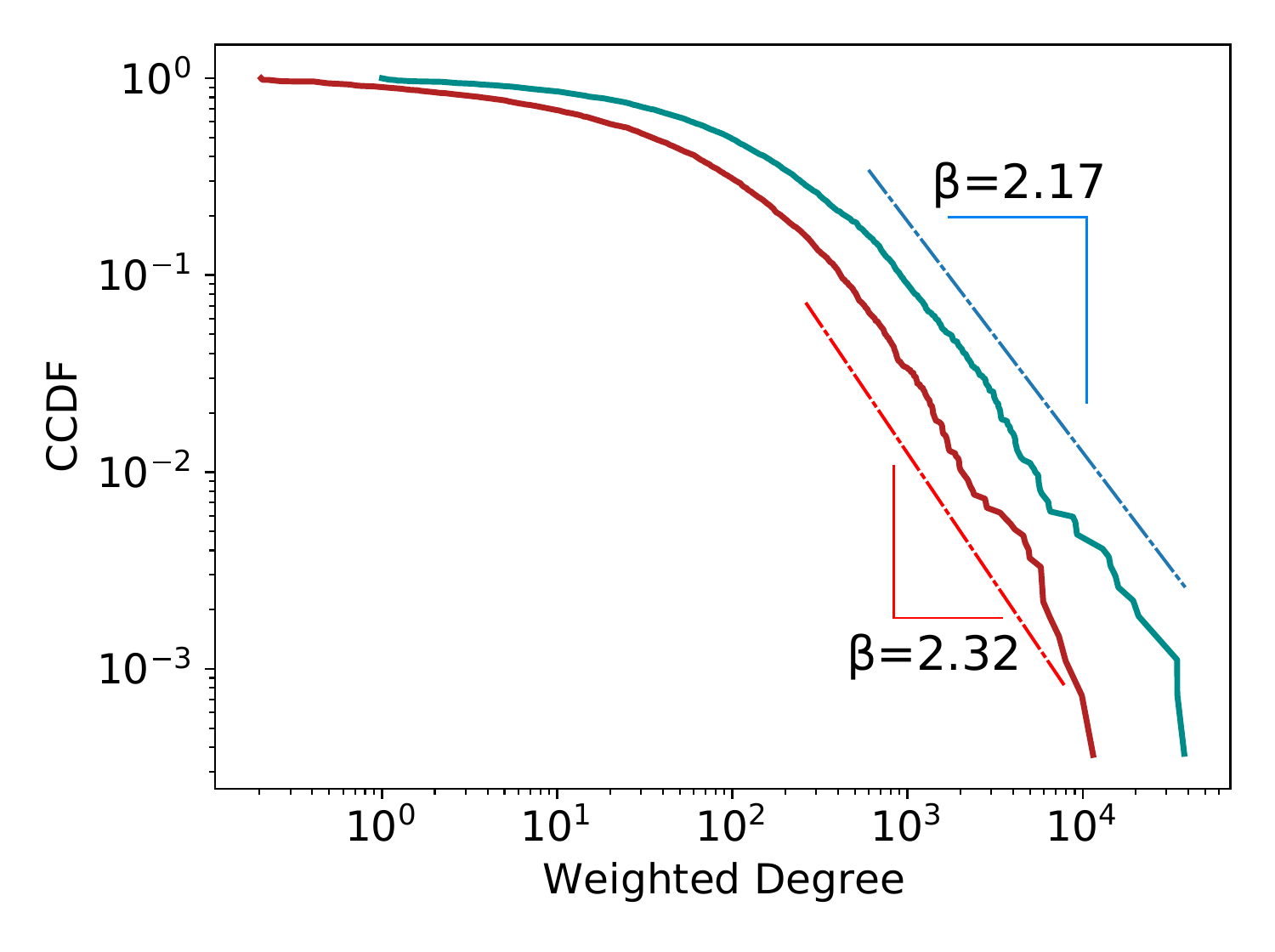}\label{2a}}
   \hspace{0.00mm}
   \subfloat[][]{\includegraphics[width=.48\textwidth,valign=t]{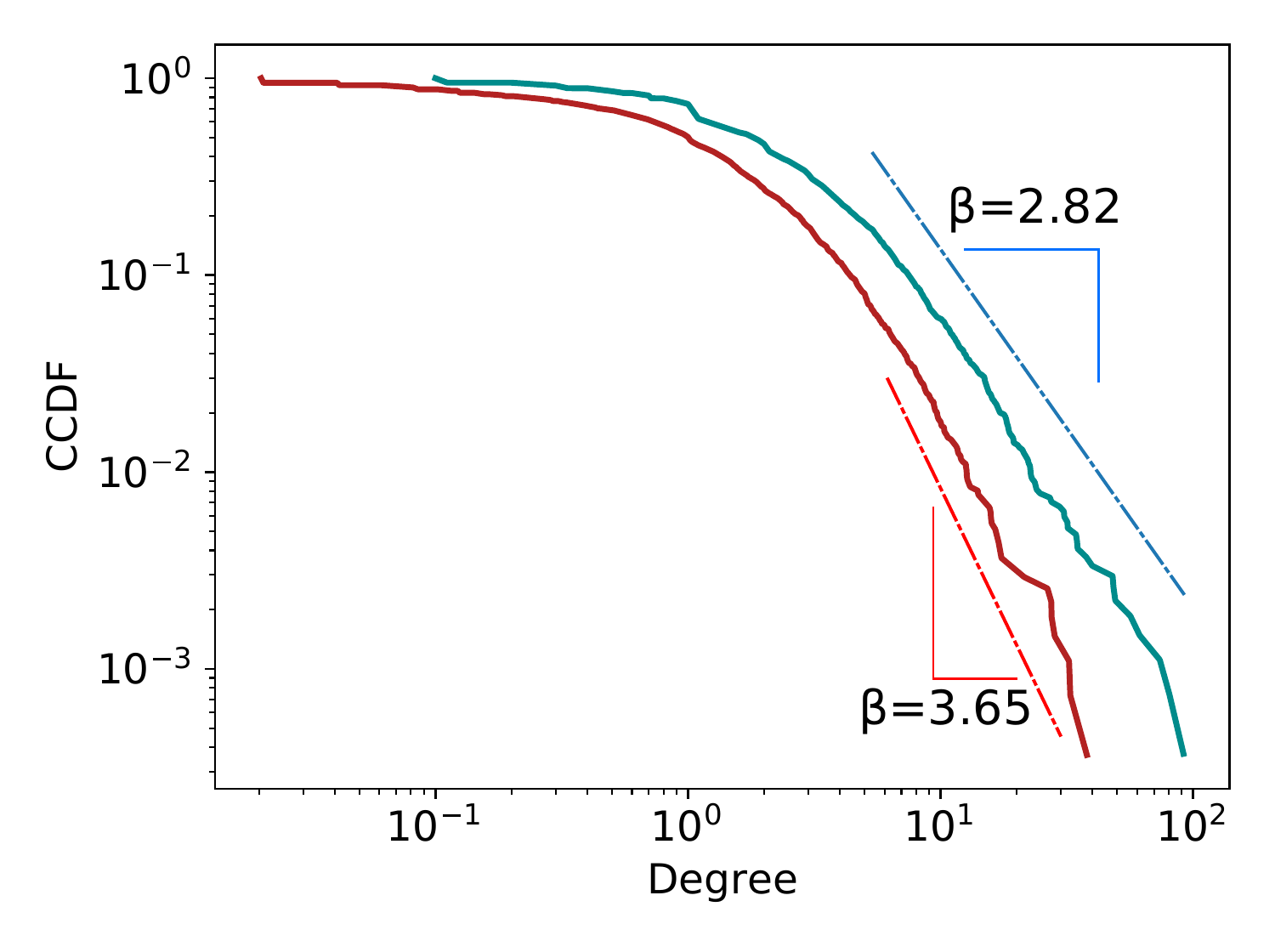}\label{2b}}
   \hspace{0.00mm}
   \subfloat[][]{\includegraphics[width=.48\textwidth,valign=t]{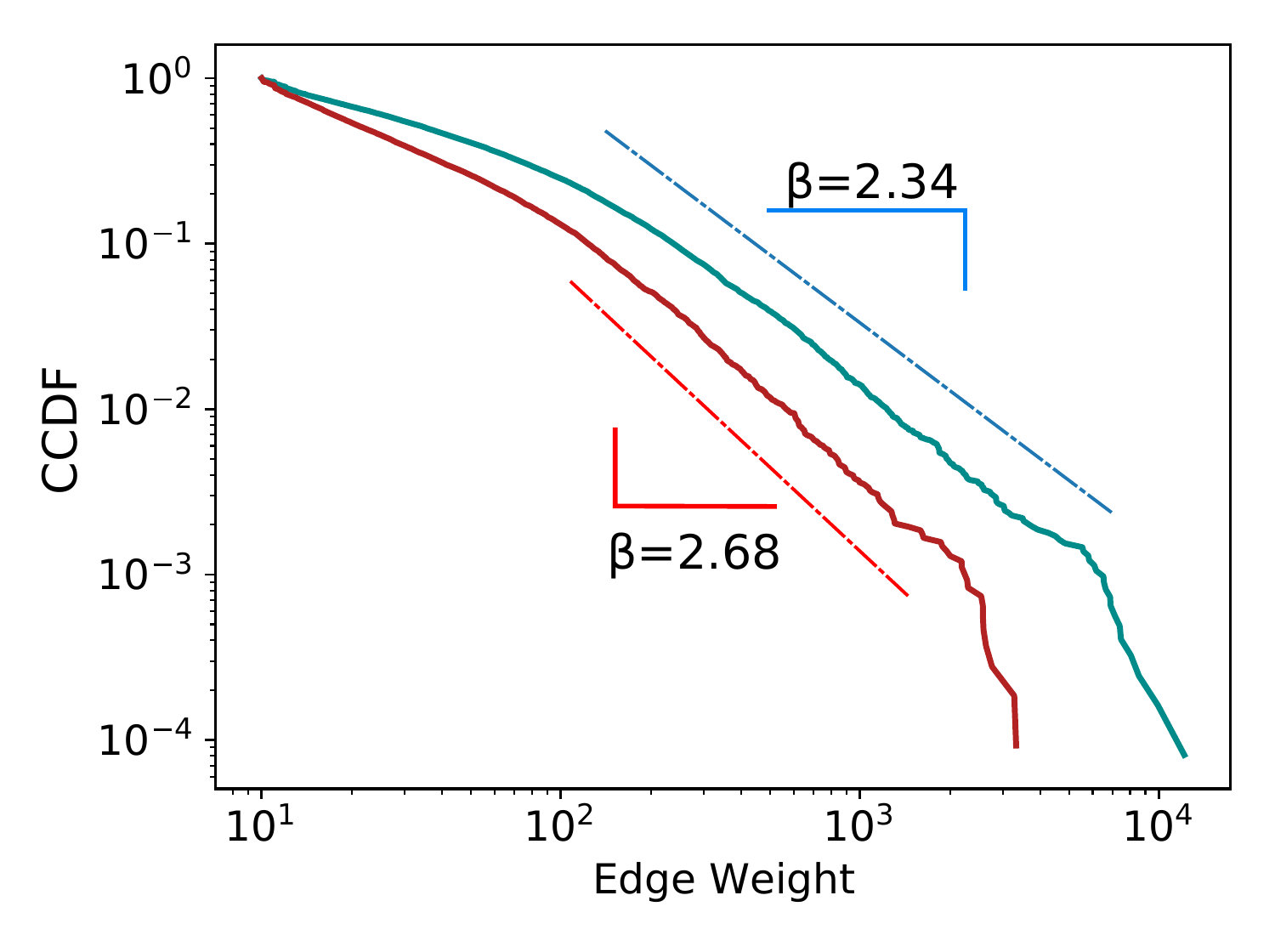}\label{2c}}
   \hspace{0.00mm}
   \subfloat[][]{\includegraphics[width=.48\textwidth,valign=t]{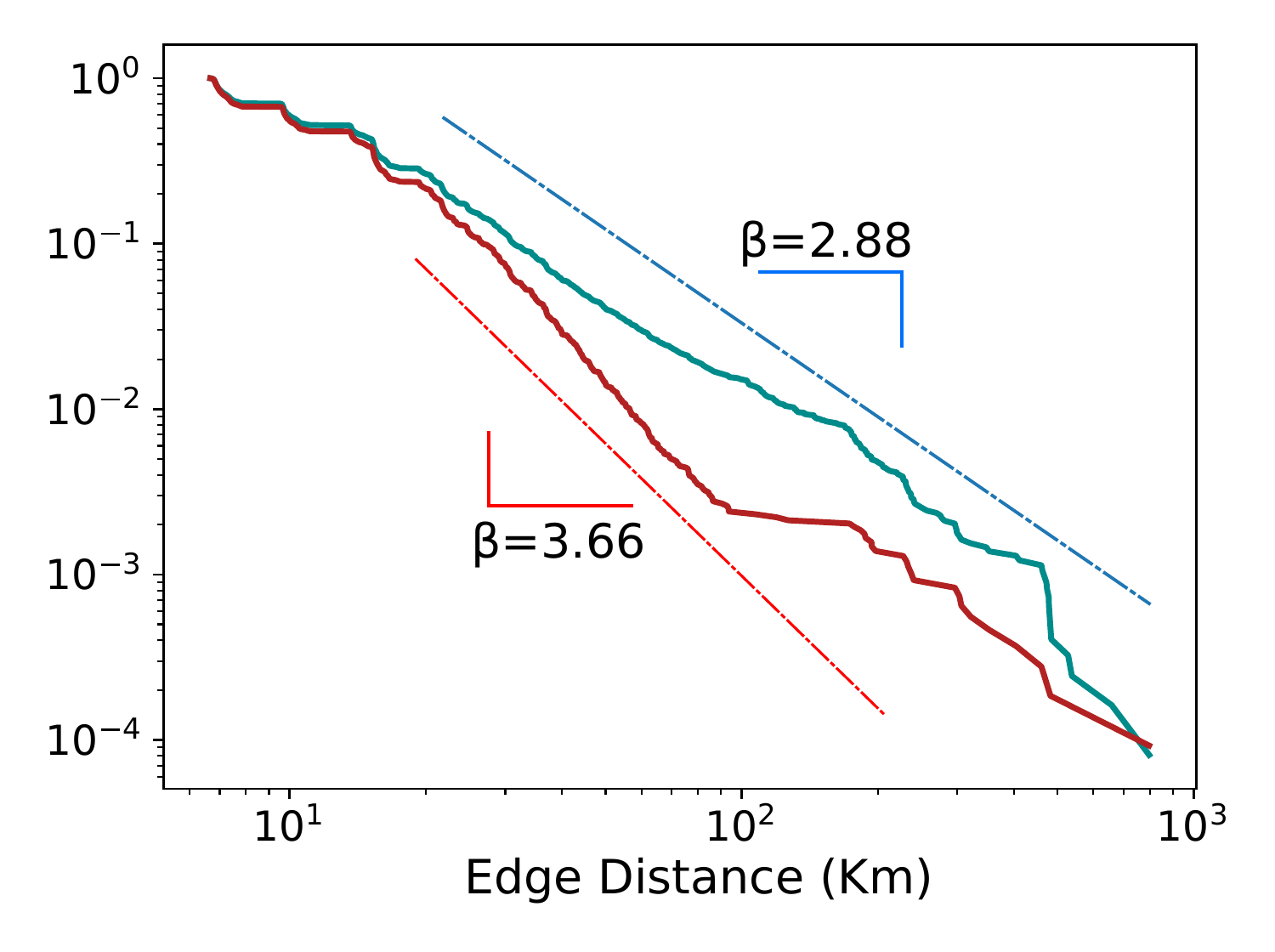}\label{2d}}

   \caption{\footnotesize Comparison between pre-lockdown and lockdown of (a) weighted degrees, (b) Unweighted degree (for the degree distribution we also show a stretched exponential fit in the SI), (c) edge weights and (d) distances. Teal and dark red lines show the Complimentary Cumulative Distribution function (CCDF) of the weighted degrees, weights and distances before and during the lockdown, while the dashed lines show the stretched exponential (top) and power law fits (bottom) to the data.}
   \label{fig:Scaling}
\end{figure*}

Next, we examine how some of our tile-level network's characteristics behave before vs during lockdown. We analyze the PDFs of some network characteristics and find that they can be approximated by power-laws~\cite{alstott2014powerlaw} (see Sec.~\ref{sec:theofit} in the SI for additional possible fits). The probability distribution of the data, together with the fitted power-law functions is shown in Fig.~\ref{fig:Scaling}. Starting with the degree distributions, we note the larger exponents during lockdown (Fig.~\ref{2a} and~\ref{2b}). Both the weighted and simple degree distribution contract which means less people (weighted degree) travel to fewer places. That is backed up by examining the distributions of the edge weights (Fig.~\ref{2c}) and distances traveled  (Fig.~\ref{2d}).

Weight distribution tells probably the most important story of the positive and negative aspects of the lockdown. The reduction of people traveling along edges assists in reducing contagion~\cite{schlosser2020covid, jia_population_2020, kissler_reductions_2020, nouvelle_reduction_2021} but also induces a significant economic impact ~\cite{Bartik17656,  spelta2020after, chang_mobility_2020, chetty_economic_nodate}. We note here the entire distribution shrinks, both the typical and extreme values. The relative ubiquity of the impact is elaborated in Sec.~\ref{sec:Impact}. The extent to which reduction in mobility affects the economy as a whole is discussed in Sec.~\ref{sec:Econ}. 

To conclude our exploration of scaling behavior we revisit an often-measured characteristic of human mobility, namely, laws governing distances traveled~\cite{brockmann2006scaling,gonzalez2008understanding,liang2012scaling,alessandretti2020scales}. Two points are worth noting here. The exponents observed here are noticeably larger than the findings on human traveling in Ref.~\cite{gonzalez2008understanding}. Several sources of the difference may be identified. One is the specific geometry of Italy, characterized mostly by relatively short distances. The typical width of the country is about 300km, while the distance between Naples and Milan, two economic centers relatively far from one another, is close to 900km. Due to that, we do not expect to find enough distances much greater than 300km. Indeed, we can see in Fig.~\ref{2d} that before the lockdown the slope changes and become larger at about 200km. Another important factor comes from our data limitations relating to our sampling period being 8 hours and the lower bound on number of at least 10 people to establish an edge between two tiles (see~\nameref{sec:MatMet} for details). Taking those limitations into account, however, the scale-free behavior of human mobility is approximately reproduced in our analysis. More importantly, we find that under the lockdown restrictions, the scaling laws for weights and distances still hold, while the exponents increase further, shrinking the covered distances. For the distances, as opposed to weights, where the entire distribution shrank, the range of the distribution does not change much, with the farthest points observed before the lockdown being reachable during lockdown as well. Moreover, the most prominent change takes place in the medium distances, between roughly twenty and two hundred kilometers, the distances that cover (with some excess) what would typically be the daily commute. Workers related to essential activities, however, are probably still present and travel to distances comparable to those before the lockdown \cite{spelta2020after, di_porto_partial_nodate}.\\

\section{Local Impact and Recovery}\label{sec:Impact}
\begin{figure*}[h!]
   \centering
   \hspace{0.00mm}
   \subfloat{\includegraphics[width=.45\textwidth,valign=t]{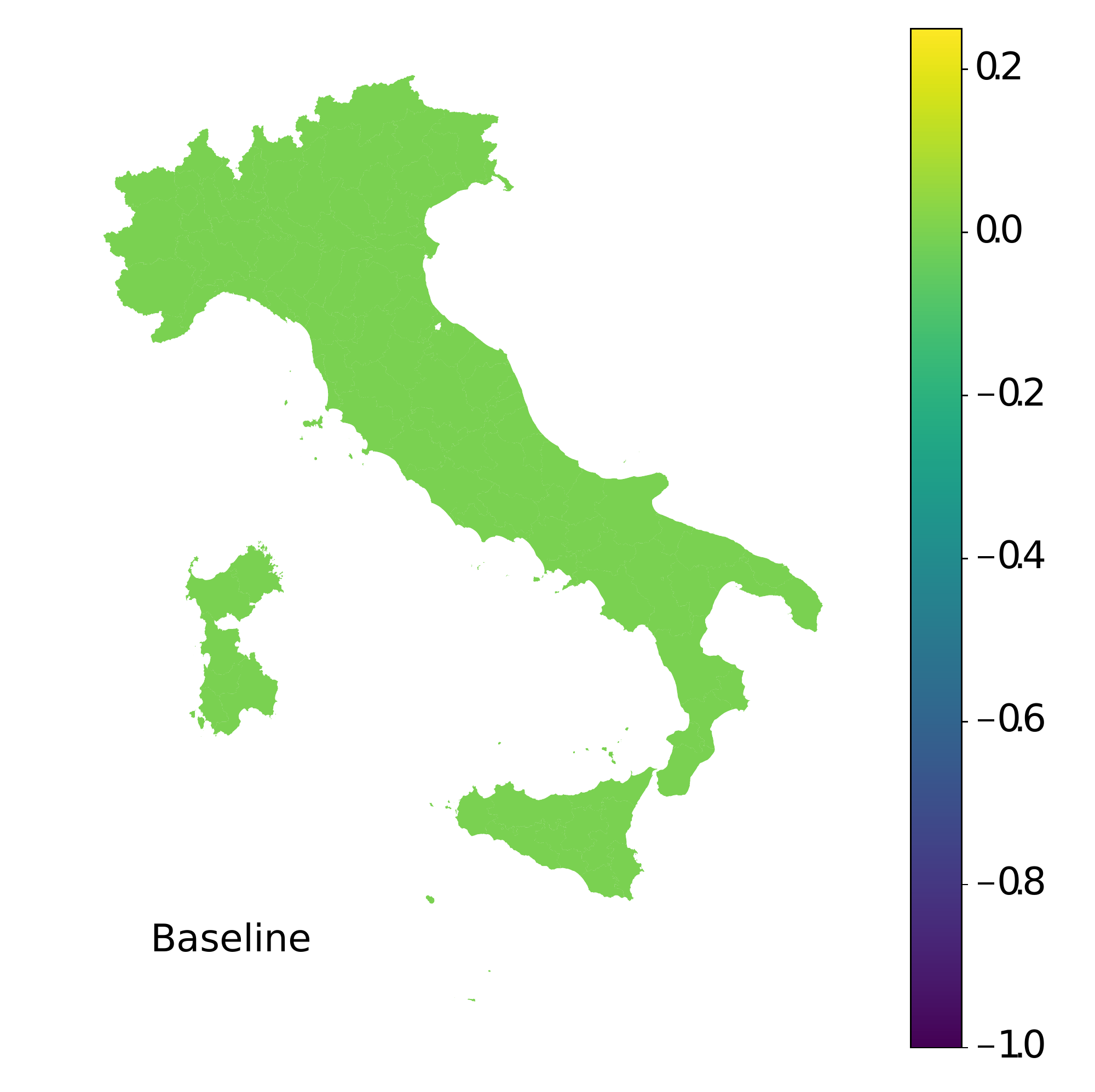}\label{3a}}
   \hspace{0.00mm}
   \subfloat{\includegraphics[width=.45\textwidth,valign=t]{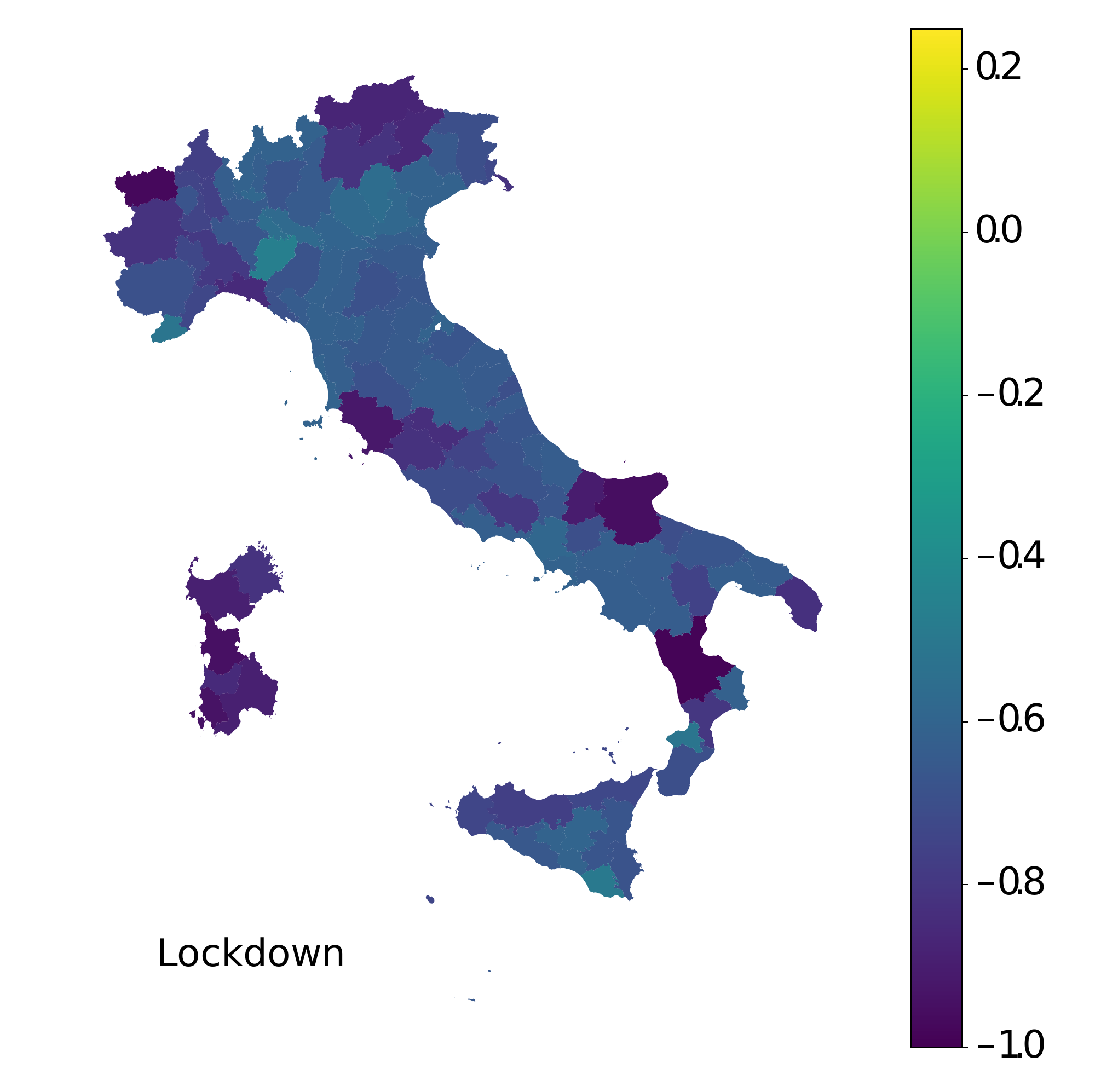}\label{3b}}
       \hspace{0.00mm}
   \subfloat{\includegraphics[width=.45\textwidth,valign=t]{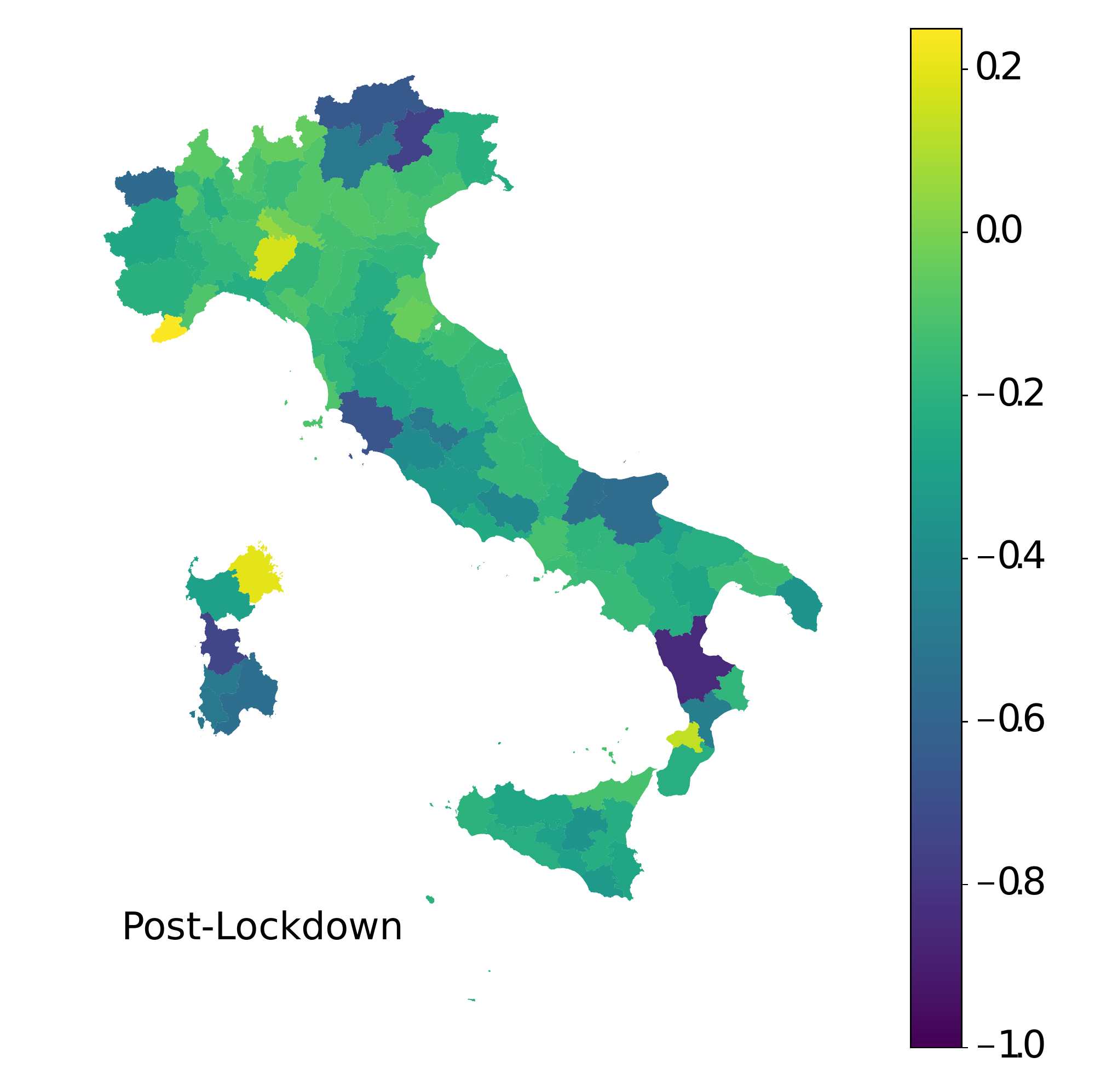}\label{3c}}
      \hspace{0.00mm}
   \subfloat{\includegraphics[width=.45\textwidth,valign=t]{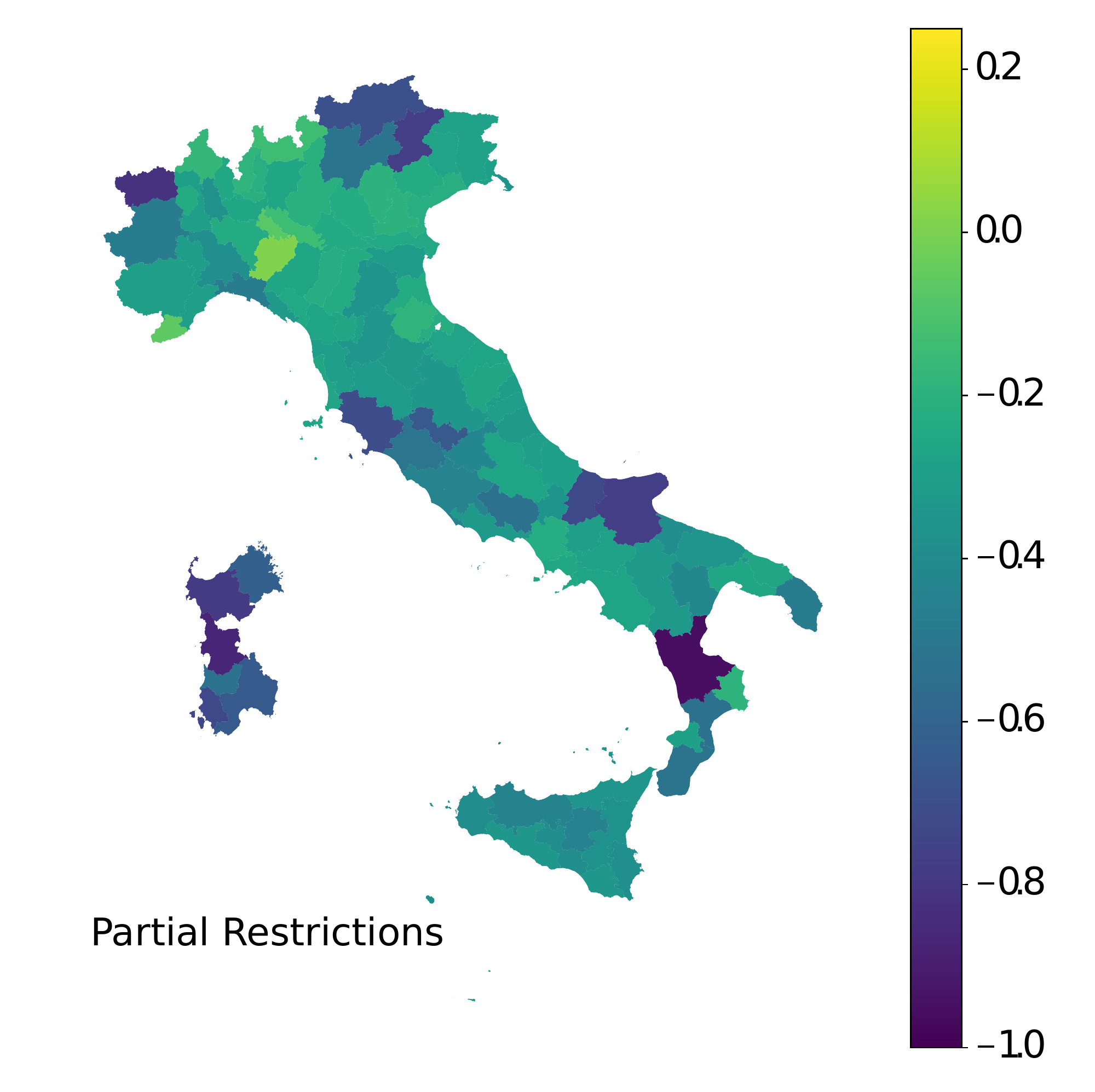}\label{3d}}

   \caption{\footnotesize Clockwise from top-left: Baseline map of Italian Provinces; Reduction in mobility during Lockdown; Partial lift of limitations; All limitations lifted. The color bar indicates the reduction in weighted degree for each Province, with darker blue meaning stronger reduction. The lower panels show the gradual ``return to normal'', with the majority of Provinces behaving similarly, but a few noticeable outliers as described in the main text and Fig.~\ref{fig:ImpactRegion}}
   \label{fig:ImpactMap}
\end{figure*}

\begin{figure*}[h!]
   \centering
   \hspace{0.00mm}
   \subfloat{\includegraphics[width=.95\textwidth,valign=t]{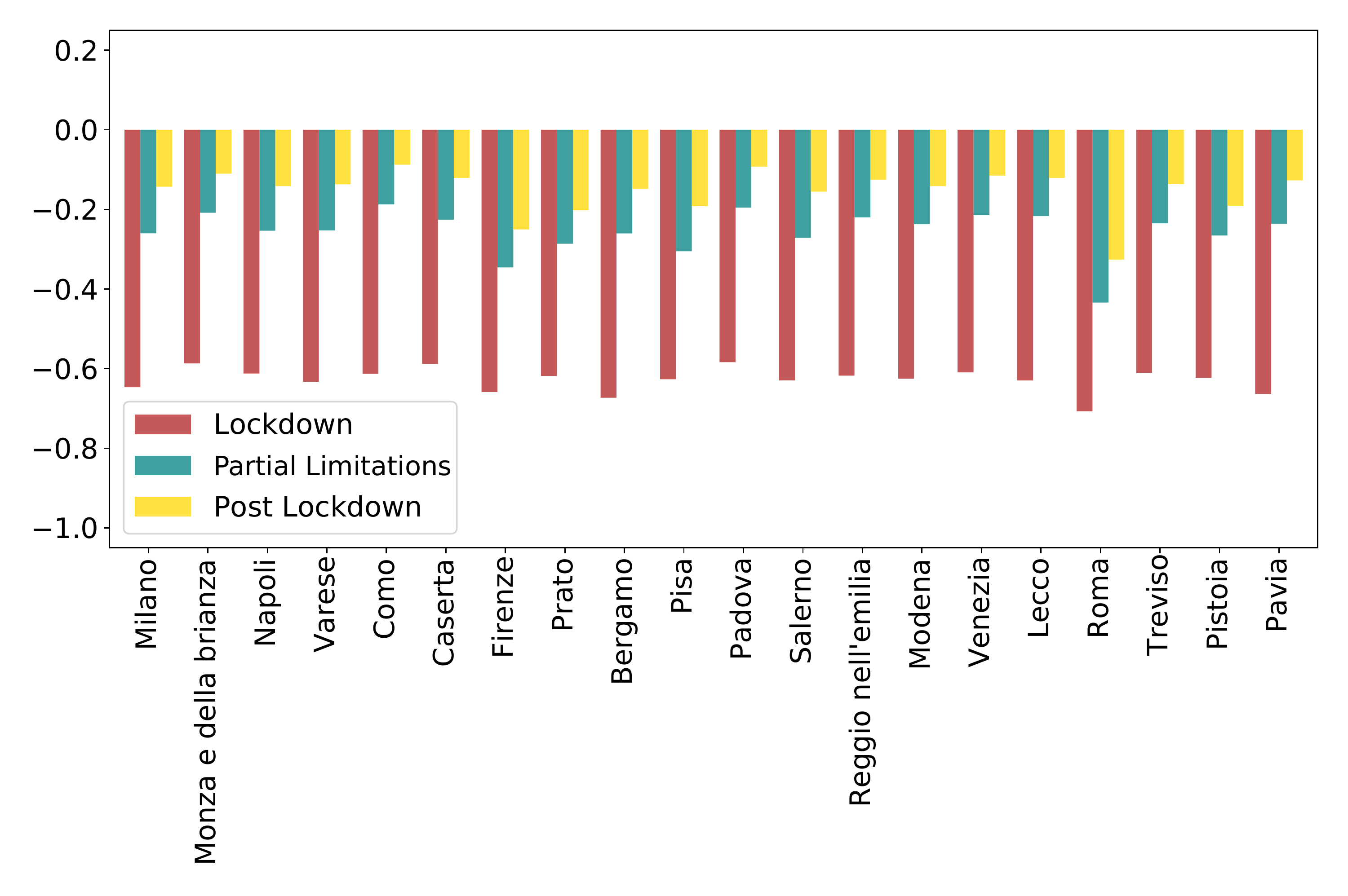}\label{4a}}
   \hspace{0.00mm}
   \subfloat{\includegraphics[width=.95\textwidth,valign=t]{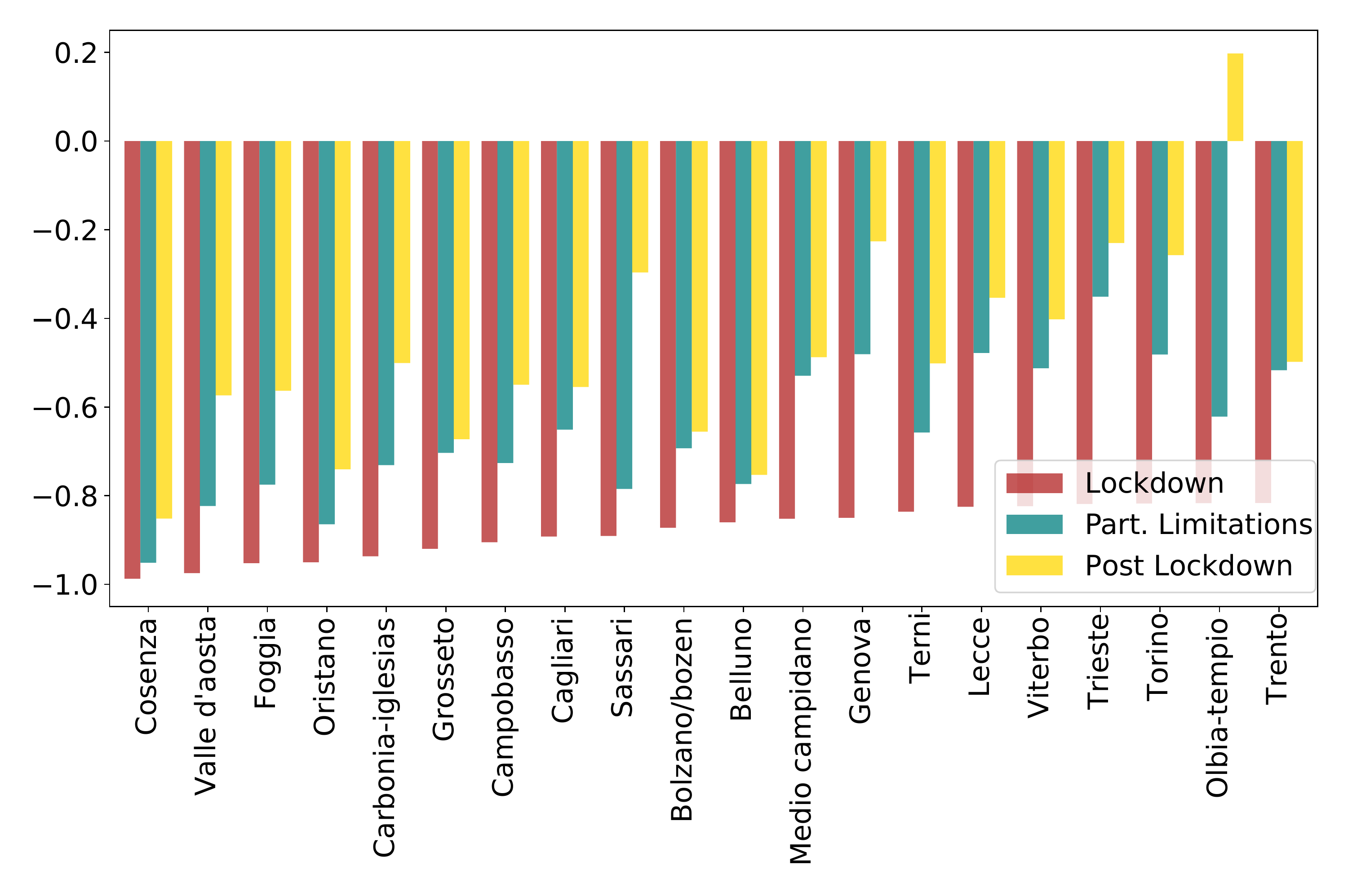}\label{4b}}
   \caption{\footnotesize(Top) Most mobile - Provinces with the highest weighted degree, and (Bottom) most impacted Provinces and their recovery through partial restrictions to complete lift of limitations.}
   \label{fig:ImpactRegion}
\end{figure*}

Previous section analyzes mobility between individual tiles, whereas here, in order to take a step toward assessing potential economic implications, we aggregate mobility across Provinces - a more granular perspective than the Regions in Fig.~\ref{fig:NetworkBehavior} but containing economic meaning as compared to the individual tiles. The lockdown (top-right) panel of Fig.~\ref{fig:ImpactMap} shows the full impact of the restrictions. For non-zero distances (i.e., excluding people who stay within the same tile for the measurement period, as explained in~\nameref{sec:MatMet}), mobility decreased dramatically, with Provinces dropping between 50\% and 90\%. The lower two panels, perhaps, hold the more interesting and unexpected effects of the lockdown. While some Provinces (as will be detailed in Fig.~\ref{fig:ImpactRegion}) rebound relatively quickly, even when only some of the limitations are lifted, other Provinces remain far below their initial levels throughout the time frame shown.

Indeed, Fig.~\ref{4a} highlights the relative similarity of the impact suffered by the most mobile Provinces, those with the highest weighted degree. Those Provinces show a consistent, close to 60\%, reduction of mobility during lockdown and an equally consistent recovery when restrictions are gradually lifted. This is contrasted by the higher variation seen in the lower panels of Fig~\ref{fig:ImpactMap} and detailed by Fig.~\ref{fig:ImpactRegion} for Provinces with lower initial mobility (weighted degree). A noticeable difference exists in the level of recovery, with some of the most impacted Provinces, such as the southern Province of Cosenza and the northern Aosta Valley, remaining well below initial levels after restrictions are lifted, while others, such as the coastal Provinces of Genoa and Olbia-Tempio recovering close to, and above, their initial levels.

\begin{figure*}[h!]
   \centering
   \hspace{0.00mm}
   \subfloat[][]{\includegraphics[width=.31\textwidth,valign=t]{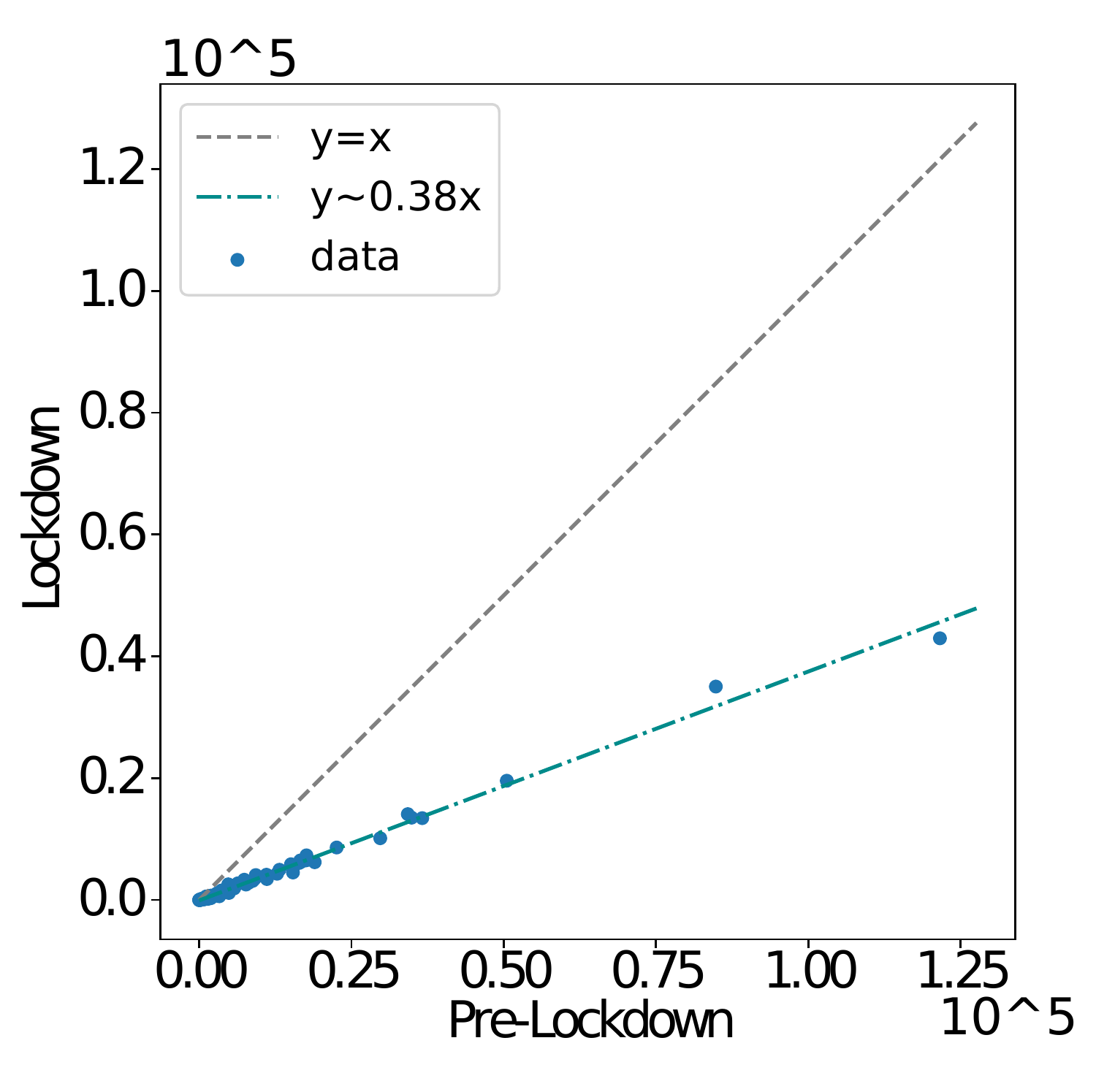}\label{5a}}
   \hspace{0.00mm}
   \subfloat[][]{\includegraphics[width=.31\textwidth,valign=t]{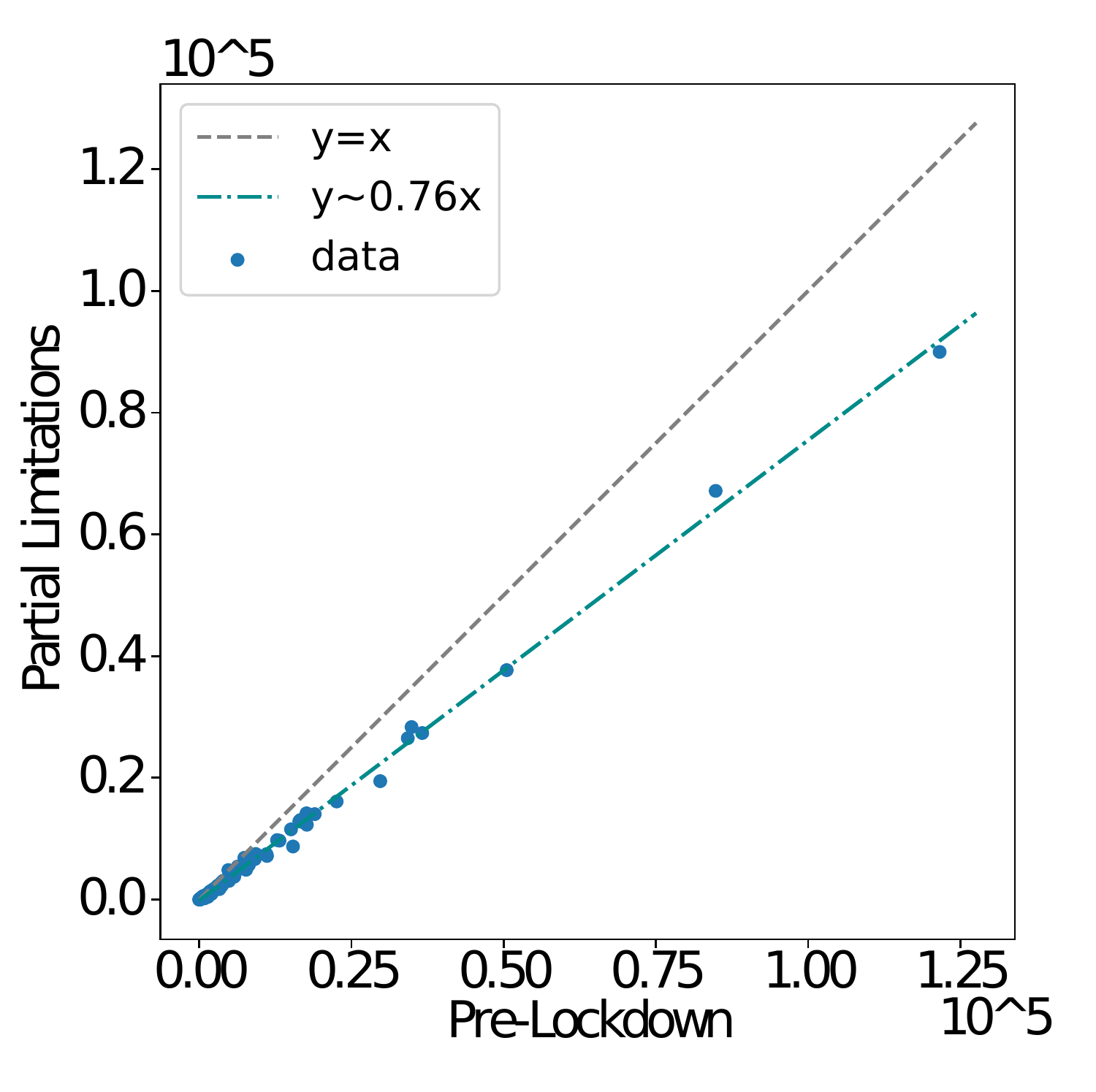}\label{5b}}
     \hspace{0.00mm}
   \subfloat[][]{\includegraphics[width=.31\textwidth,valign=t]{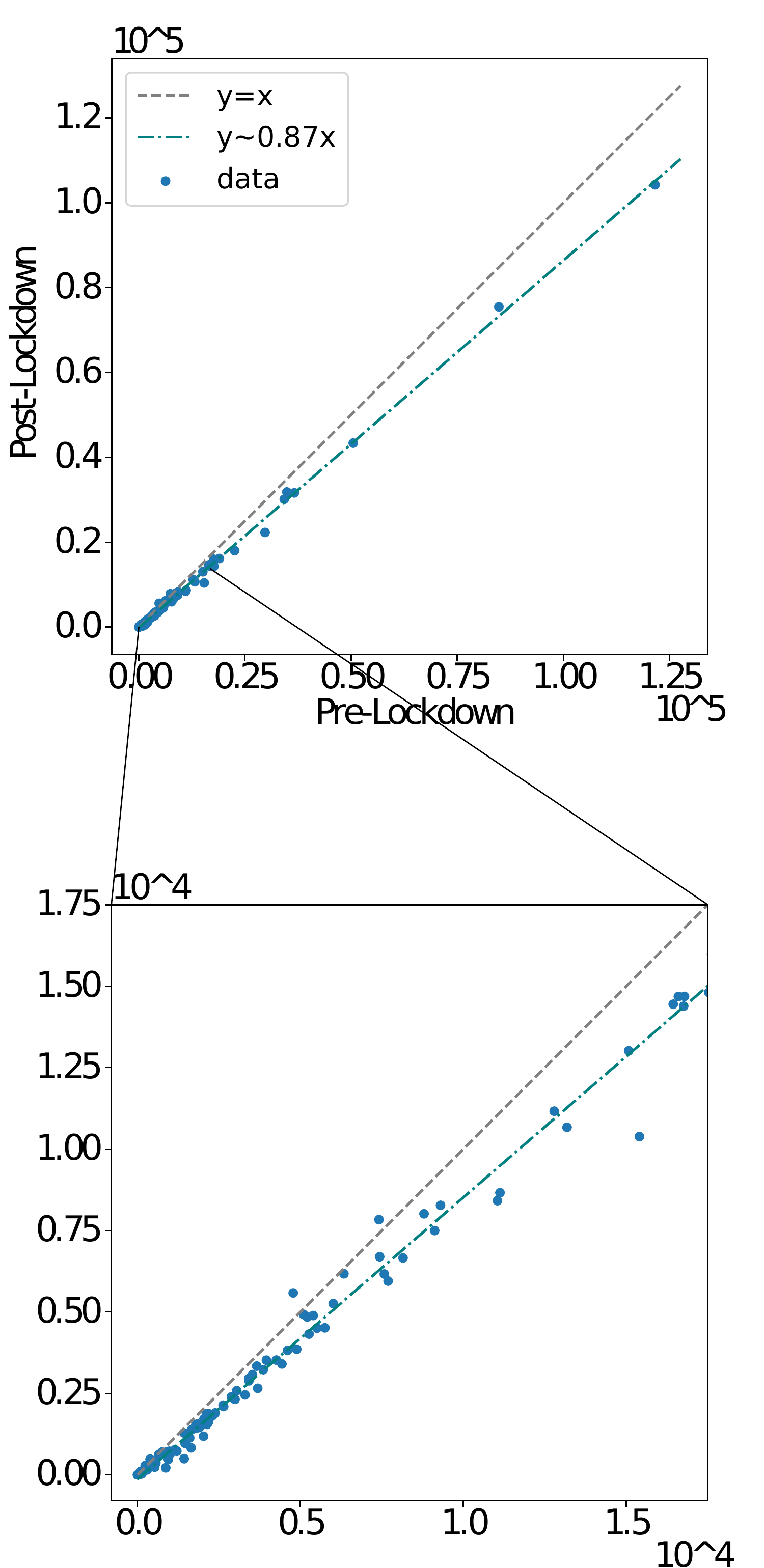}\label{5c}}
        \vspace{-59.00mm}
        \hspace{-160mm}
   \subfloat[][]{\includegraphics[width=.62\textwidth,valign=t,left]{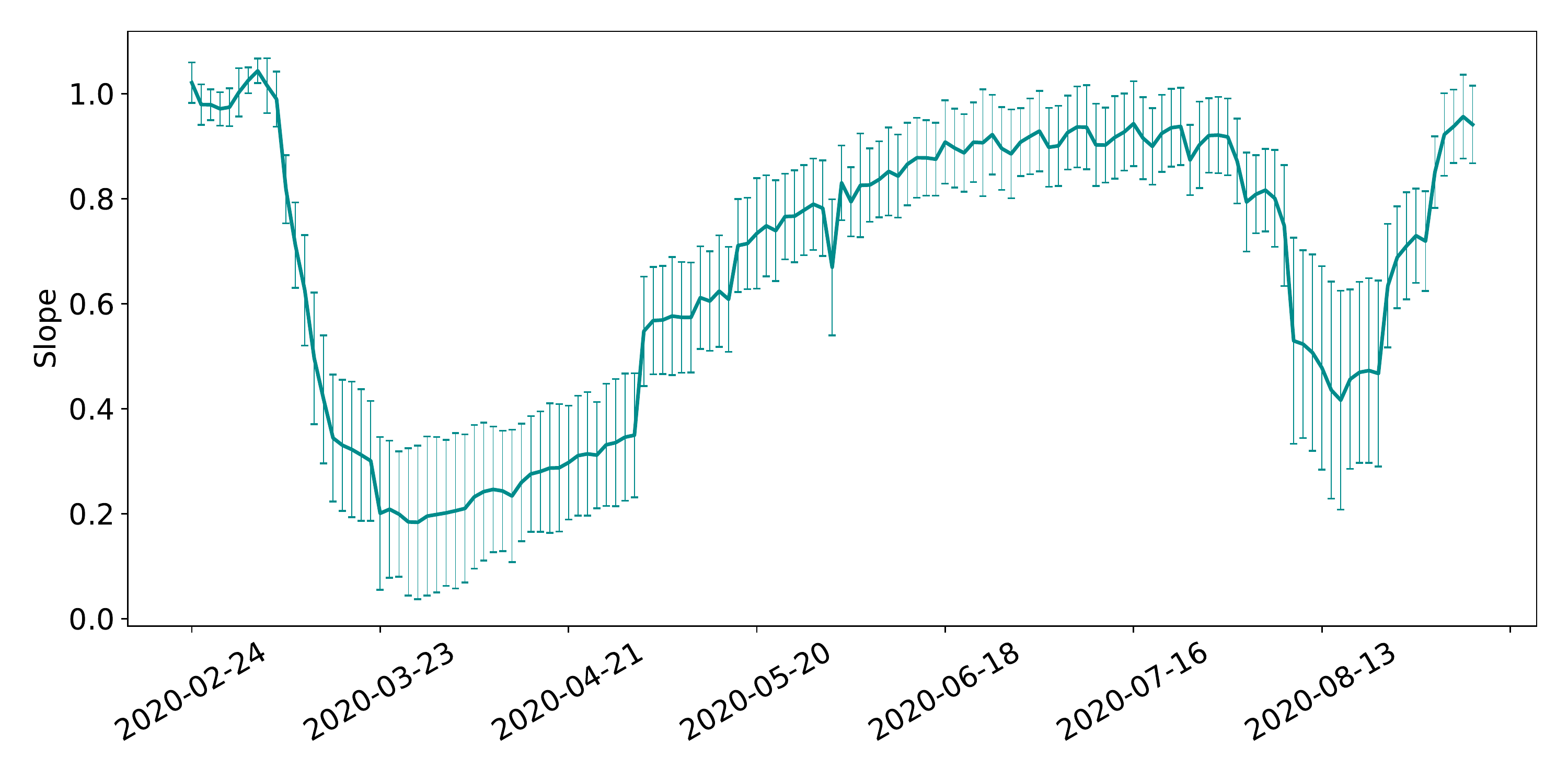}\label{5d}}
   \caption{\footnotesize Uniformity in damage. (a) A linear relation between the lockdown levels of mobility and the initial ones shows a 60\% drop, as seen via the regression coefficient, consistent with~\ref{4a}. (b), (c) show the relations remain very closely linear and uniform in the following stages of partial (b) and full (c) lift of limitations. We note that following the removal of all limitations, overall mobility is still about 15\% below its initial levels. Zooming into the smaller Provinces we see the variation mentioned in Fig.~\ref{4b} while still largely following the overall trend (d) traces the slope of impact vs. the averages pre-lockdown levels over time. We see again the shape encountered in Figs.~\ref{1c} and~\ref{1f}. Insert - the distributions of impact magnitude, i.e. weighted degree drop per time frame.}
   
   \label{fig:ImpactScatter}
\end{figure*}

To generalize those statements we compare mobility levels at the three stages, lockdown, partial restrictions, and lifting of restrictions to the initial levels (Fig.~\ref{fig:ImpactScatter}). The level to which the relations are linear over the various phases of mobility limitations is very surprising. This evidence will be exploited in Sec.~\ref{sec:Econ} to assess the economic impact of mobility restrictions. Moreover, we note that overall mobility had not returned to pre-lockdown levels in the post-lockdown phase, but roughly to 87\% of those levels, which may thus suggest the presence of long-lasting negative effects in the structure of the Italian mobility network persisting the removal of mobility restrictions.

\section{A Resilience Perspective}\label{sec:Resilinece}
Figs.~\ref{1c} and~\ref{fig:ImpactMap} may lead us to assume things were, to some extent, almost back to normal when restrictions were lifted. We wish, however, to employ a methodology taken from traffic jam and service quality domain~\cite{Zhang8673} to show this is not, in fact, the case. For that we return briefly to the tile perspective taken in Sec.~\ref{sec:Scaling}. For every edge between two tiles we calculate the distribution of its weight across time. We then determine a maximal value for that weight, $W_{max}$, as each edge's 95th percentile. Now we introduce a parameter $q$, the fraction of that maximal weight $q=W_q/W_{max}$ and we state that for a given level of $q$, an edge whose current weight is above $W_q$ is considered functional and below it - the edge does not exist. This may be motivated from several perspectives. The obvious approach is to say that if an edge exists but it is only a small fraction of what that edge had been previously - we cannot state that connectivity between the two nodes the edge joins had really been restored. An argument closer to our economic analysis in the previous and following sections may say that there exists a non-linear relation between the number of people traveling along an edge and the value the edge carries. That relation may come from a skewed distribution of people's income, the non-uniformity of economic activity of the travelers or other similar notions. Because of that relation we want to see not only which edges remain, but which edges carry sufficient weight to carry economic value.

\begin{figure*}[h!]
   \centering
   \hspace{0.00mm}
   \subfloat[][]{\includegraphics[width=.75\textwidth,valign=t]{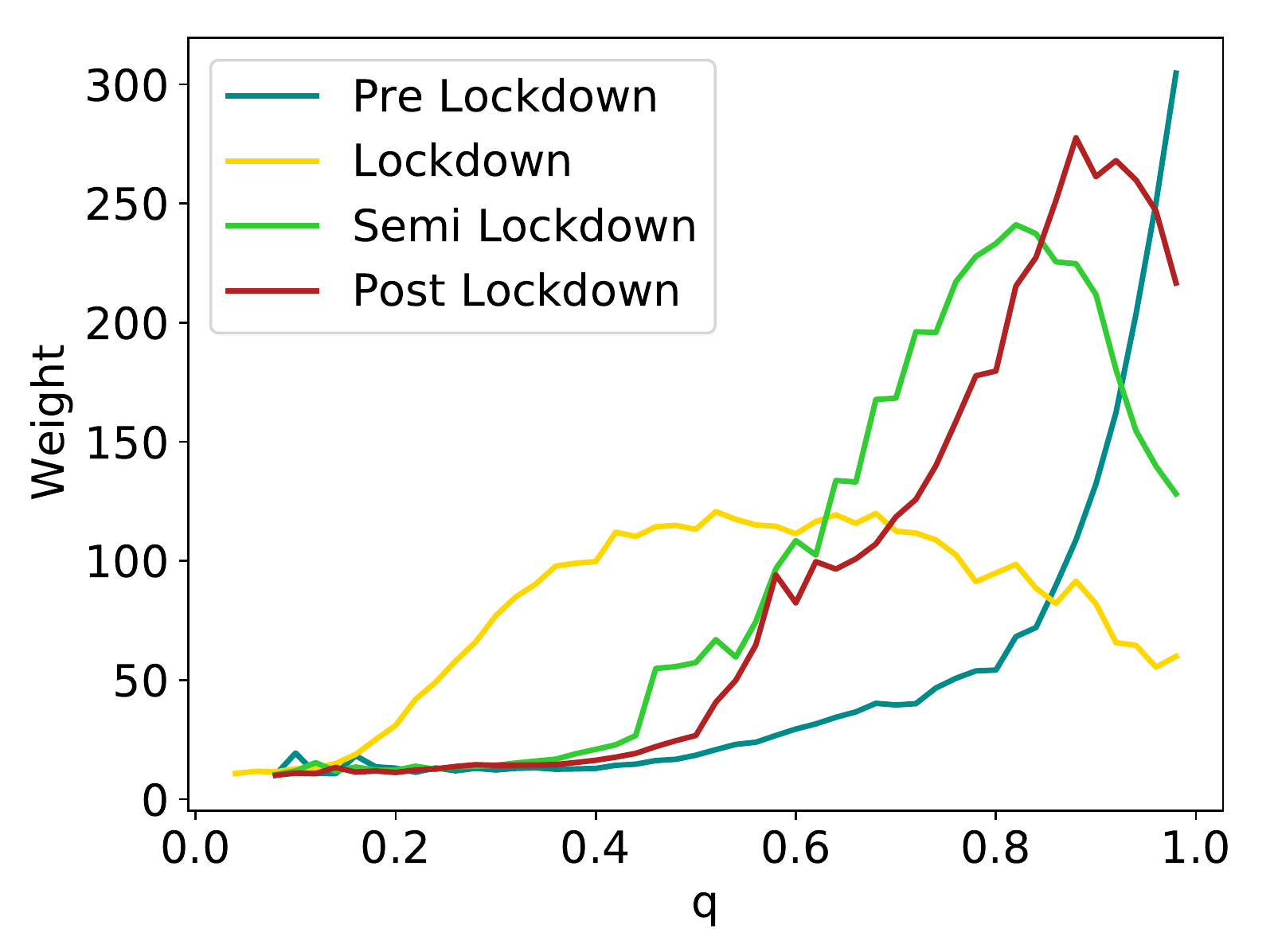}\label{6a}}
   \hspace{0.00mm}
   \subfloat[][]{\includegraphics[width=.45\textwidth,valign=t]{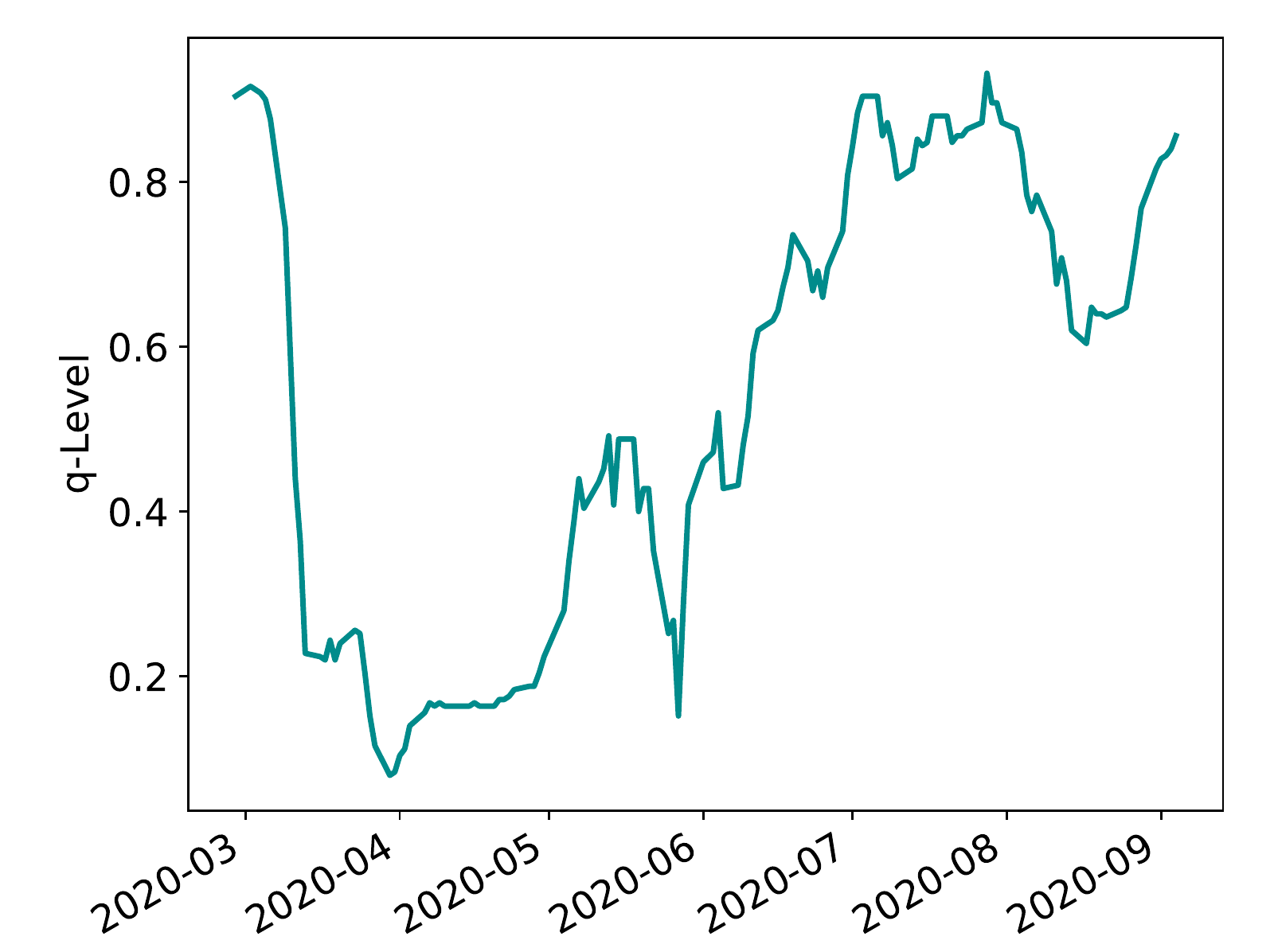}\label{6b}}
     \hspace{0.00mm}
   \subfloat[][]{\includegraphics[width=.45\textwidth,valign=t]{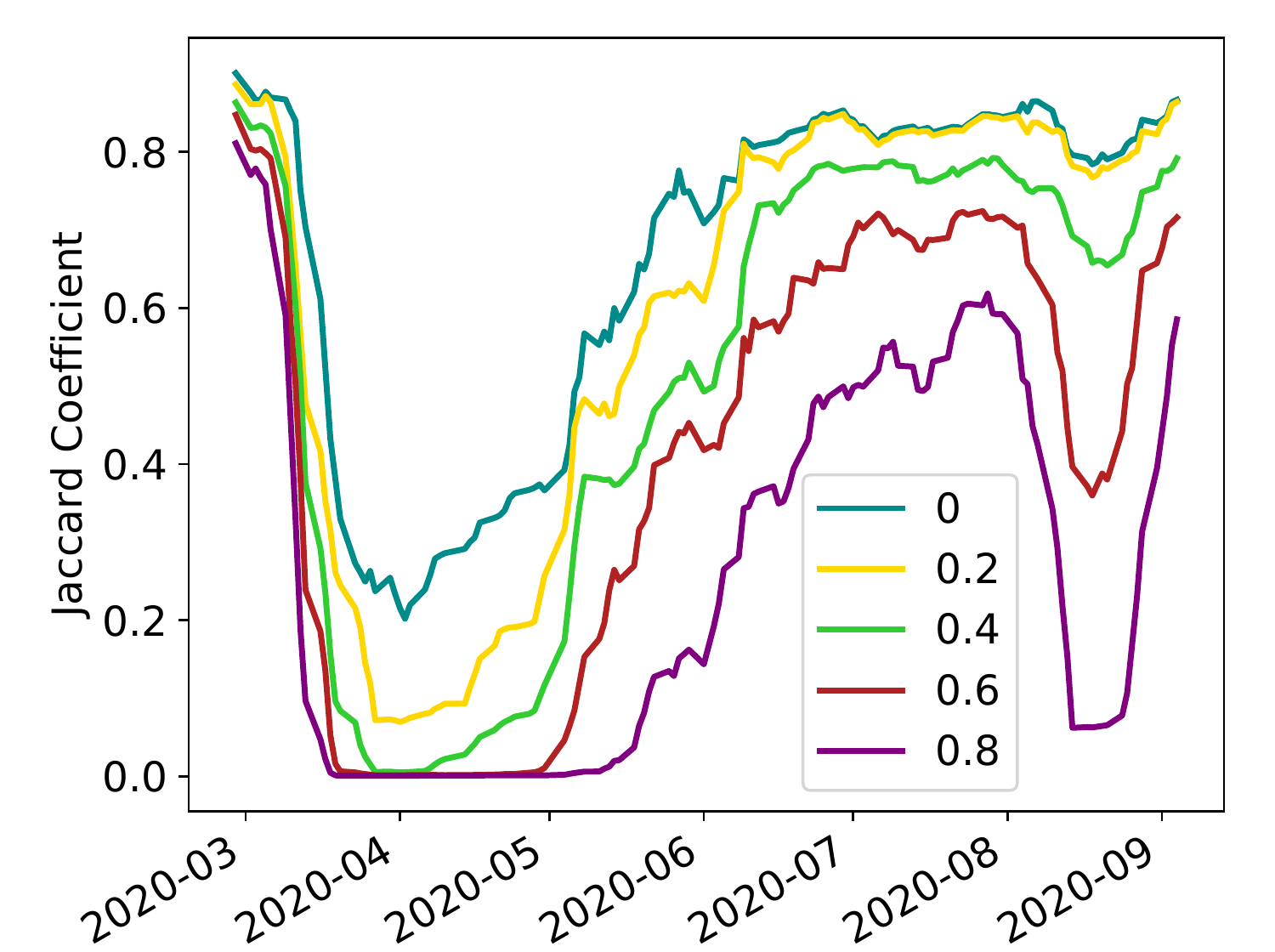}\label{6c}}
     
   \caption{\footnotesize Network under Q-Resilience. (a) The distinct characteristics of failing edges during lockdown vs the other periods. In particular, during the lockdown the network is particularly brittle, and moreover, the edges lost at relatively low Q's carry comparatively high weights, reflecting the stronger impact, (b) Fracturing of the GCC at various q-levels over time. As can be seen, during, but more importantly, after the lockdown the network remains susceptible to failure for only very slight perturbations - the links that are present are roughly 20\% of the original weights. The GCC is in effect very weakly connected. (c) Jaccard similarity for different levels of Q. Here the fragility of the network is even more prominent - if we ask when the edges regain their original strength - the answer is they do not, and many of them remain at around 60\% of their initial levels }
   \label{fig:Resilience}
\end{figure*}

Armed with this approach we now modify the parameter $q$ and analyze the changes in network connectivity. We start by asking whether or not the edges that fail under our $q$-perturbation differ during changing conditions. The result is shown in Fig.~\ref{6a}. We find that initially (before the lockdown) the network is insensitive to the removal of low-value edges. That is, only at high levels of $q$ do significant edges fail. During lockdown, however, the picture changes. Now we see that many edges are at 20\%-60\% of their maximal levels and the weight they carry is very substantial. Thus the network is much more brittle than it was initially and edges carry only a small fraction of their maximum historic capacity. We can ask, then, how is this under-the-surface fragmentation reflect on the GCC, which is often considered the main functioning component of the network. The standard presentation~\cite{Zhang8673} relates the level at which the GCC fragments, termed $q_c$, to network strength and resilience, showing that high $q_c$ corresponds to a robust network, while low levels of $q_c$ characterize a weak, fragile network. Fig.~\ref{6b} shows at which value of $q=q_c$ the GCC breaks down, i.e. split into noticeably smaller components. We note, again, that before the lockdown it happens only close to $q=1$, that is, when the edges are removed only when they are marginally below their maximal value. What this means, simply, is that before the lockdown, the edge weights are relatively stable and close to their typical peaks. During lockdown the network becomes much more fragile and breaks down at much lower levels, at $q_c$ close to 0.2. That means the edges that are present - are nearly five times weaker than in normal times. On one hand, that is to be expected. We claimed throughout the analysis mobility had been highly impacted. On the other hand, we saw in Fig.~\ref{5a} that node weighs dropped by about 60\% on average - but the surprising detail here is that the global mobility is actually more fragile, there are many critical edges that break under very light pressure. But perhaps most interesting is the situation post lockdown, when everything, presumably, goes back to normal. Looking at nominal edges, and even to some extent node degrees and distributions, it may be assumed that this is indeed the case. However, under the magnifying glass of $q$-analysis we see that picture is not so clear. Below we explore the extent to which the network remains damaged.

In order to further understand the lasting impact of the lockdown, we compare the nodes that form the GCC at a given $q$-level over time. To this end we calculate what is known as the Jaccard coefficient, which is a straight-forward metric to measure node similarity. It is defined as the size of intersection of two sets divided by its union: 

\begin{equation}J(q)=\frac{GCC_1(q)\cap GCC_2(q)}{GCC_1(q)\cup GCC_2(q)}\end{equation}

We perform this calculation for a range of $q$ values and discover (Fig.~\ref{6c}) that the GCC, as constructed at various levels of $q$, 40\% and up, retains only about 50\%-70\% of the initial nodes. Moreover, during the lockdown itself, when the simple intersection shows a decrease of 60\%-70\% as seen by other metrics in Figs.~\ref{fig:NetworkBehavior}-~\ref{fig:ImpactScatter}, even for very low levels of $q$, the GCC is almost entirely gone, bearing only about 10\% similarity to the initial structure. We now want to use this understanding to tie together the levels of mobility and observed decline of network functionality to the actual levels of economic activity.

\section{Economic Impact}
\label{sec:Econ} 
Reliable and accurate data on economic activity from official sources takes a long time to measure and evaluate. There are many reasons for that, some include multiple sources of information that need to be gathered and unified. Others just due to the fact that economic and financial information is reported long after the occurrence of the phenomenon. For instance, a company would file its fiscal year reports only towards the end of the following year's first quarter, and similarly national accounts are typically released with some months of delay. 

However, mobility patterns can be readily observed in near-real-time via various methods. One such method is employed in this manuscript, relying on cellphone data. Other providers of such data exist and often share or sell this data so it can be observed in very high spatio-temporal resolution. Other measures include public transportation usage, sensors that count numbers of vehicles on the roads and more. In short, mobility is relatively easy to observe. 

\begin{figure*}[h!]
   \centering
   \hspace{0.00mm}
   \subfloat[][]{\includegraphics[width=.45\textwidth,valign=t]{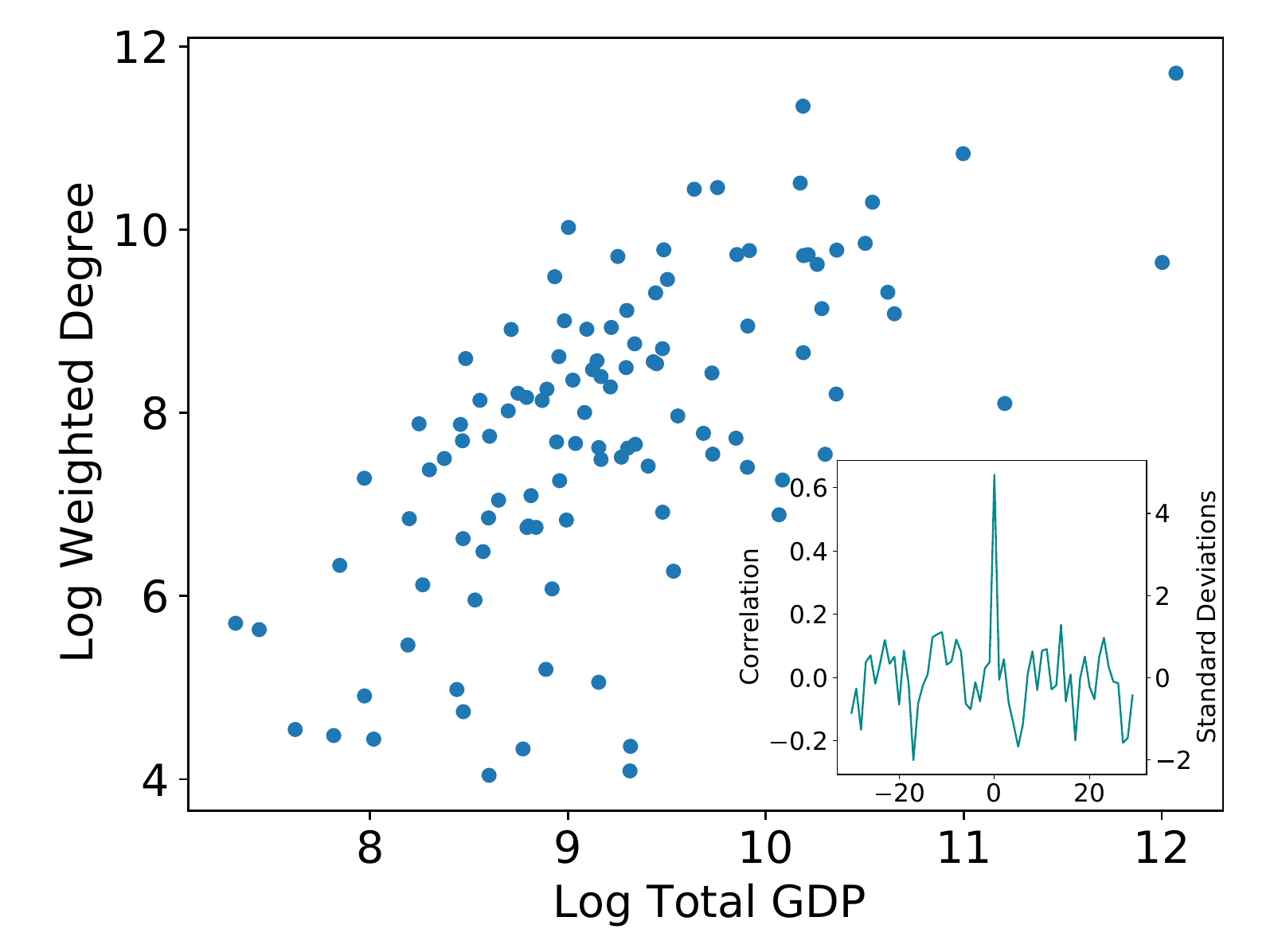}\label{7a}}
   \hspace{0.00mm}
   \subfloat[][]{\includegraphics[width=.45\textwidth,valign=t]{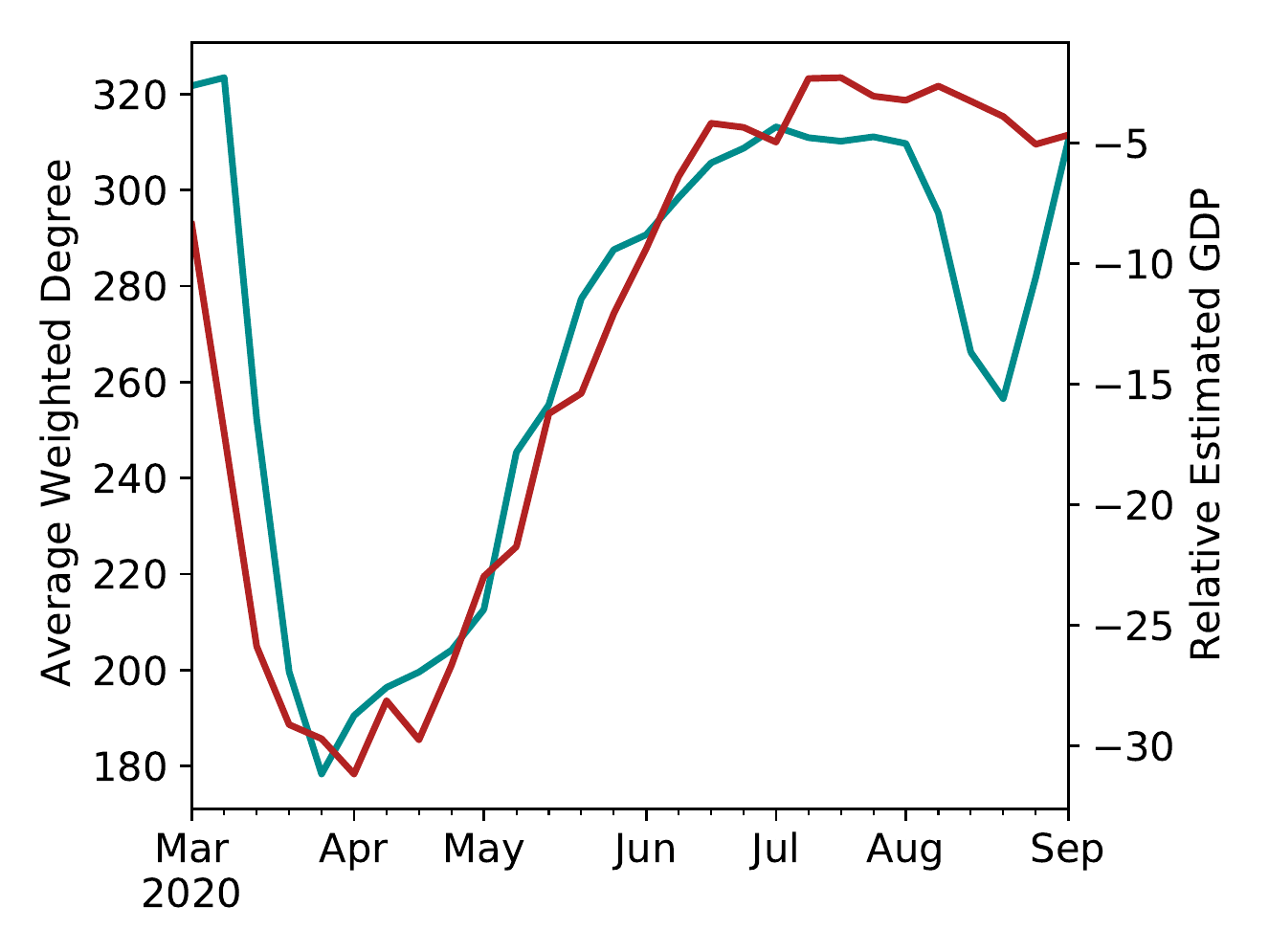}\label{7b}}
      \hspace{0.00mm}
   \subfloat[][]{\includegraphics[width=.45\textwidth,valign=t]{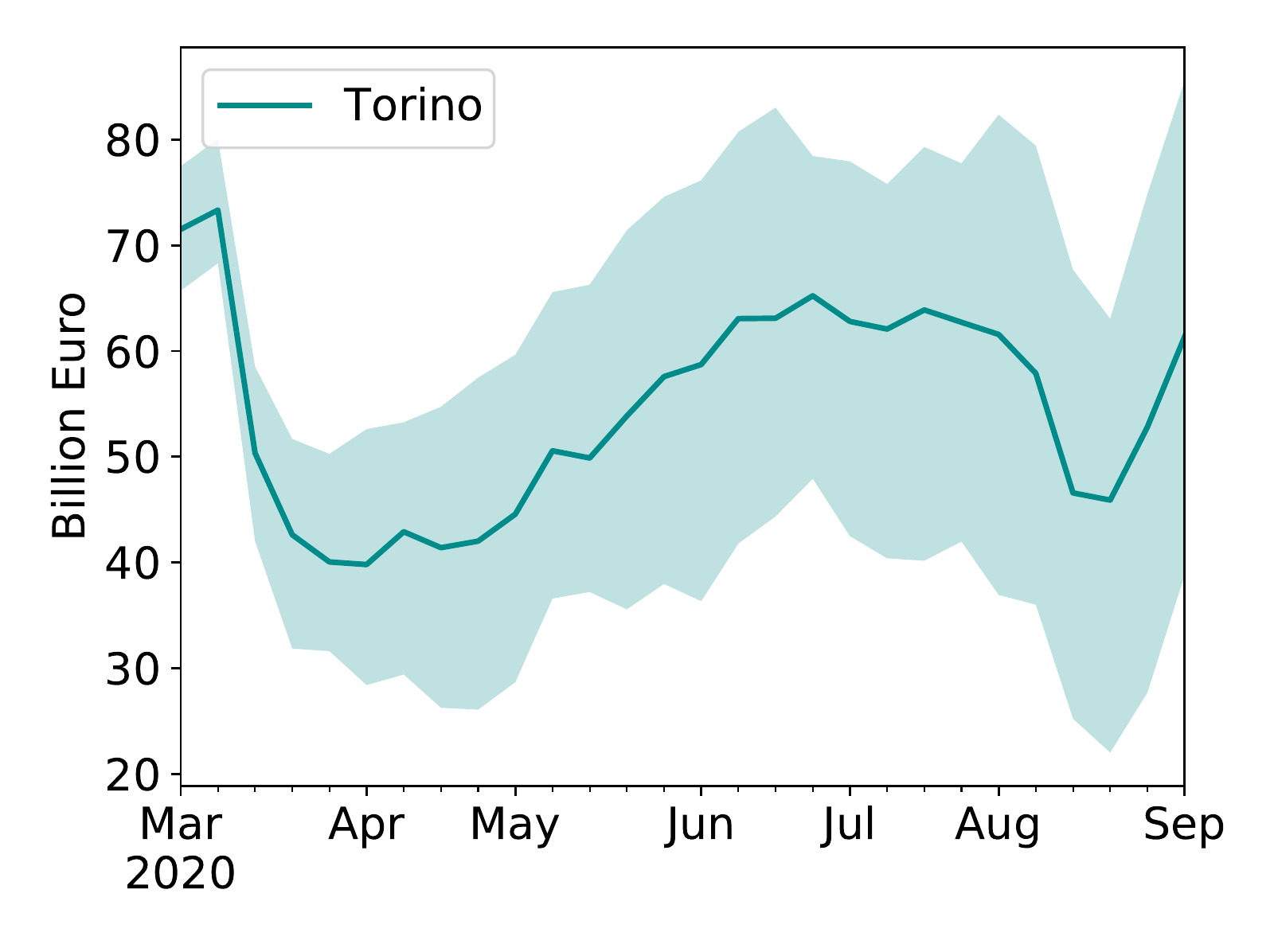}\label{7c}}   
         \hspace{0.00mm}
   \subfloat[][]{\includegraphics[width=.45\textwidth,valign=t]{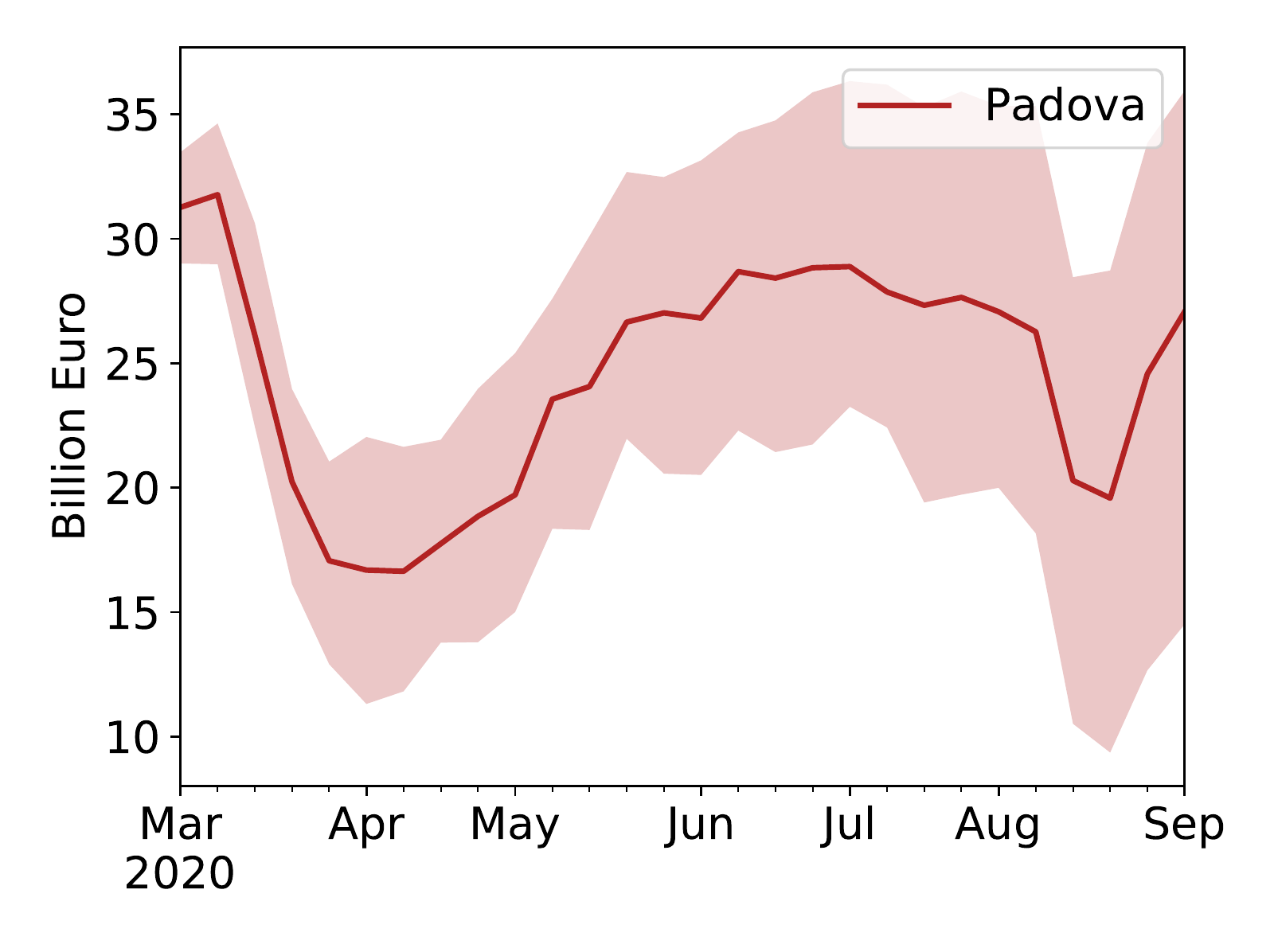}\label{7d}}   
         \hspace{0.00mm}
   \subfloat[][]{\includegraphics[width=.45\textwidth,valign=t]{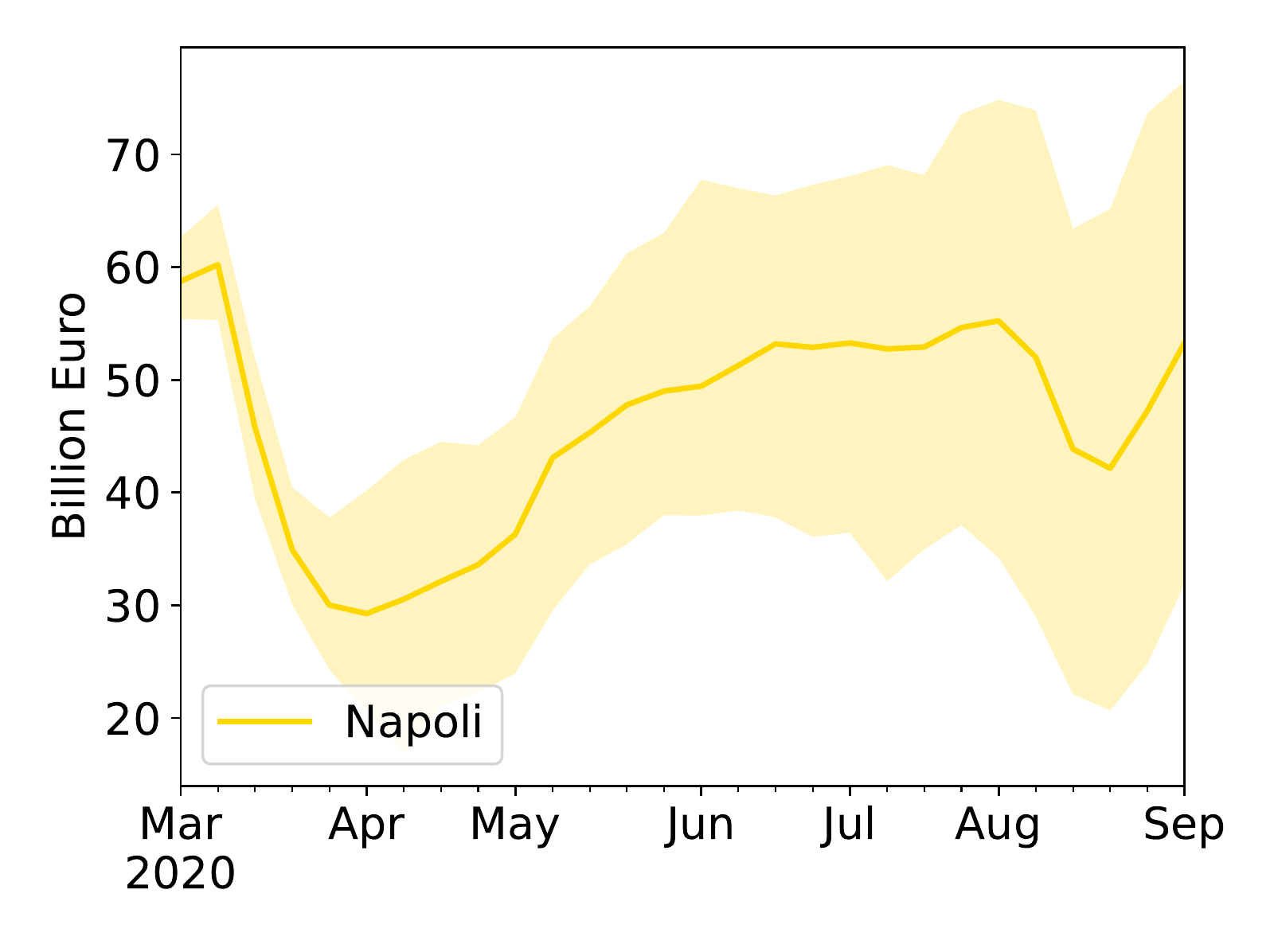}\label{7e}}
         \hspace{0.00mm}
   \subfloat[][]{\includegraphics[width=.45\textwidth,valign=t]{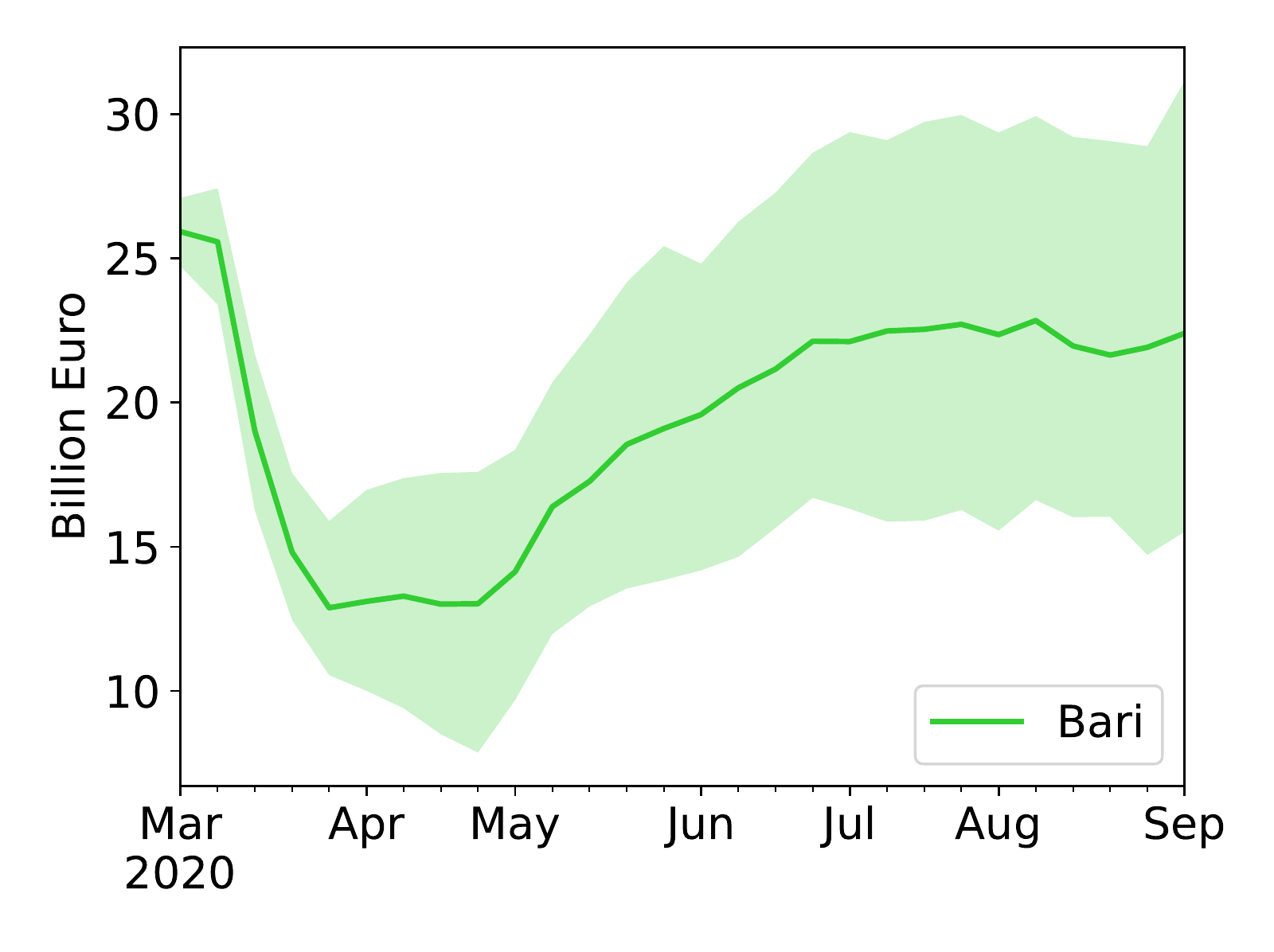}\label{7f}}   

   \caption{\footnotesize Inferring the economy from mobility. (a) The correlation between the GDP levels of individual Provinces for 2017 and the weighted Province degree. Log of values is shown to highlight the relation holds regardless the strongest Province economies. Inset - shifted correlation to validate the significance of the relationship (b) The Average weighted degree, teal, left axis; and the EGDP estimates, red, right axis. Here, as in the right panel of Fig.	~\ref{fig:NetworkBehavior} we see the similarity in dynamics. (c)-(f) The Province-based GDP forecast based on mobility data for Turin, Padua, Como and Pisa Provinces. The thick line is the average forecast, with the shaded area showing the 25-75 percentile range.}
   \label{fig:Econ}
\end{figure*}

Here we propose the use of mobility data as a proxy for the economic performance. As an example, Fig.~\ref{7a} shows the correlation between Italian Provinces' Gross Domestic Product (GDP) and their mobility-based weighted degree. The official economic data used in this figure is only available until 2017. Newer data is available only for the regional level, for 2018, while if we are interested in quarterly data of 2019 and onward, we can only look at the country level. This creates a difficult situation for policy makers and stakeholders when economic impact or distribution of financial aid need to be determined at a granular spatial level of intervention. 

Thus, we come to the concluding section of our analysis, bringing together the various network insights gathered so far into an effective fine-grained decision-aiding tool (see \nameref{sec:MatMet} for further details). This tool is made available due to two interesting facts. One, as seen in Fig.~\ref{7a}, there is a close relation between a Province's mobility performance and its GDP. In fact, the levels of mobility correlate with the GDP more than 60\%: $corr(GDP, \langle K_w \rangle) = 0.64$ (we present additional metrics tying together GDP per capita to average degree, as well as ``normalized'' average degree, in the SI). This means that mobility is a reasonable proxy for the economy. We next compare OECD data which calculates economic activity based on multiple attainable sources on the national level~\cite{/content/paper/6b9c7518-en} to produce an estimated GDP (EGDP) with our mobility metrics. While this is an approximate metric, it analyzes various Internet search patterns related to economic activity and shows a close prediction (to within its margin of error) to the officially published quarterly national GDP~\cite{oecdTracking}. Fig.~\ref{7b} shows how closely the average mobility weighted degree (teal, left axis) follows the proposed economic metric (red, right axis). The correlation here is even higher, $\rho (EGDP, \langle K_{w,p} \rangle) = 0.89$, and the weekly updates $\Delta EGDP = EGDP(i) - EGDP(i-1), \Delta \langle K_w\rangle =\langle  K_w(i)\rangle - \langle K_w(i-i)\rangle$, are also very closely tied,  $\rho (\Delta EGDP, \Delta\langle K_w \rangle) = 0.58$.

While the causal relations between affected mobility and economic impact may be far from trivial (although the intuition that people, when traveling, conduct various business activities, is straightforward), we now have a means to offer both a national and a regional or local estimate of impact to the economy. Specifically, the remarkably linear relations of Fig.~\ref{fig:ImpactScatter} can help us give a rough estimate of a uniform impact to Provinces' GDPs. However, \textit{the deviations from that linear relationship,} as seen from the mobility patterns of each individual Province, as shown above (Sec.~\ref{sec:Impact}), together with the close relations between GDP and mobility as shown here, can allow for a finer-grained, Province-specific analysis, and give in-depth expectations about heterogeneous local economic impact from the observed dynamics of mobility. Figs.~\ref{7c}-\ref{7f} show a sample of such forecasts. The process to obtain the estimates is detailed in~\nameref{sec:MatMet}.

\begin{figure*}[h!]
   \centering
   \hspace{0.00mm}
   \subfloat{\includegraphics[width=\textwidth,valign=t]{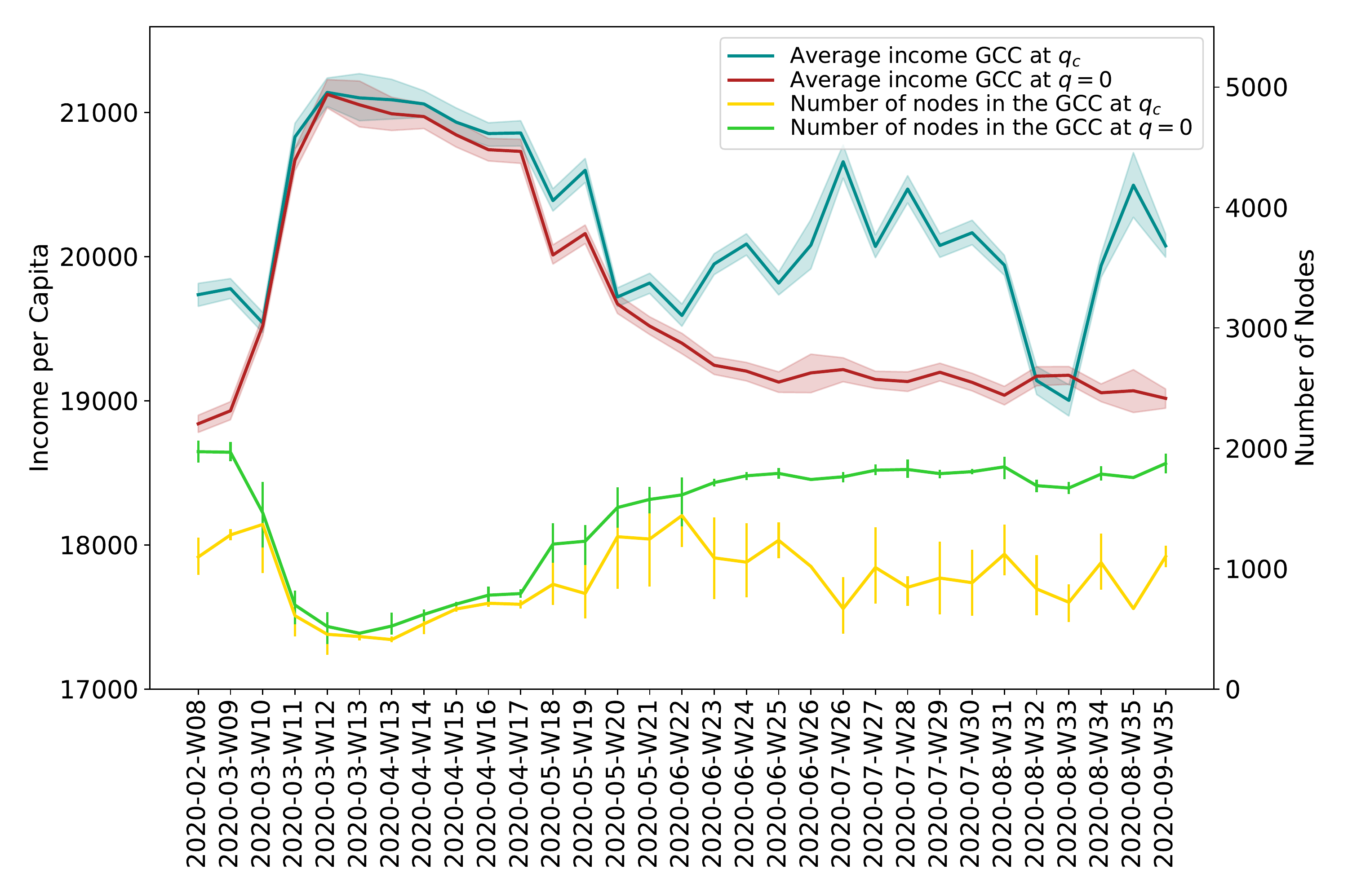}\label{8}}
   \caption{\footnotesize Average income per capita and total number of nodes in the GCC for every week of the period of observation. The GCC is either calculated at the start of the percolation process, i. e. without removing any nodes (GCC at $q=0$, red and green lines), or it is obtained by removing edges incrementally by weight until the optimal $q$ is reached but before the GCC is dissolved (GCC at $q_c$, blue and yellow lines). The lines represent the mean value of income per capita of the nodes or the total number of nodes in the GGC, averaged over one week of observations. 
   }
   \label{fig:Econ_2}
\end{figure*}

Finally, in order to uncover heterogeneous responses of Italian territories to mobility restrictions which may signal competing economic trajectories of local economic systems, we show in Fig.~\ref{fig:Econ_2} how the network of Italian mobility relates to average income levels. We consider the network of mobility constructed at the Municipality level and income per capita referred to 2019, hence before the outbreak of the pandemic. The red line in Fig.~\ref{fig:Econ_2} shows the dynamics of GCC, while in green we report its number of nodes. The Italian mobility network is composed by a main connected group of Municipalities which are on average associated with higher levels of income per capita than those more peripheral in the network. Interestingly, after the deployment of lockdown restrictions we note a remarkable reduction in the size of GCC, while the economic well-being of the remaining population, as seen through the average income per capita, increases. Hence, a relevant effect of mobility restrictions put in place to contain the spread of the virus is the emergence of an enforced economic segregation, where Municipalities with high-income population tend to be connected while Municipalities with poorer economic conditions of their population are excluded from the main component of the Italian mobility network. This effect vanishes once mobility restrictions are lifted, with a rapid reversal to the pre-lockdown configuration. In addition, the percolation process is informative to identify those Municipalities which may exit GCC when the network starts deteriorating. As shown in Fig.~\ref{fig:Resilience}, lockdown restrictions determine a sharp decrease of the parameter $q$. As a consequence, Fig.~\ref{fig:Econ_2} highlights how during the lockdown on average the configuration reached before the dissipation of GCC is very similar to the one at $q=0$. Nevertheless, during business as usual periods (pre-lockdown phase) or after the removal of lockdown restrictions, we notice that progressively nodes that are removed have lower average income per capita than those that remain in the GCC. Once again, the core of the GCC seems to point to the presence of highly connected nodes that have on average also better economic conditions of their population.

\section{Conclusion}\label{sec:Conclusion}
We analyze in detail the changes in mobility patterns in Italy induced by the lockdown measures. We show the fragmentation and disconnection they brought. While it is without a doubt the intention of those measures, we show also, using concepts borrowed from percolation theory, that the impact goes much deeper than is seen superficially. In fact, the network after the removal of limitations is still far more fragile than it was initially. We estimate the overall impact and the deviations from that uniformity for various Regions. Combining those deviations with close relations between mobility and economic conditions, we present a novel approach to estimate hard-to-observe economic activity from easily observable mobility patterns. These perspectives on the network resilience and local economic impact of lockdown measures may be used by policy makers to estimate more accurately and promptly the costs and benefits of potential additional restrictive measures.

\section*{Materials and Methods}\label{sec:MatMet}
\subsection*{Data}
The data used for the analysis is provided under academic license agreement with Facebook through the ``Data for Good'' program and is available at \url{https://dataforgood.fb.com/tools/disease-prevention-maps/}. Economic data is taken from the Istituto Nazionale di Statistica (ISTAT) \url{http://dati.istat.it/} and the Ministero dell'Economia e delle Finanze (MEF) \url{https://www.mef.gov.it/documenti-pubblicazioni/open-data/index.html}.\\

\subsection*{Network structure}
As described in Sec.~\ref{struct}, we construct our network by observing people move between various locations. For ease of processing, continuous space is divided by Facebook into tiles, approximately 7x7km each, covering the entire country. When calculating overall statistics, such as degree distributions, component sizes and shortest paths we analyze the network using those elements. However, in order to make our research more meaningful from a geographic and economic perspective, we also map those tiles to Italy's Regions and Provinces based on the the location of a tile's center. For that we aggregate all relevant node information into a higher-level node (Province or Region) by summing all the edge weights incident on the appropriate nodes.

Edges represent the number of people traveling within an eight-hour period between two nodes. That is, if over a period of eight hours, $N > 10$ people were located at nodes \textit{u} and \textit{v} then and edge will exist between the two nodes and its weight will be \textit{N}. The reason for this limitation lies in privacy restrictions, where a smaller number of people may make it possible to identify specific individuals, so information is only exposed where such deanonymization is unlikely. 

Those two limitations carry noteworthy meaning to our analysis. For one, the relatively short time period of eight hours is shorter than the typical time frames in human mobility research. That leads directly to shorter typical distances traveled. Secondly, the limitation on quantity can also directly lead to the rarer but longer edges being omitted from our data set.

Self-loops reflect a situation where, throughout the measurement period, a person is not recorded outside their initial tile. Those events are frequent in our analysis, more so at the large aggregation levels. Those loops effectively mean people stay in the initial location, whether a tile or a Region. Since we are interested in connectivity between various locations, those loops are removed from our analysis.

Information is only available about the presence of people in two locations, without time ordering, so our edges are undirected. This precludes us from performing Origin-Destination analysis, or to follow explicitly the direction of flow of people\\

\subsection*{Percolation and Resilience analysis}
Percolation takes its origins in the analysis of disordered systems, where matter is modeled as sites that may or may not be occupied with probability $p_n$ (node percolation) or sites between which edges are present or absent with probability $p_e$. The question is then asked whether or not a Giant Connected Component exists, that is, whether the largest component is extensive, i.e., proportional to the original system size. Originally developed around regular systems, such as two-dimensional or three-dimensional lattices, network theory expanded the notion to random networks, where each ``site'' has potentially a random degree taken from a given degree distribution. Specifically, important differences appear when the degrees are drawn from a Poisson distribution (Erdos-Renyi, (ER)  networks, where every two nodes are connected by an edge with probability $p$) or a scale free (SF) degree distribution, where node degrees are distributed according to a power law, $P(x) \propto x^{-\beta}$ with usually $2 < \beta < 3$. In the former case the network experiences a \textit{percolation transition} when the average degreee, $\langle k \rangle$ crosses the threshold of 1, where below that value the GCC increases slowly with network size, $N$, and its relative fraction of the entire network goes to zero as $N$ goes to infinity~\cite{newman2018networks,cohen2010complex}. SF networks, on the other hand, do not experience such a transition, and rather have a non-zero GCC for any nonzero average degree.
It is often assumed for real-world networks that in order for a node to be functional, whether the node is a financial institution, a geographical location or an infrastructure component such as a power plant~\cite{buldyrev2010catastrophic}, it needs to be part of the GCC. While we do not suggest Regions, Provinces and Municipalities disconnected from the GCC cease to function, we do suggest their participation in economic activity is impaired. Since there can only be one GCC in a network~\cite{newman2018networks,cohen2010complex}, we assume only nodes connected to it operate at their fullest capacity. Thus, as described in Sec.~\ref{sec:Resilinece}, we define a critical level of $q$ at which the reduction of the GCC is largest, thereby reflecting the most significant impact to the economic activity.

\subsection*{Province-based GDP forecast}
To generate GDP forecasts per Province we combine the close relations between mobility and GDP, both statically (Fig.~\ref{7a}) and dynamically (Fig.~\ref{7b}). For that we rewrite the GDP over time for a Region as 
\begin{equation} GDP_t = GDP_0 + \sum_{i=1}^{t} \Delta GDP_i \label{eq:summoves}\end{equation} 
where $GDP_0$ is the initial level (here taken as the value at 2017 as the latest published data point for Provinces, but may be replaced with a more recent estimate), and the weekly updates $GDP_i$ are estimated using the dynamic relations. We may also rewrite the process as a multiplication of percentage changes, more in keeping with economic literature, but no additional insight is gained by it and the equations become more cumbersome. We now regress the weekly GDP changes against the mobility updates as 

\begin{equation} \Delta GDP_i = \alpha \cdot \Delta \langle K_i \rangle + \beta \label{eq:linfit} \end{equation} for the initial estimate. Then, based on the correlation between $GDP_i$ and $\Delta \langle K_i \rangle$ we generate correlated random variables from the latter and substitute~\ref{eq:linfit} into~\ref{eq:summoves}. The last part is done so that statistics can be generated around the average levels, that may be derived directly by plugging the relevant Province's $\langle K_i \rangle$ into~\ref{eq:linfit}. We now have all the components to calculate both the expected $GDP_t$ and the range it may take.

\section*{Acknowledgements}
We thank the 
Israel Science Foundation, the Binational Israel-China
Science Foundation (Grant No. 3132/19), the  NSF-BSF (Grant No. 2019740), the EU
H2020 project RISE (Project No. 821115), the EU H2020
DIT4TRAM, and DTRA (Grant No. HDTRA-1-19-1-
0016) for financial support. 

\section*{Author contribution}
All authors conceived and designed the research. A.S. and G.B. carried out the numerical experiments and analysed them. A.S., G.B., A.F and S.H. interpreted the simulation results. All authors wrote the manuscript.

\section*{Competing interests}
The authors declare no competing interests

\clearpage
\printbibliography 

@MISC{scrollin,
   author =       {Venkataramakrishnan, Rohan},
   title =        {Coronavirus: Did India rush into a full lockdown without planning? Or did it have little choice?},
   month =    {03},
   year =         {2020},
   url =         {https://scroll.in/article/957101/coronavirus-did-india-rush-into-a-lockdown-without-planning-or-did-it-have-little-choice},

 }

@MISC{guardian,
   author =       {Ward, Helen},
   title =        {We scientists said lock down. But UK politicians refused to listen},
   month =    {04},
   year =         {2020},
   url =         {https://www.theguardian.com/commentisfree/2020/apr/15/uk-government-coronavirus-science-who-advice},
 }

@MISC{oecd,
   author =       {OECD},
   title =        {E-commerce in the time of COVID-19},
   month =    {11},
   year =         {2020},
   url =         {http://www.oecd.org/coronavirus/policy-responses/e-commerce-in-the-time-of-covid-19-3a2b78e8/},
 }

@MISC{toi,
   author =       {Ahren, Raphael},
   title =        {Complete lockdown would cause more harm than good, dissenting top doctor claims},
   month =    {03},
   year =         {2020},
   url =         {https://www.timesofisrael.com/complete-lockdown-would-cause-more-harm-than-good-dissenting-top-doctor-claims/},
 }

@book{newman2018networks,
  title={Networks},
  author={Newman, Mark},
  year={2018},
  publisher={Oxford university press}
}

@book{barabasi2016network,
  title={Network science},
  author={Barab{\'a}si, Albert-L{\'a}szl{\'o} and others},
  year={2016},
  publisher={Cambridge university press}
}

@book{cohen2010complex,
  title={Complex networks: structure, robustness and function},
  author={Cohen, Reuven and Havlin, Shlomo},
  year={2010},
  publisher={Cambridge university press}
}

@article {Zhang8673,
	author = {Zhang, Limiao and Zeng, Guanwen and Li, Daqing and Huang, Hai-Jun and Stanley, H. Eugene and Havlin, Shlomo},
	title = {Scale-free resilience of real traffic jams},
	journal={Proceedings of the National Academy of Sciences},
	volume = {116},
	number = {18},
	pages = {8673--8678},
	year = {2019},
	doi = {10.1073/pnas.1814982116},
	publisher = {National Academy of Sciences},
}

@book{bollobas2013modern,
  title={Modern graph theory},
  author={Bollob{\'a}s, B{\'e}la},
  volume={184},
  year={2013},
  publisher={Springer Science \& Business Media}
}

@article{albert2002statistical,
  title={Statistical mechanics of complex networks},
  author={Albert, R{\'e}ka and Barab{\'a}si, Albert-L{\'a}szl{\'o}},
  journal={Reviews of modern physics},
  volume={74},
  number={1},
  pages={47},
  year={2002},
  publisher={APS}
}

@article {Bartik17656,
	author = {Bartik, Alexander W. and Bertrand, Marianne and Cullen, Zoe and Glaeser, Edward L. and Luca, Michael and Stanton, Christopher},
	title = {The impact of COVID-19 on small business outcomes and expectations},
	volume = {117},
	number = {30},
	pages = {17656--17666},
	year = {2020},
	doi = {10.1073/pnas.2006991117},
	publisher = {National Academy of Sciences},
	issn = {0027-8424},
	url = {https://www.pnas.org/content/117/30/17656},
	eprint = {https://www.pnas.org/content/117/30/17656.full.pdf},
	journal = {Proceedings of the National Academy of Sciences}
}

@article{gross2020spatio,
  title={Spatio-temporal propagation of COVID-19 epidemics},
  author={Gross, Bnaya and Zheng, Zhiguo and Liu, Shiyan and Chen, Xiaoqi and Sela, Alon and Li, Jianxin and Li, Daqing and Havlin, Shlomo},
  journal={Europhysics Letters 131, 58003-58008},
  year={2020}
}

@article{bascompte2003nested,
  title={The nested assembly of plant--animal mutualistic networks},
  author={Bascompte, Jordi and Jordano, Pedro and Meli{\'a}n, Carlos J and Olesen, Jens M},
  journal={Proceedings of the National Academy of Sciences},
  volume={100},
  number={16},
  pages={9383--9387},
  year={2003},
  publisher={National Acad Sciences}
}

@article{williams2000simple,
  title={Simple rules yield complex food webs},
  author={Williams, Richard J and Martinez, Neo D},
  journal={Nature},
  volume={404},
  number={6774},
  pages={180--183},
  year={2000},
  publisher={Nature Publishing Group}
}

@article{gao2016universal,
  title={Universal resilience patterns in complex networks},
  author={Gao, Jianxi and Barzel, Baruch and Barab{\'a}si, Albert-L{\'a}szl{\'o}},
  journal={Nature},
  volume={530},
  number={7590},
  pages={307--312},
  year={2016},
  publisher={Nature Publishing Group}
}

@article{jeong2000large,
  title={The large-scale organization of metabolic networks},
  author={Jeong, Hawoong and Tombor, B{\'a}lint and Albert, R{\'e}ka and Oltvai, Zoltan N and Barab{\'a}si, A-L},
  journal={Nature},
  volume={407},
  number={6804},
  pages={651--654},
  year={2000},
  publisher={Nature Publishing Group}
}

@article{barabasi2004network,
  title={Network biology: understanding the cell's functional organization},
  author={Barabasi, Albert-Laszlo and Oltvai, Zoltan N},
  journal={Nature reviews genetics},
  volume={5},
  number={2},
  pages={101--113},
  year={2004},
  publisher={Nature Publishing Group}
}

@article{li2015percolation,
  title={Percolation transition in dynamical traffic network with evolving critical bottlenecks},
  author={Li, Daqing and Fu, Bowen and Wang, Yunpeng and Lu, Guangquan and Berezin, Yehiel and Stanley, H Eugene and Havlin, Shlomo},
  journal={Proceedings of the National Academy of Sciences},
  volume={112},
  number={3},
  pages={669--672},
  year={2015},
  publisher={National Acad Sciences}
}

@article{buldyrev2010catastrophic,
  title={Catastrophic cascade of failures in interdependent networks},
  author={Buldyrev, Sergey V and Parshani, Roni and Paul, Gerald and Stanley, H Eugene and Havlin, Shlomo},
  journal={Nature},
  volume={464},
  number={7291},
  pages={1025},
  year={2010},
  publisher={Nature Publishing Group}
}

@article{shao2015percolation,
  title={Percolation of localized attack on complex networks},
  author={Shao, Shuai and Huang, Xuqing and Stanley, H Eugene and Havlin, Shlomo},
  journal={New Journal of Physics},
  volume={17},
  number={2},
  pages={023049},
  year={2015},
  publisher={IOP Publishing}
}

@article{gai2010contagion,
  title={Contagion in financial networks},
  author={Gai, Prasanna and Kapadia, Sujit},
  journal={Proceedings of the Royal Society A: Mathematical, Physical and Engineering Sciences},
  volume={466},
  number={2120},
  pages={2401--2423},
  year={2010},
  publisher={The Royal Society Publishing}
}

@article{battiston2012debtrank,
  title={Debtrank: Too central to fail? financial networks, the fed and systemic risk},
  author={Battiston, Stefano and Puliga, Michelangelo and Kaushik, Rahul and Tasca, Paolo and Caldarelli, Guido},
  journal={Scientific Reports},
  volume={2},
  pages={541},
  year={2012},
  publisher={Nature Publishing Group}
}

@article{smolyak2020mitigation,
  title={Mitigation of cascading failures in complex networks},
  author={Smolyak, Alex and Levy, Orr and Vodenska, Irena and Buldyrev, Sergey and Havlin, Shlomo},
  journal={Scientific reports},
  volume={10},
  number={1},
  pages={1--12},
  year={2020},
  publisher={Nature Publishing Group}
}

@article{bonaccorsi2020economic,
  title={Economic and social consequences of human mobility restrictions under COVID-19},
  author={Bonaccorsi, Giovanni and Pierri, Francesco and Cinelli, Matteo and Flori, Andrea and Galeazzi, Alessandro and Porcelli, Francesco and Schmidt, Ana Lucia and Valensise, Carlo Michele and Scala, Antonio and Quattrociocchi, Walter and others},
  journal={Proceedings of the National Academy of Sciences},
  volume={117},
  number={27},
  pages={15530--15535},
  year={2020},
  publisher={National Acad Sciences}
}

@article{brockmann2006scaling,
  title={The scaling laws of human travel},
  author={Brockmann, Dirk and Hufnagel, Lars and Geisel, Theo},
  journal={Nature},
  volume={439},
  number={7075},
  pages={462--465},
  year={2006},
  publisher={Nature Publishing Group}
}

@article{gonzalez2008understanding,
  title={Understanding individual human mobility patterns},
  author={Gonzalez, Marta C and Hidalgo, Cesar A and Barabasi, Albert-Laszlo},
  journal={nature},
  volume={453},
  number={7196},
  pages={779--782},
  year={2008},
  publisher={Nature publishing group}
}

@article{schlosser2020covid,
  title={COVID-19 lockdown induces disease-mitigating structural changes in mobility networks},
  author={Schlosser, Frank and Maier, Benjamin F and Jack, Olivia and Hinrichs, David and Zachariae, Adrian and Brockmann, Dirk},
  journal={Proceedings of the National Academy of Sciences},
  volume={117},
  number={52},
  pages={32883--32890},
  year={2020},
  publisher={National Acad Sciences}
}

@article{haug2020ranking,
  title={Ranking the effectiveness of worldwide COVID-19 government interventions},
  author={Haug, Nils and Geyrhofer, Lukas and Londei, Alessandro and Dervic, Elma and Desvars-Larrive, Am{\'e}lie and Loreto, Vittorio and Pinior, Beate and Thurner, Stefan and Klimek, Peter},
  journal={Nature human behaviour},
  pages={1--10},
  year={2020},
  publisher={Nature Publishing Group}
}

@article{vespignani2020modelling,
  title={Modelling COVID-19},
  author={Vespignani, Alessandro and Tian, Huaiyu and Dye, Christopher and Lloyd-Smith, James O and Eggo, Rosalind M and Shrestha, Munik and Scarpino, Samuel V and Gutierrez, Bernardo and Kraemer, Moritz UG and Wu, Joseph and others},
  journal={Nature Reviews Physics},
  pages={1--3},
  year={2020},
  publisher={Nature Publishing Group}
}

@article{liu2020new,
  title={A new SAIR model on complex networks for analysing the 2019 novel coronavirus (COVID-19)},
  author={Liu, Congying and Wu, Xiaoqun and Niu, Riuwu and Wu, Xiuqi and Fan, Ruguo},
  journal={Nonlinear Dynamics},
  volume={101},
  number={3},
  pages={1777--1787},
  year={2020},
  publisher={Springer}
}

@article{alstott2014powerlaw,
  title={powerlaw: a Python package for analysis of heavy-tailed distributions},
  author={Alstott, Jeff and Bullmore, Ed and Plenz, Dietmar},
  journal={PloS one},
  volume={9},
  number={1},
  pages={e85777},
  year={2014},
  publisher={Public Library of Science}
}

@article{liang2012scaling,
  title={The scaling of human mobility by taxis is exponential},
  author={Liang, Xiao and Zheng, Xudong and Lv, Weifeng and Zhu, Tongyu and Xu, Ke},
  journal={Physica A: Statistical Mechanics and its Applications},
  volume={391},
  number={5},
  pages={2135--2144},
  year={2012},
  publisher={Elsevier}
}

@article{alessandretti2020scales,
  title={The scales of human mobility},
  author={Alessandretti, Laura and Aslak, Ulf and Lehmann, Sune},
  journal={Nature},
  volume={587},
  number={7834},
  pages={402--407},
  year={2020},
  publisher={Nature Publishing Group}
}

@article{/content/paper/6b9c7518-en,
   author = "Nicolas Woloszko",
   title = "Tracking activity in real time with Google Trends",
   year = "2020",
   number = "1634", 
   url = {https://www.oecd-ilibrary.org/content/paper/6b9c7518-en},
   doi = {https://doi.org/https://doi.org/10.1787/6b9c7518-en}, 
}

@MISC{oecdTracking,
   author =       {OECD},
   title =        {Tracking GDP growth in real time},
   month =    {12},
   year =         {2020},
   url =         {http://www.oecd.org/economy/weekly-tracker-of-gdp-growth/},
 }

@article{laherrere1998stretched,
  title={Stretched exponential distributions in nature and economy:“fat tails” with characteristic scales},
  author={Laherrere, Jean and Sornette, Didier},
  journal={The European Physical Journal B-Condensed Matter and Complex Systems},
  volume={2},
  number={4},
  pages={525--539},
  year={1998},
  publisher={Springer}
}

@article{frisch1997extreme,
  title={Extreme deviations and applications},
  author={Frisch, Uriel and Sornette, Didier},
  journal={Journal de Physique I},
  volume={7},
  number={9},
  pages={1155--1171},
  year={1997},
  publisher={EDP Sciences}
}

@article {chinazzi_2020_science,
	author = {Chinazzi, Matteo and Davis, Jessica T. and Ajelli, Marco and Gioannini, Corrado and Litvinova, Maria and Merler, Stefano and Pastore y Piontti, Ana and Mu, Kunpeng and Rossi, Luca and Sun, Kaiyuan and Viboud, C{\'e}cile and Xiong, Xinyue and Yu, Hongjie and Halloran, M. Elizabeth and Longini, Ira M. and Vespignani, Alessandro},
	title = {The effect of travel restrictions on the spread of the 2019 novel coronavirus ({{COVID}}-19) outbreak},
	elocation-id = {eaba9757},
	year = {2020},
	doi = {10.1126/science.aba9757},
	publisher = {American Association for the Advancement of Science},
	issn = {0036-8075},
	url = {https://science.sciencemag.org/content/early/2020/03/05/science.aba9757},
	eprint = {https://science.sciencemag.org/content/early/2020/03/05/science.aba9757.full.pdf},
	journal = {Science}
}

@article {kraemer_2020_effect,
	author = {Kraemer, Moritz U. G. and Yang, Chia-Hung and Gutierrez, Bernardo and Wu, Chieh-Hsi and Klein, Brennan and Pigott, David M.  and du Plessis, Louis and Faria, Nuno R. and Li, Ruoran and Hanage, William P. and Brownstein, John S. and Layan, Maylis and Vespignani, Alessandro and Tian, Huaiyu and Dye, Christopher and Pybus, Oliver G. and Scarpino, Samuel V.},
	title = {The effect of human mobility and control measures on the  {{COVID}}-19 epidemic in China},
	elocation-id = {eabb4218},
	year = {2020},
	doi = {10.1126/science.abb4218},
	publisher = {American Association for the Advancement of Science},
	issn = {0036-8075},
	url = {https://science.sciencemag.org/content/early/2020/03/25/science.abb4218},
	eprint = {https://science.sciencemag.org/content/early/2020/03/25/science.abb4218.full.pdf},
	journal = {Science}
}

@article{li2021temporal,
  title={The temporal association of introducing and lifting non-pharmaceutical interventions with the time-varying reproduction number (R) of SARS-CoV-2: a modelling study across 131 countries},
  author={Li, You and Campbell, Harry and Kulkarni, Durga and Harpur, Alice and Nundy, Madhurima and Wang, Xin and Nair, Harish and for COVID, Usher Network and others},
  journal={The Lancet Infectious Diseases},
  volume={21},
  number={2},
  pages={193--202},
  year={2021},
  publisher={Elsevier}
}

@article{dehning_inferring_2020,
  title = {Inferring Change Points in the Spread of {{COVID}}-19 Reveals the Effectiveness of Interventions},
  author = {Dehning, Jonas and Zierenberg, Johannes and Spitzner, F. Paul and Wibral, Michael and Neto, Joao Pinheiro and Wilczek, Michael and Priesemann, Viola},
  year = {2020},
  journal = {Science},
  volume = {369},
  publisher = {{American Association for the Advancement of Science}},
  issn = {0036-8075, 1095-9203},
  doi = {10.1126/science.abb9789},
  url = {https://science.sciencemag.org/content/369/6500/eabb9789},
  urldate = {2021-03-09},
  eprint = {32414780},
  eprinttype = {pmid},
  langid = {english},
  number = {6500}
}

@article{brauner_inferring_2021,
  title = {Inferring the Effectiveness of Government Interventions against {{COVID}}-19},
  author = {Brauner, Jan M. and Mindermann, S\"oren and Sharma, Mrinank and Johnston, David and Salvatier, John and Gaven\v{c}iak, Tom\'a\v{s} and Stephenson, Anna B. and Leech, Gavin and Altman, George and Mikulik, Vladimir and Norman, Alexander John and Monrad, Joshua Teperowski and Besiroglu, Tamay and Ge, Hong and Hartwick, Meghan A. and Teh, Yee Whye and Chindelevitch, Leonid and Gal, Yarin and Kulveit, Jan},
  year = {2020},
  journal = {Science},
  shortjournal = {Science},
  volume = {371},
  pages = {eabd9338},
  issn = {0036-8075, 1095-9203},
  doi = {10.1126/science.abd9338},
  url = {https://www.sciencemag.org/lookup/doi/10.1126/science.abd9338},
  urldate = {2021-03-08},
  langid = {english},
  number = {6531}
}

@book{buldyrev2020rise,
  title={The rise and fall of business firms: A stochastic framework on innovation, creative destruction and growth},
  author={Buldyrev, SV and Pammolli, Fabio and Riccaboni, Massimo and Stanley, Harry Eugene},
  year={2020},
  publisher={Cambridge University Press}
}

@article{guan_global_2020,
  title = {Global Supply-Chain Effects of {{COVID}}-19 Control Measures},
  author = {Guan, Dabo and Wang, Daoping and Hallegatte, Stephane and Davis, Steven J. and Huo, Jingwen and Li, Shuping and Bai, Yangchun and Lei, Tianyang and Xue, Qianyu and Coffman, D'Maris and Cheng, Danyang and Chen, Peipei and Liang, Xi and Xu, Bing and Lu, Xiaosheng and Wang, Shouyang and Hubacek, Klaus and Gong, Peng},
  date = {2020-06},
  journal = {Nature Human Behaviour},
  shortjournal = {Nat Hum Behav},
  volume = {4},
  pages = {577--587},
  issn = {2397-3374},
  doi = {10.1038/s41562-020-0896-8},
  url = {http://www.nature.com/articles/s41562-020-0896-8},
  urldate = {2020-11-12},
  langid = {english},
  number = {6}
}

@article{chang_mobility_2020,
  title = {Mobility Network Models of {{COVID}}-19 Explain Inequities and Inform Reopening},
  author = {Chang, Serina and Pierson, Emma and Koh, Pang Wei and Gerardin, Jaline and Redbird, Beth and Grusky, David and Leskovec, Jure},
  date = {2020-11-10},
  journal = {Nature},
  shortjournal = {Nature},
  issn = {0028-0836, 1476-4687},
  doi = {10.1038/s41586-020-2923-3},
  url = {http://www.nature.com/articles/s41586-020-2923-3},
  urldate = {2020-11-12},
  langid = {english}
}

@article{cox_initial_nodate,
  title = {Initial {{Impacts}} of the {{Pandemic}} on {{Consumer Behavior}}: {{Evidence}} from {{Linked Income}}, {{Spending}}, and {{Savings Data}}},
  author = {Cox, Natalie and Ganong, Peter and Noel, Pascal and Vavra, Joseph and Wong, Arlene and Farrell, Diana and Greig, Fiona},
  pages = {37},
  langid = {english}
}

@techreport{chetty_economic_nodate,
  title = {The {{Economic Impacts}} of {{COVID}}-19: {{Evidence}} from a {{New Public Database Built Using Private Sector Data}}},
  author = {Chetty, Raj and Friedman, John N and Hendren, Nathaniel and Stepner, Michael},
  pages = {116},
  year={2020},
  institution = "National Bureau of Economic Research",
  langid = {english}
}

@techreport{carvalho_tracking_nodate,
  title = {Tracking the {{COVID}}-19 {{Crisis}} with {{High}}-{{Resolution Transaction Data}}},
  author = {Carvalho, Vasco M and Garcia, Juan R and Hansen, Stephen and Ortiz, \'Alvaro and Rodrigo, Tomasa and Mora, Jos\'e V Rodr\'iguez},
  pages = {49},
  year={2020},
  institution={CEPR Discussion Paper No. DP14642},
  langid = {english}
}

@article{martin2020socio,
  title={Socio-economic impacts of COVID-19 on household consumption and poverty},
  author={Martin, Amory and Markhvida, Maryia and Hallegatte, St{\'e}phane and Walsh, Brian},
  journal={Economics of disasters and climate change},
  volume={4},
  number={3},
  pages={453--479},
  year={2020},
  publisher={Springer}
}

@article{aleta_modelling_2020,
  title = {Modelling the Impact of Testing, Contact Tracing and Household Quarantine on Second Waves of {{COVID}}-19},
  author = {Aleta, Alberto and Mart\'in-Corral, David and Pastore y Piontti, Ana and Ajelli, Marco and Litvinova, Maria and Chinazzi, Matteo and Dean, Natalie E. and Halloran, M. Elizabeth and Longini Jr, Ira M. and Merler, Stefano and Pentland, Alex and Vespignani, Alessandro and Moro, Esteban and Moreno, Yamir},
  date = {2020-09},
  journal = {Nature Human Behaviour},
  volume = {4},
  pages = {964--971},
  publisher = {{Nature Publishing Group}},
  issn = {2397-3374},
  doi = {10.1038/s41562-020-0931-9},
  url = {https://www.nature.com/articles/s41562-020-0931-9},
  urldate = {2020-11-20},
  issue = {9},
  langid = {english},
  number = {9}
}

@article{block2020social,
  title={Social network-based distancing strategies to flatten the COVID-19 curve in a post-lockdown world},
  author={Block, Per and Hoffman, Marion and Raabe, Isabel J and Dowd, Jennifer Beam and Rahal, Charles and Kashyap, Ridhi and Mills, Melinda C},
  journal={Nature Human Behaviour},
  volume={4},
  number={6},
  pages={588--596},
  year={2020},
  publisher={Nature Publishing Group}
}

@article{pullano_evaluating_2020, title={Evaluating the effect of demographic factors, socioeconomic factors, and risk aversion on mobility during the {{COVID}}-19 epidemic in {F}rance under lockdown: a population-based study}, volume={0}, ISSN={2589-7500}, url={https://www.thelancet.com/journals/landig/article/PIIS2589-7500(20)30243-0/abstract}, DOI={10.1016/S2589-7500(20)30243-0}, number={0}, journal={The Lancet Digital Health}, publisher={Elsevier}, author={Pullano, Giulia and Valdano, Eugenio and Scarpa, Nicola and Rubrichi, Stefania and Colizza, Vittoria}, year={2020}, month={11} }

@article{spelta2020after,
  title={After the lockdown: simulating mobility, public health and economic recovery scenarios},
  author={Spelta, Alessandro and Flori, Andrea and Pierri, Francesco and Bonaccorsi, Giovanni and Pammolli, Fabio},
  journal={Scientific Reports},
  volume={10},
  number={1},
  pages={1--13},
  year={2020},
  publisher={Nature Publishing Group}
}

@article{jia_population_2020, title={Population flow drives spatio-temporal distribution of COVID-19 in China}, volume={582}, ISSN={1476-4687}, DOI={10.1038/s41586-020-2284-y}, abstractNote={.}, number={78127812}, journal={Nature}, publisher={Nature Publishing Group}, author={Jia, Jayson S. and Lu, Xin and Yuan, Yun and Xu, Ge and Jia, Jianmin and Christakis, Nicholas A.}, year={2020}, month={06}, pages={389–394} }

@article{kissler_reductions_2020, title={Reductions in commuting mobility correlate with geographic differences in SARS-CoV-2 prevalence in New York City}, volume={11}, ISSN={2041-1723}, DOI={10.1038/s41467-020-18271-5}, abstractNote={.}, number={1}, journal={Nature Communications}, author={Kissler, Stephen M. and Kishore, Nishant and Prabhu, Malavika and Goffman, Dena and Beilin, Yaakov and Landau, Ruth and Gyamfi-Bannerman, Cynthia and Bateman, Brian T. and Snyder, Jon and Razavi, Armin S. and et al.}, year={2020}, month={12}, pages={4674} }

@article{nouvelle_reduction_2021, title={Reduction in mobility and COVID-19 transmission}, volume={12}, ISSN={2041-1723}, DOI={10.1038/s41467-021-21358-2}, abstractNote={.}, number={1}, journal={Nature Communications}, author={Nouvellet, Pierre and Bhatia, Sangeeta and Cori, Anne and Ainslie, Kylie E. C. and Baguelin, Marc and Bhatt, Samir and Boonyasiri, Adhiratha and Brazeau, Nicholas F. and Cattarino, Lorenzo and Cooper, Laura V. and et al.}, year={2021}, month={12}, pages={1090} }

@article{di_porto_partial_nodate,
  title = {Partial Lockdown and the Spread of {{Covid}}-19: Lessons from the {{Italian}} Case},
  author = {di Porto, Edoardo and Naticchioni, Paolo and Scrutinio, Vincenzo},
  pages = {27},
  langid = {english},
  options = {useprefix=true}
}

\end{document}


\begin{center}
\Large{\sc{Effects of mobility restrictions during COVID19 in Italy - Supplementary}}
\vspace{0.3cm}

\small{Alex Smolyak$^{1,*}$, Giovanni Bonaccorsi$^{2}$, Andrea Flori$^{2}$, Fabio Pammolli$^{2,3}$  Shlomo Havlin$^1$}

\vspace{0.3cm}

\footnotesize{$^1$ Department of Physics, Bar-Ilan University, Ramat-Gan 52900, Israel; \\
\footnotesize{$^2$ Impact, Department of Management, Economics and Industrial Engineering, Politecnico di Milano}\\
\footnotesize{$^3$ SIT, Schaffhausen Institute of Technology, Schaffhausen}\\
                    $^*$ Correspondence to alex.smolyak@gmail.com\\}

\end{center}

\setcounter{page}{1} \renewcommand{\thepage}{\arabic{chapter}.\arabic{page}}  
\renewcommand{\thesection}{\arabic{chapter}.\arabic{section}}   
\renewcommand{\thetable}{\arabic{chapter}.\arabic{table}}   
\setcounter{figure}{0} \renewcommand{\thefigure}{\arabic{chapter}.\arabic{figure}}

\renewcommand{\thepage}{S\arabic{page}}  
\renewcommand{\thesection}{S\arabic{section}}   
\renewcommand{\thetable}{S\arabic{table}}   
\renewcommand{\thefigure}{S\arabic{figure}}

\begin{enumerate}[wide, labelwidth=!, labelindent=0pt]
\item {\bf{Additional Network statistics}\label{sec:netstats}}\\

Our data consists of tiles, as defined by Facebook which we identify as nodes, and the people traveling between those tiles form the edges. There are app. 5000 nodes or tiles throughout our data set, covering about 80\% of Italy's nominal territory. Of those, approximately 3700 take part in non-local interactions (i.e. where there is mobility between the tiles and not only within). Fig.~\ref{fig:JointWeightDist} details the joint distributions of the edge weights and distances aggregated for the periods defined in the text.

\begin{figure*}[h!]
   \centering
   \hspace{0.00mm}
   \subfloat{\includegraphics[width=.48\textwidth,valign=t]{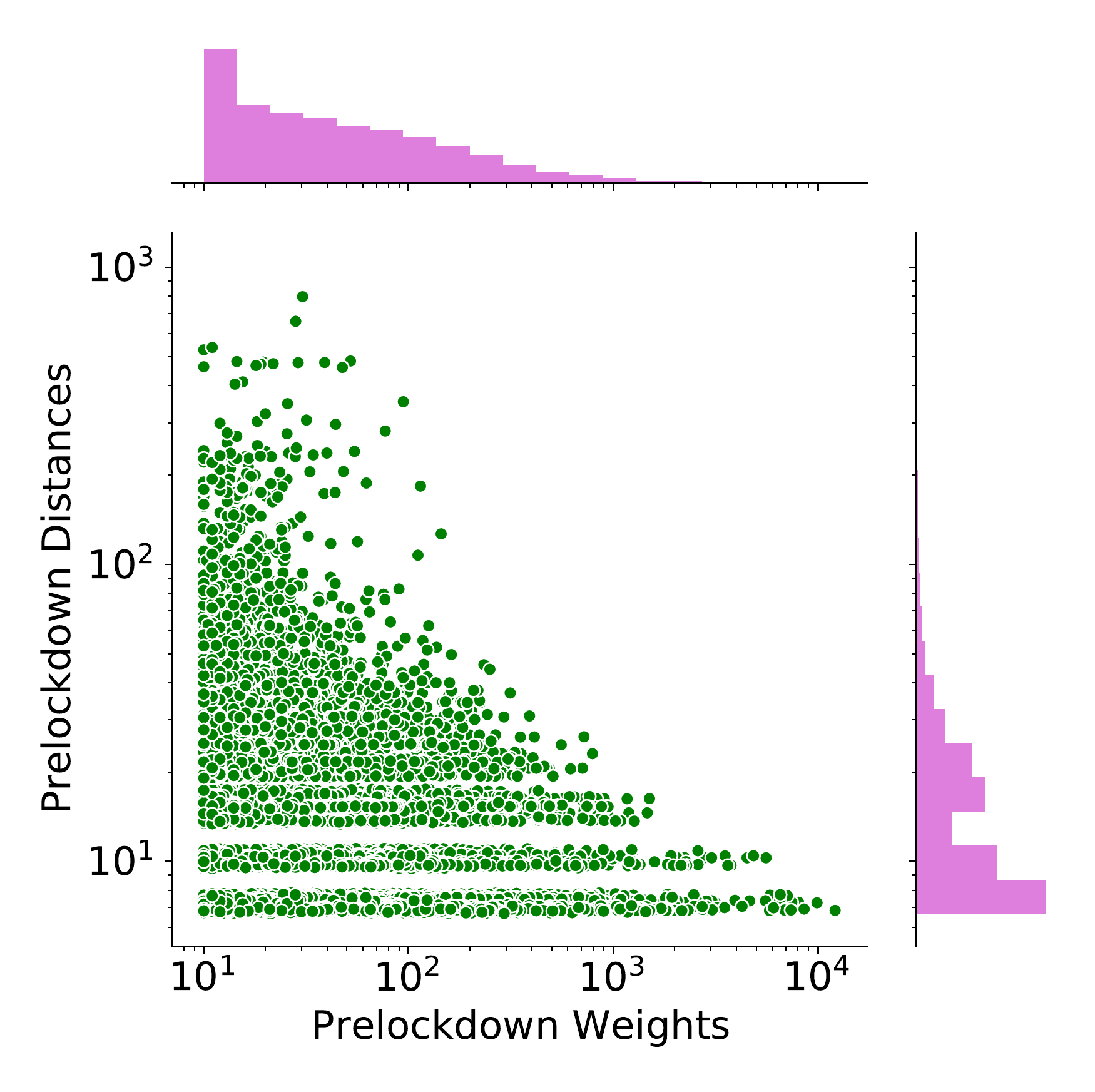}\label{s11}}
   \hspace{0.00mm}
   \subfloat{\includegraphics[width=.48\textwidth,valign=t]{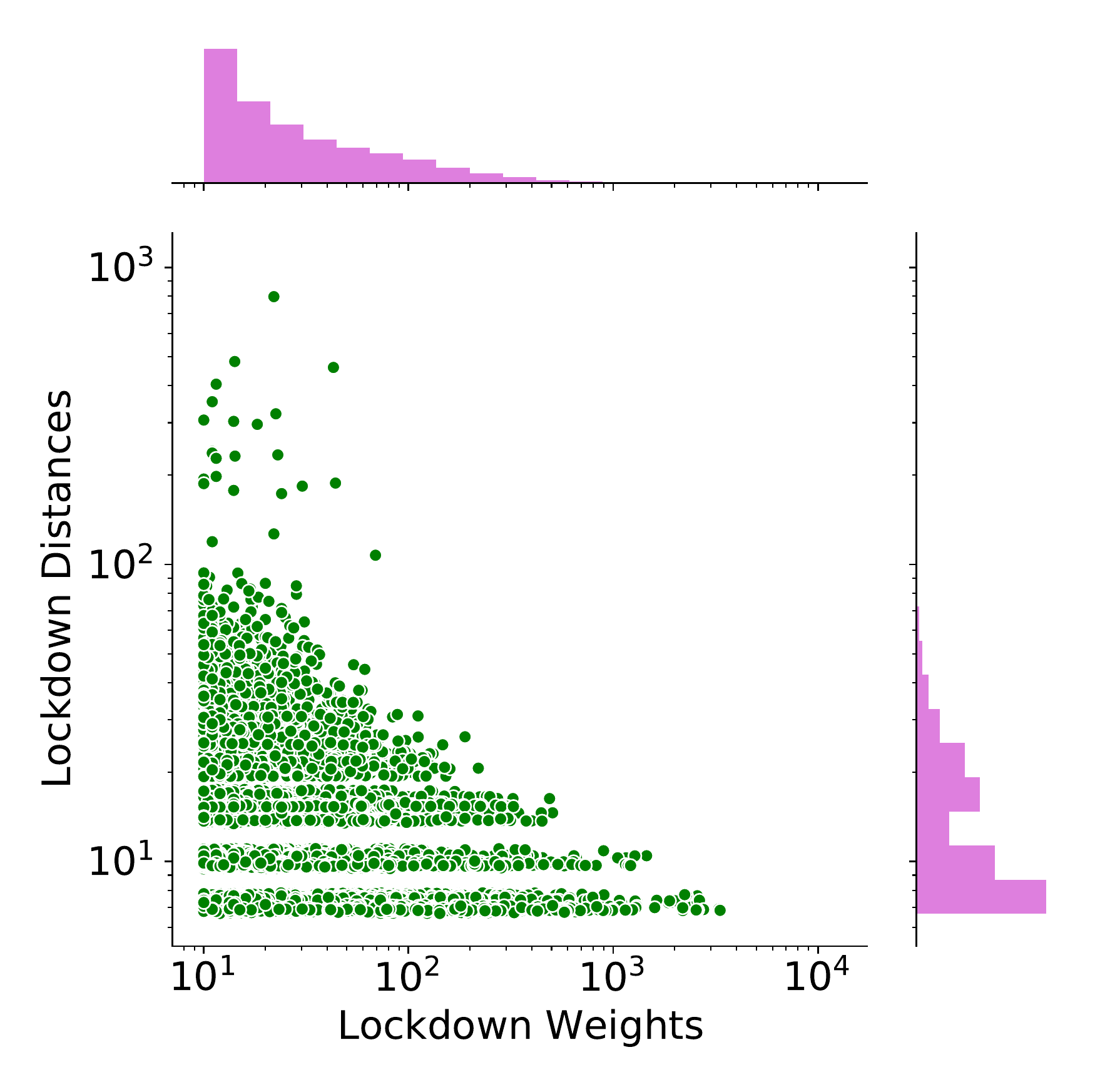}\label{s12}}   
   \hspace{0.00mm}
   \subfloat{\includegraphics[width=.48\textwidth,valign=t]{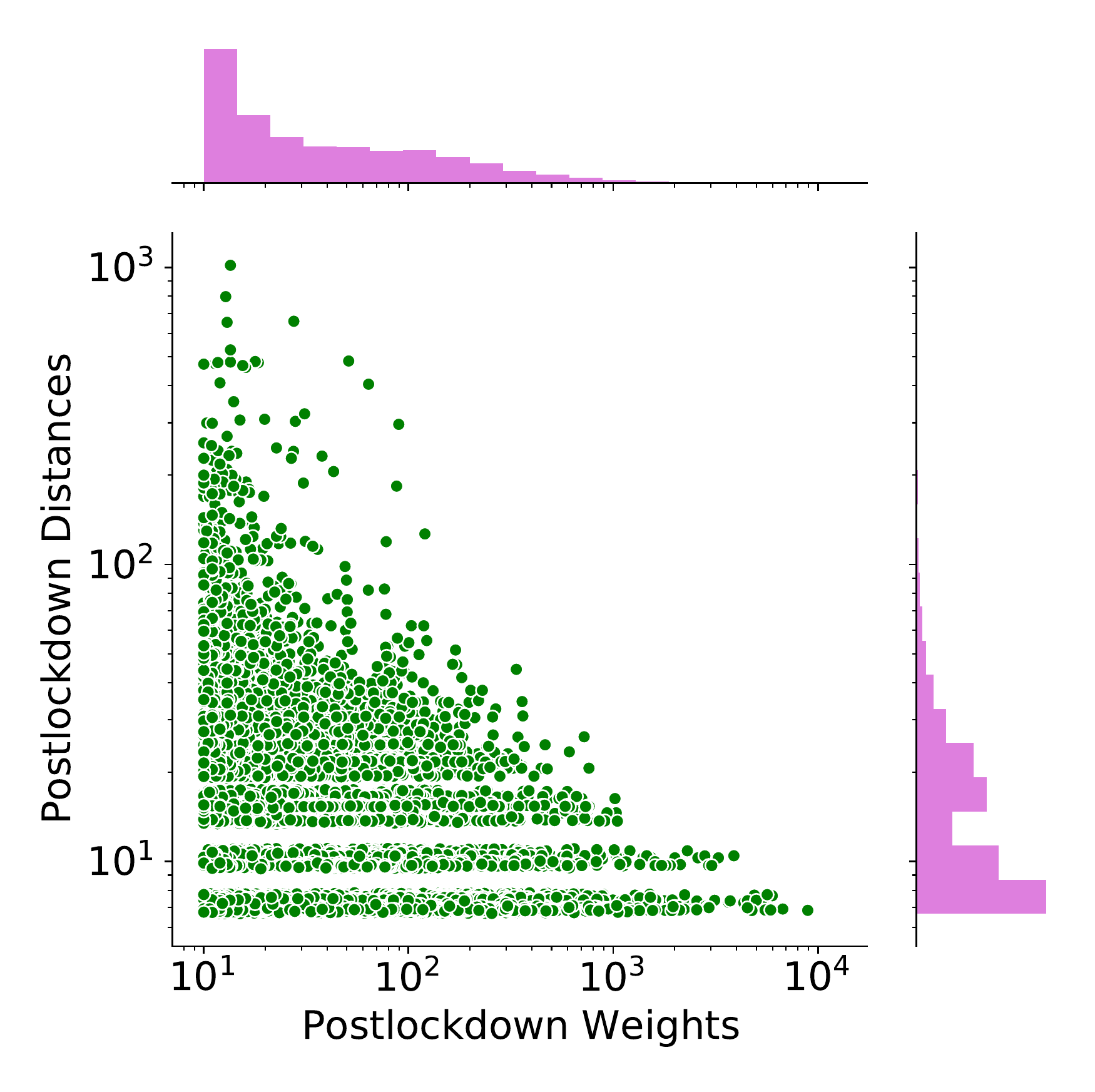}\label{s13}}
   \hspace{0.00mm}
   \subfloat{\includegraphics[width=.48\textwidth,valign=t]{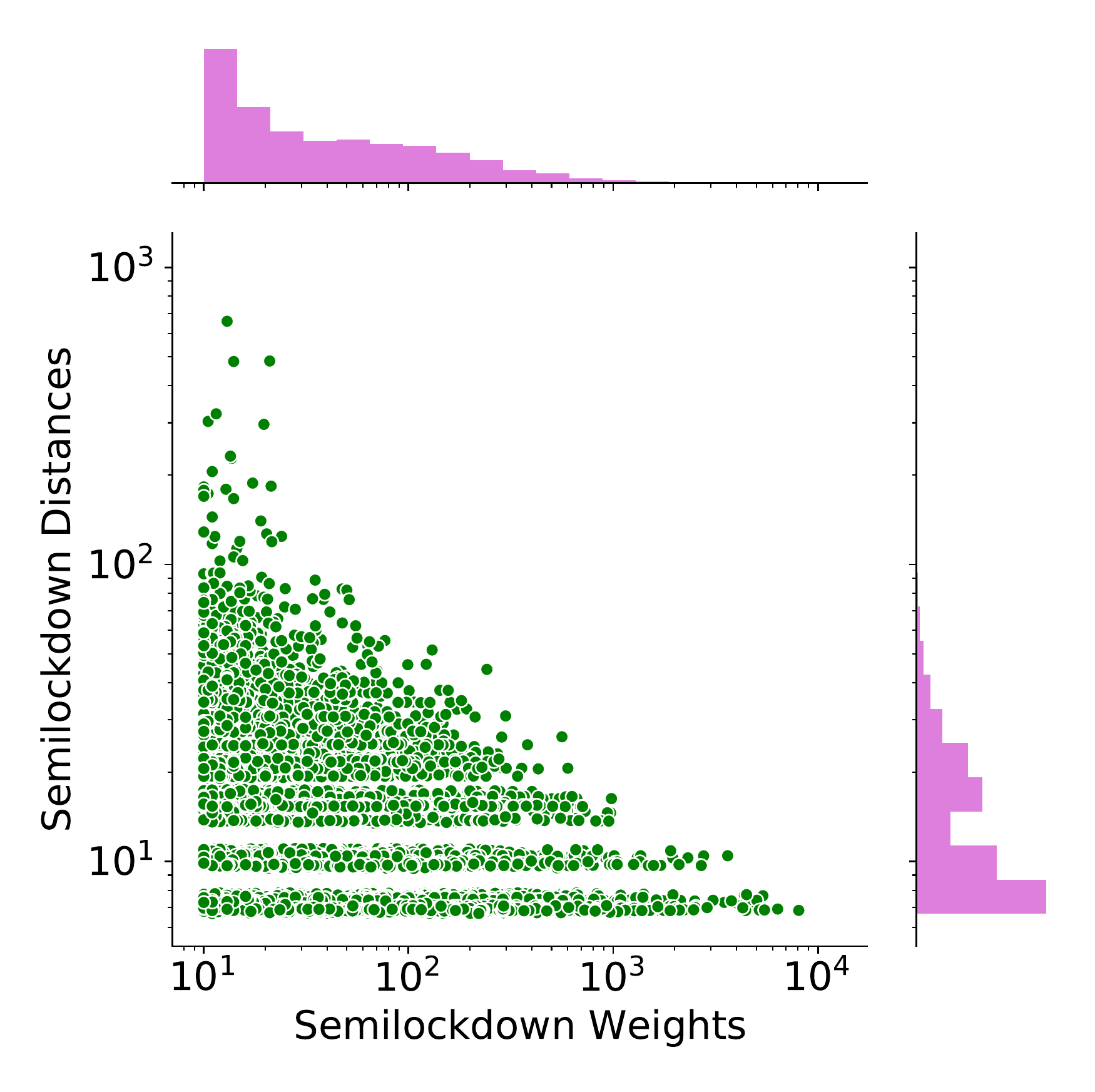}\label{s14}}

   \caption{\footnotesize Joint distributions of weights and distances. Clockwise from top left, the distribution before lockdown, under full restrictions, with partial restrictions lifted and after the full lift of limitations.}
   \label{fig:JointWeightDist}
\end{figure*}

\item {\bf{Numerical fits}\label{sec:theofit}}\\

While the distribution of the distances shows an approximate fit to a power law distribution over several orders of magnitude, the case for the weighted degree distribution is not as evident. Another possible candidate is the Stretched Exponential (Weibull) distribution, with the CDF parameterised as $F=1-e^{-(\lambda x)^\beta}$. Previous research suggested mechanisms for the existence of such distributions in empirical data (c.f. Refs.~\cite{laherrere1998stretched,frisch1997extreme}, and empirical experiments in Ref.~\cite{alstott2014powerlaw})

\begin{figure*}[h!]
   \centering
   \hspace{0.00mm}
   \subfloat{\includegraphics[width=.75\textwidth,valign=t]{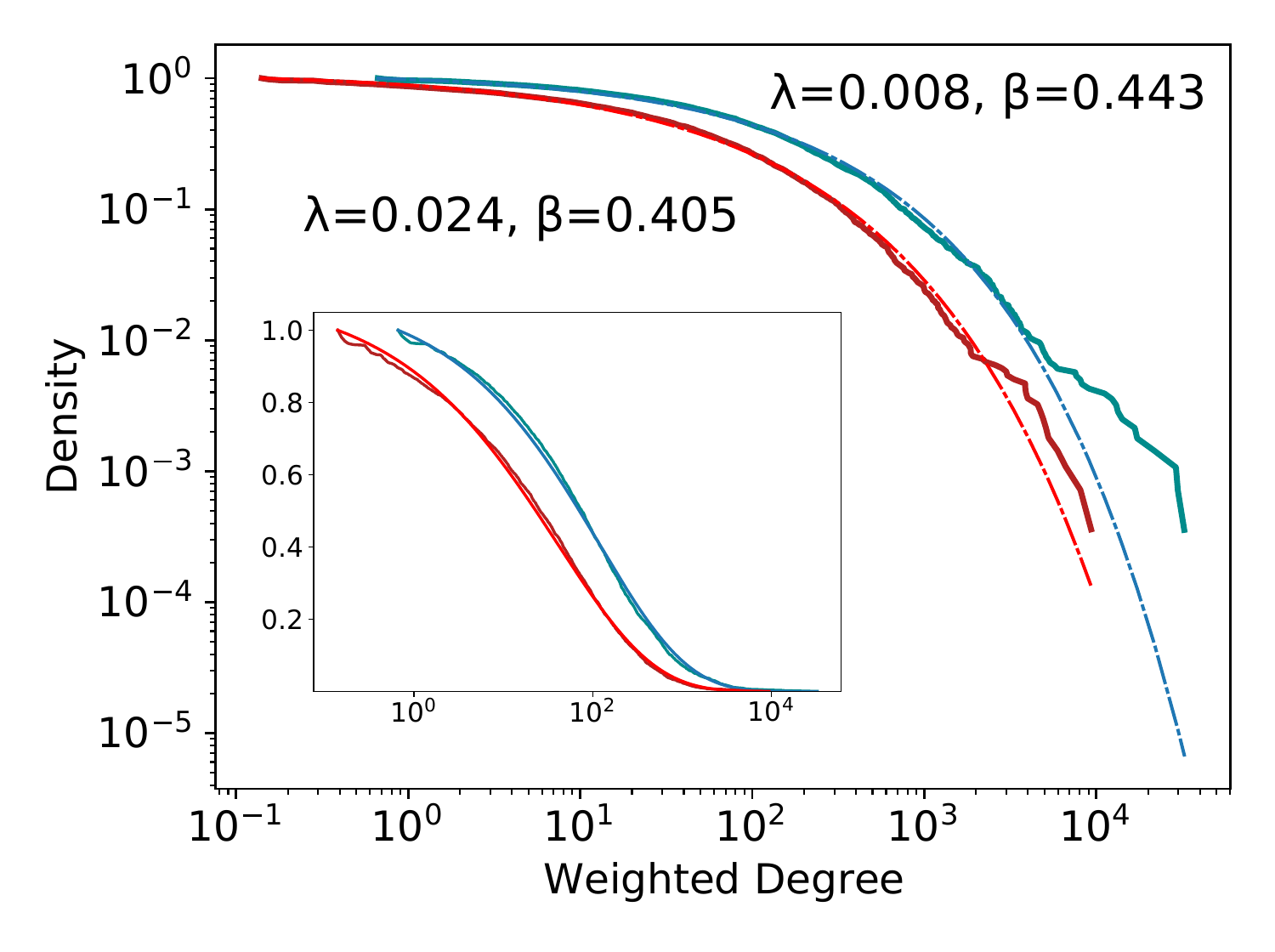}\label{s21}}
   \caption{\footnotesize The log-log Complementary CDF plot of the weighted degree distribution with a stretched exponential fit (insert - semilog plot of the same distribution, showing the quality of the fit for the low values as well)}
   \label{fig:StretchedExpFit}
\end{figure*}

\item {\bf{Impact distribution}\label{sec:impdist}}\\

The panels in Fig.~\ref{fig:ImpactScatter} in the main text show the quality of the linear fit for the impact incurred by all provinces. In order to expand on the inhomogeneity that \textit{is} present, we bring here (Fig.~\ref{fig:ImpactDist}) the distributions of each of the periods' impact. A typical impact is seen in all three relatively narrow distributions, however, values around the peak show the provinces deviating from the linear relation. \\

\begin{figure*}[h!]
   \centering
   \hspace{0.00mm}
   \subfloat{\includegraphics[width=.75\textwidth,valign=t]{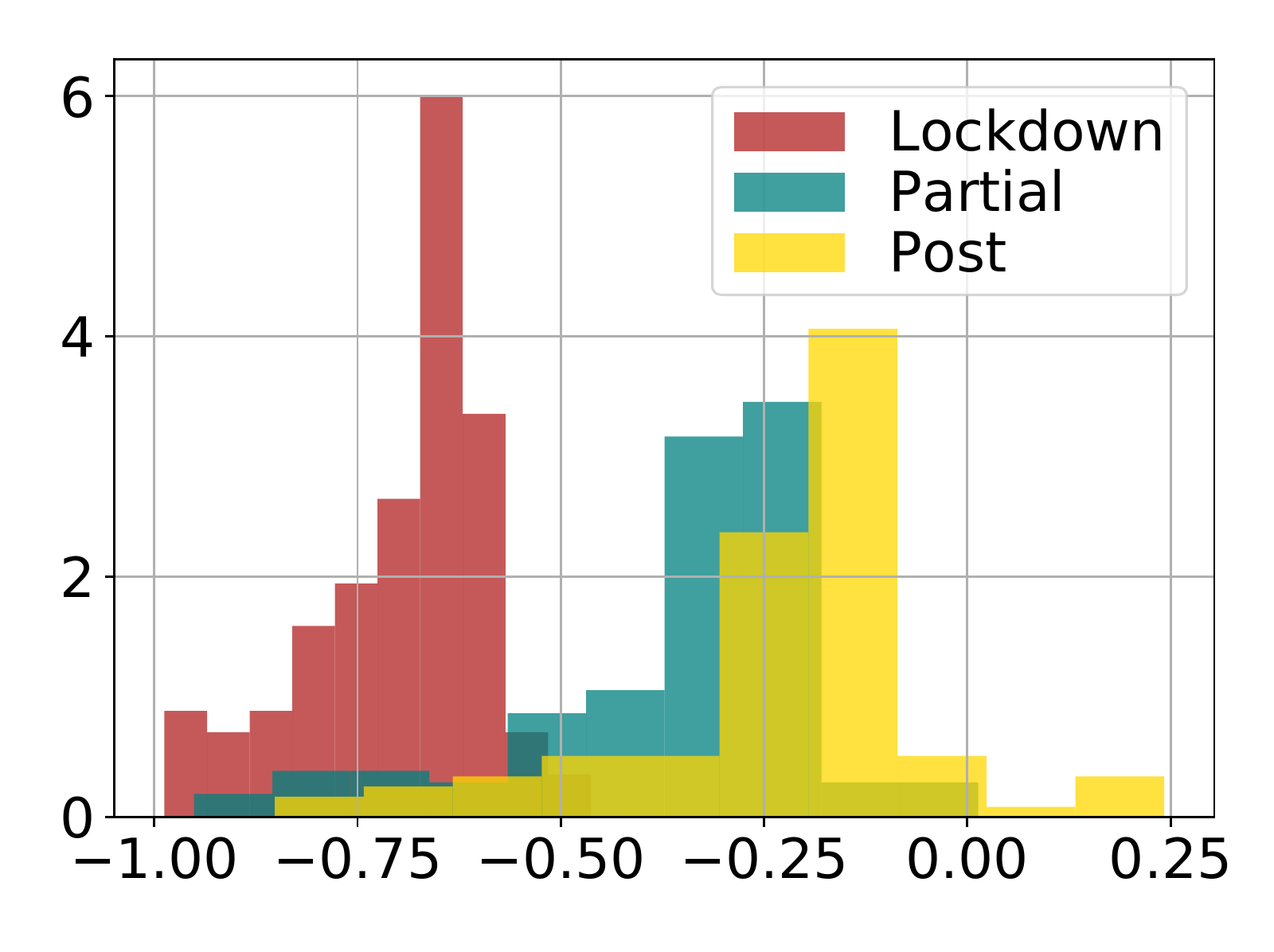}\label{s3}}
   \caption{\footnotesize The colors, matching the color scheme in the main text, show the impact is indeed centered around similar values for the lockdown and following stages, but those values are not the whole story. Specifically, for the post-lockdown stages, the distribution is broader, and thus, as the main text explains, is the variation in the economic impact.}
   \label{fig:ImpactDist}
\end{figure*}

\item {\bf{Failing edges as criticality}\label{sec:critics}}\\

As was shown in Fig.~\ref{sec:Resilinece} in the main text, the critical levels $q_c$ for for the fragmentation of the GCC during lockdown are around the value of 0.2. Here we show the failing edges and their respective weights during each of the reported periods. As Fig.~\ref{fig:FailingEdges} shows, both the number of nodes for which some of the edges fail, and the magnitude of failure (i.e. how many edges each node loses) highlight the loss of edges leading to the fragmentation. Top left panel (pre-lockdown) reveals a very strong network where the relatively low levels of $q_c$ inflict almost no damage. During lockdown, as can be seen in the top right figure, and summarized in panel~\subref{6a} of Fig.~\ref{fig:Resilience}, multiple edges fail, many of them carrying substantial weight. Lastly, we note again the qualitative difference between the post-lockdown state as compared to the pre-lockdown, where we see many more edges failing at the same, previously almost imperceptible, level.

\begin{figure*}[h!]
   \centering
   \hspace{0.00mm}
   \subfloat{\includegraphics[width=.45\textwidth,valign=t]{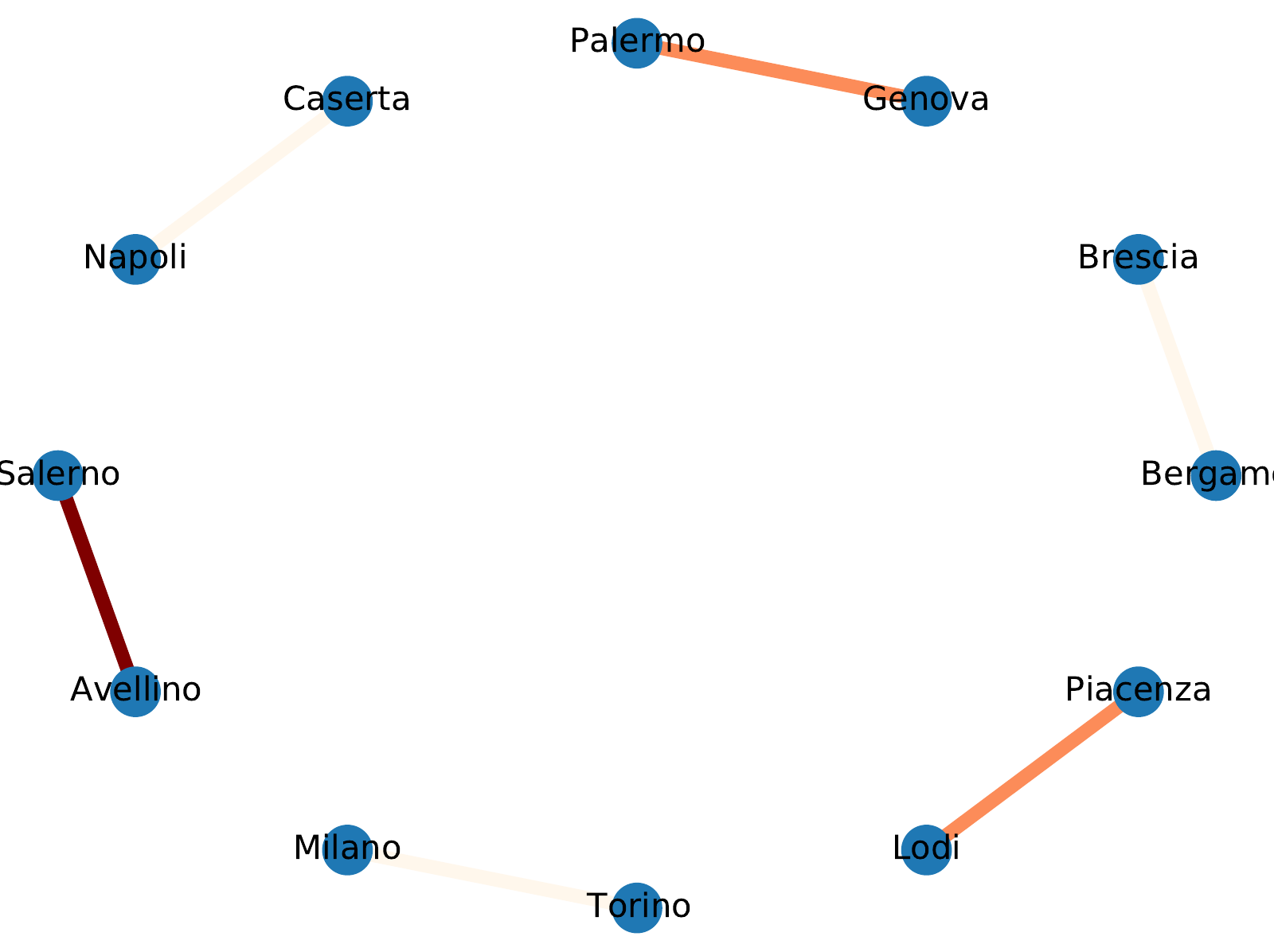}\label{s41}}   
   \hspace{0.00mm}
   \subfloat{\includegraphics[width=.45\textwidth,valign=t]{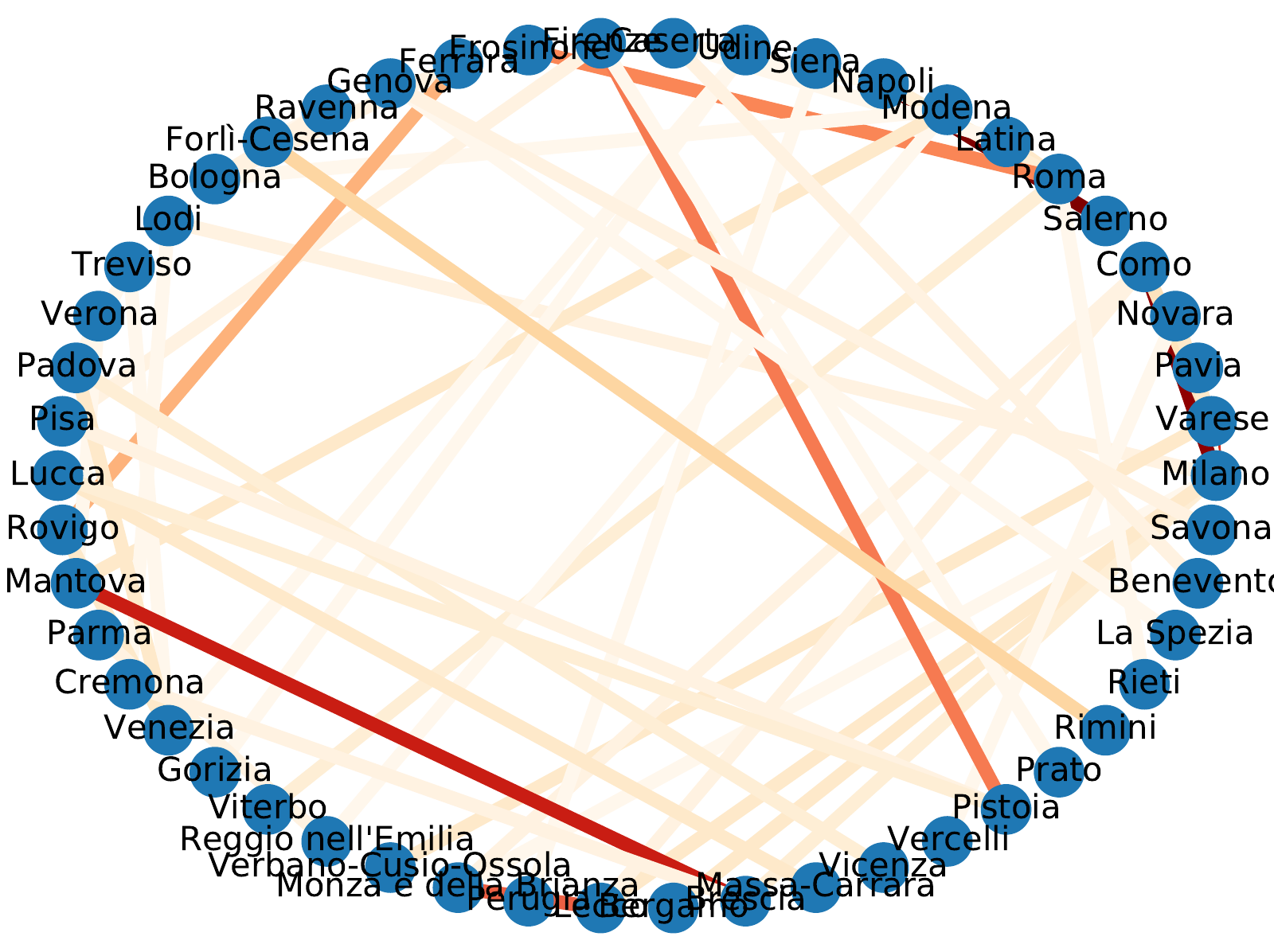}\label{s42}}
   \hspace{0.00mm}
   \subfloat{\includegraphics[width=.45\textwidth,valign=t]{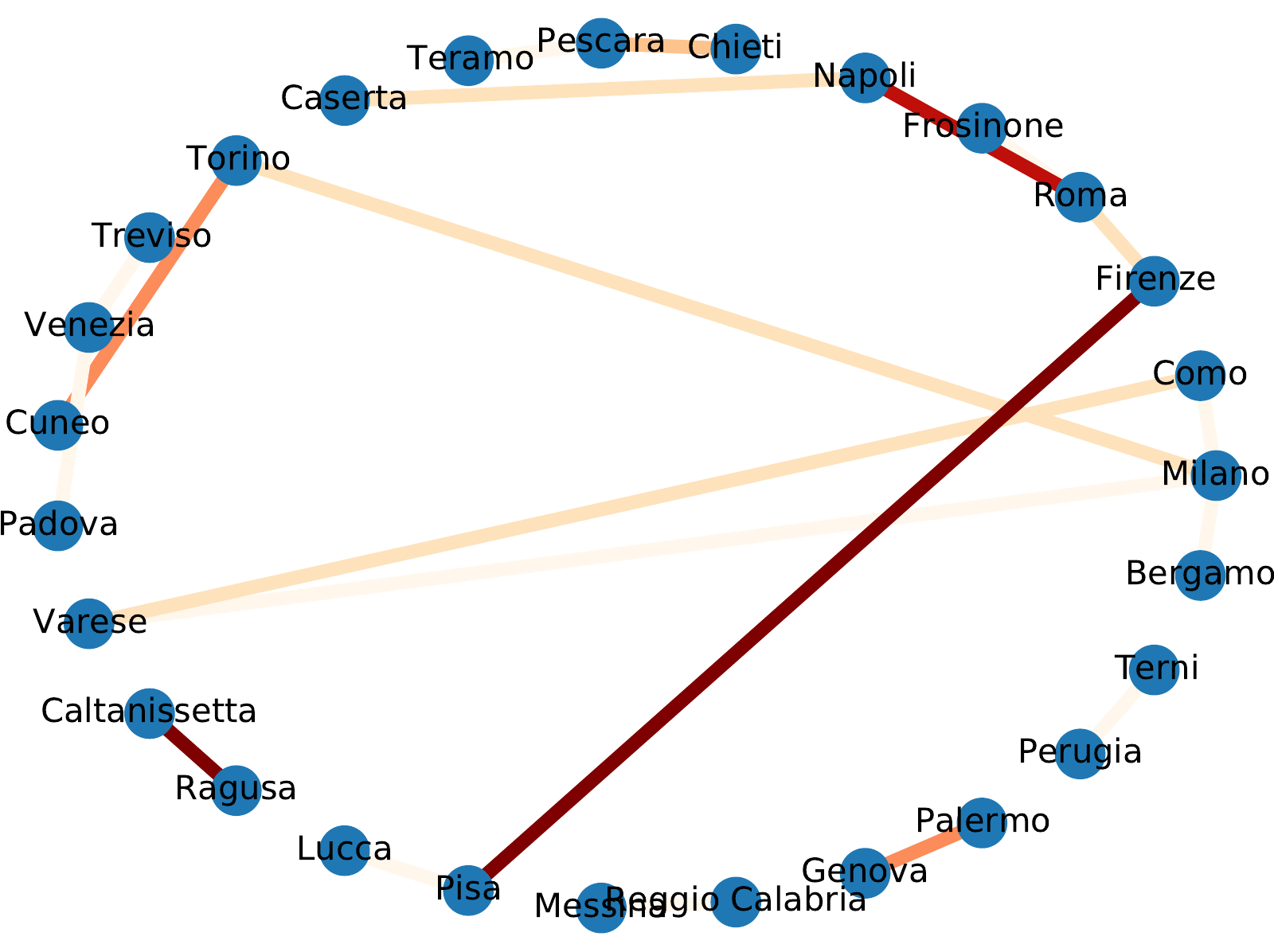}\label{s43}}
   \hspace{0.00mm}
   \subfloat{\includegraphics[width=.45\textwidth,valign=t]{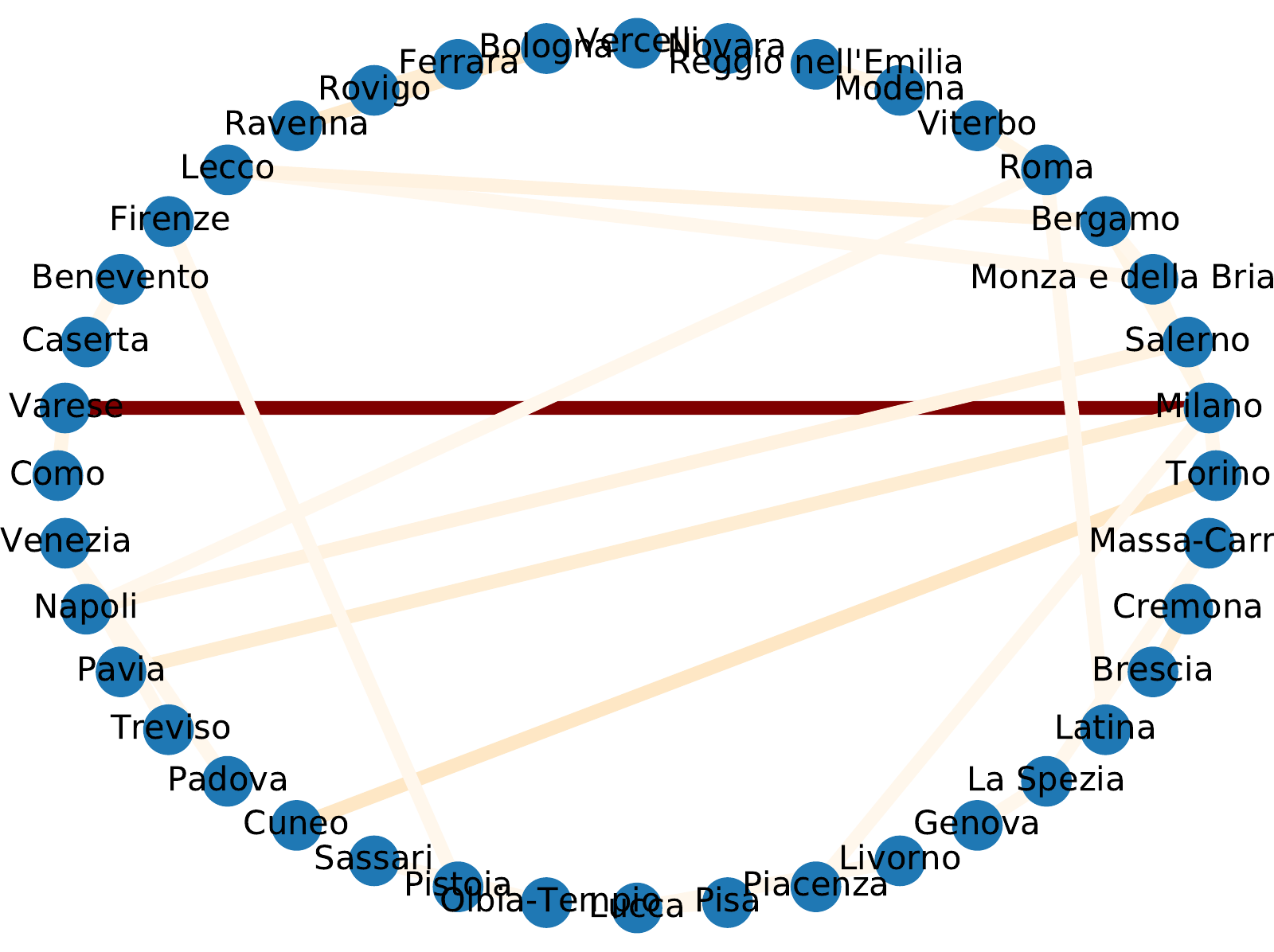}\label{s44}}   	
   \caption{\footnotesize As before, clockwise from top left, the pre-lockdown, lockdown, partial limitations and post lockdown states. The edges in the figure are the ones removed at this level of $q_c=0.22$. Edge color is relative to the other edges for the same period. }
   \label{fig:FailingEdges}
\end{figure*}
\clearpage
\item \printbibliography 
\end{enumerate}